\documentclass[11pt, a4paper]{article}

\makeatletter%

\usepackage{ifthen}



\makeatletter%

\newcommand{\MyUniPat}{lsdfgkhjvrkjlhmisdlcjn}

\makeatletter

\newcommand{\ifndef}[2]{\@ifundefined{#1}{#2}{}}

\newcommand{\NewcommandThreeArgsTwoOpt}[5]{
\DeclareRobustCommand#1{\@ifnextchar[%
{\csname\expandafter\@gobble\string#1@presq\endcsname}%
{\csname\expandafter\@gobble\string#1@nopresq\endcsname}}
\expandafter\def\csname\expandafter\@gobble\string#1@nopresq\endcsname##1{\@ifnextchar[%
{\csname\expandafter\@gobble\string#1@nopresq@postsq\endcsname[]{##1}}%
{\csname\expandafter\@gobble\string#1@nopresq@nopostsq\endcsname[]{##1}}}
\expandafter\def\csname\expandafter\@gobble\string#1@presq\endcsname[##1]##2{\@ifnextchar[%
{\csname\expandafter\@gobble\string#1@presq@postsq\endcsname[{##1}]{##2}}%
{\csname\expandafter\@gobble\string#1@presq@nopostsq\endcsname[{##1}]{##2}}}
\expandafter\def\csname\expandafter\@gobble\string#1@nopresq@nopostsq\endcsname[##1]##2{#2}
\expandafter\def\csname\expandafter\@gobble\string#1@presq@nopostsq\endcsname[##1]##2{#3}
\expandafter\def\csname\expandafter\@gobble\string#1@nopresq@postsq\endcsname[##1]##2[##3]{#4}
\expandafter\def\csname\expandafter\@gobble\string#1@presq@postsq\endcsname[##1]##2[##3]{#5}
}

\newcommand{\NewcommandFourArgsTwoOpt}[5]{
\DeclareRobustCommand#1{\@ifnextchar[%
{\csname\expandafter\@gobble\string#1@presq\endcsname}%
{\csname\expandafter\@gobble\string#1@nopresq\endcsname}}
\expandafter\def\csname\expandafter\@gobble\string#1@nopresq\endcsname##1##2{\@ifnextchar[%
{\csname\expandafter\@gobble\string#1@nopresq@postsq\endcsname[]{##1}{##2}}%
{\csname\expandafter\@gobble\string#1@nopresq@nopostsq\endcsname[]{##1}{##2}}}
\expandafter\def\csname\expandafter\@gobble\string#1@presq\endcsname[##1]##2##3{\@ifnextchar[%
{\csname\expandafter\@gobble\string#1@presq@postsq\endcsname[{##1}]{##2}{##3}}%
{\csname\expandafter\@gobble\string#1@presq@nopostsq\endcsname[{##1}]{##2}{##3}}}
\expandafter\def\csname\expandafter\@gobble\string#1@nopresq@nopostsq\endcsname[##1]##2##3{#2}
\expandafter\def\csname\expandafter\@gobble\string#1@presq@nopostsq\endcsname[##1]##2##3{#3}
\expandafter\def\csname\expandafter\@gobble\string#1@nopresq@postsq\endcsname[##1]##2##3[##4]{#4}
\expandafter\def\csname\expandafter\@gobble\string#1@presq@postsq\endcsname[##1]##2##3[##4]{#5}
}

\makeatother

\newcommand{\nospell}[1]{#1}  %

\makeatletter
\NewcommandThreeArgsTwoOpt{\myusepackage}
{\@ifpackageloaded{#2}{}{\usepackage{#2}}}
{\@ifpackageloaded{#2}{}{\usepackage[#1]{#2}}}
{\@ifpackageloaded{#2}{}{#3 \usepackage{#2}}}
{\@ifpackageloaded{#2}{}{#3 \usepackage[#1]{#2}}}

\makeatother

\myusepackage{amssymb}
\myusepackage{amsmath}  %

\myusepackage{amsthm}[ ]  %

{}

{}\myusepackage[T1,OT2,T2A]{fontenc}{}
\DeclareTextSymbolDefault{\CYRYAT}{OT2}
\DeclareTextSymbolDefault{\cyryat}{OT2}
\DeclareTextSymbolDefault{\CYRFITA}{OT2}
\DeclareTextSymbolDefault{\cyrfita}{OT2}
\DeclareTextSymbolDefault{\CYRIZH}{OT2}
\DeclareTextSymbolDefault{\cyrizh}{OT2}
\myusepackage[utf8]{inputenc}
\let\f\relax

{}
{}
\usepackage[greek,ukrainian,russian,english]{babel}
{}
{} %

\myusepackage{soul}      %
\setul{1.25pt}{.16pt}  %
\myusepackage{latexsym}
\myusepackage{amsfonts}
\myusepackage{units}    %
\newcommand{\dr}{\nicefrac}
\myusepackage{psfrag}   %
\myusepackage[hyphens,spaces,obeyspaces]{url}   %
{}{}
\myusepackage{hyperref}  %
\setpdflinkmargin{0.75pt}  %

\myusepackage[usenames,dvipsnames]{xcolor}
\hypersetup{
colorlinks,
linkcolor={red!55!black},
citecolor={blue!55!black},
urlcolor={blue!35!black}
}

\myusepackage{hyphenat}  %
\myusepackage{caption}   %
\myusepackage{microtype}   %
\myusepackage{marginnote}  %
\myusepackage{calc}  %
\myusepackage{mathrsfs}   %
{}  %
\myusepackage{enumitem}  %
\myusepackage{graphicx}
{}  %
{} %

{}     %
\newcommand{\dgCapDefinition}{Definition}
\newcommand{\dgCapDefinitions}{Definitions}
\newcommand{\dgCapPostulate}{Postulate}
\newcommand{\dgCapPostulates}{Postulates}
\newcommand{\dgCapExample}{Example}

\newcommand{\dgCapFact}{Fact}
\newcommand{\dgCapFacts}{Facts}
\newcommand{\dgCapQuestion}{Question}
\newcommand{\dgCapQuestions}{Questions}
\newcommand{\dgCapLemma}{Lemma}
\newcommand{\dgCapLemmas}{Lemmas}
\newcommand{\dgCapNotation}{Notation}
\newcommand{\dgCapCorollary}{Corollary}
\newcommand{\dgCapCorollaries}{Corollaries}
\newcommand{\dgCapProposition}{Proposition}
\newcommand{\dgCapPropositions}{Propositions}
\newcommand{\dgCapClaim}{Claim}
\newcommand{\dgCapClaims}{Claims}
\newcommand{\dgCapTheorem}{Theorem}
\newcommand{\dgCapTheorems}{Theorems}
\newcommand{\dgCapProblem}{Problem}
\newcommand{\dgCapProblems}{Problems}
\newcommand{\dgCapRemark}{Remark}
\newcommand{\dgCapRemarks}{Remarks}
\newcommand{\dgCapConjecture}{Conjecture}
\newcommand{\dgCapConjectures}{Conjectures}
\newcommand{\dgCapResult}{Result}

\newcommand{\dgCapChapter}{Chapter}
\newcommand{\dgCapChapters}{Chapters}
\newcommand{\dgCapSection}{Section}
\newcommand{\dgCapSections}{Sections}
\newcommand{\dgCapSubsection}{Subsection}
\newcommand{\dgCapSubsections}{Subsections}
\newcommand{\dgCapFigure}{Figure}
\newcommand{\dgCapFigures}{Figures}
\newcommand{\dgCapEquation}{Equation}
\newcommand{\dgCapEquations}{Equations}
\newcommand{\dgCapExpression}{Expression}
\newcommand{\dgCapExpressions}{Expressions}
\newcommand{\dgCapInequality}{Inequality}
\newcommand{\dgCapInequalities}{Inequalities}
\newcommand{\dgProofOf}{\proofname\ of}
{}
\newcommand{\dgDefinition}{Definition}
\newcommand{\dgDefinitions}{Definitions}
\newcommand{\dgPostulate}{Postulate}
\newcommand{\dgPostulates}{Postulates}

\newcommand{\dgFact}{Fact}
\newcommand{\dgFacts}{Facts}
\newcommand{\dgQuestion}{Question}
\newcommand{\dgQuestions}{Questions}
\newcommand{\dgLemma}{Lemma}
\newcommand{\dgLemmas}{Lemmas}
\newcommand{\dgCorollary}{Corollary}
\newcommand{\dgCorollaries}{Corollaries}
\newcommand{\dgProposition}{Proposition}
\newcommand{\dgPropositions}{Propositions}
\newcommand{\dgClaim}{Claim}
\newcommand{\dgClaims}{Claims}
\newcommand{\dgTheorem}{Theorem}
\newcommand{\dgTheorems}{Theorems}
\newcommand{\dgProblem}{Problem}
\newcommand{\dgProblems}{Problems}
\newcommand{\dgRemark}{Remark}
\newcommand{\dgRemarks}{Remarks}
\newcommand{\dgConjecture}{Conjecture}
\newcommand{\dgConjectures}{Conjectures}

\newcommand{\dgChapter}{Chapter}
\newcommand{\dgChapters}{Chapters}
\newcommand{\dgSection}{Section}
\newcommand{\dgSections}{Sections}
\newcommand{\dgSubsection}{Subsection}
\newcommand{\dgSubsections}{Subsections}
\newcommand{\dgFigure}{Figure}
\newcommand{\dgFigures}{Figures}
\newcommand{\dgEquation}{Equation}
\newcommand{\dgEquations}{Equations}
\newcommand{\dgExpression}{Expression}
\newcommand{\dgExpressions}{Expressions}
\newcommand{\dgInequality}{Inequality}
\newcommand{\dgInequalities}{Inequalities}
{}
{}

{}
\ifndef{theorem}{}
\ifndef{theorem*}{\newtheorem*{theorem*}{\dgCapTheorem}}
{}

{}  %
{}  %
\ifndef{lemma}{}
\ifndef{lemma*}{\newtheorem*{lemma*}{\dgCapLemma}}
\ifndef{corollary}{}
\ifndef{corollary*}{\newtheorem*{corollary*}{\dgCapCorollary}}
\ifndef{conjecture}{}
\ifndef{conjecture*}{\newtheorem*{conjecture*}{\dgCapConjecture}}
\ifndef{proposition}{}
\ifndef{proposition*}{\newtheorem*{proposition*}{\dgCapProposition}}
\ifndef{claim}{}
\ifndef{claim*}{\newtheorem*{claim*}{\dgCapClaim}}
\ifndef{result}{}
\ifndef{result*}{\newtheorem*{result*}{\dgCapResult}}
\ifndef{problem}{}
\ifndef{problem*}{\newtheorem*{problem*}{\dgCapProblem}}
{}  %
{}  %

\newtheoremstyle{mydefinition}  %
{\topsep}{\topsep}  %
{\slshape}  %
{}  %
{\bfseries}  %
{.}  %
{ }  %
{}  %

\newtheoremstyle{mynotation}  %
{\topsep}{\topsep}  %
{}  %
{}  %
{\bfseries\itshape}  %
{.}  %
{ }  %
{}  %

\newtheoremstyle{myremark}  %
{\topsep}{\topsep}  %
{\slshape}  %
{}  %
{\bfseries\itshape}  %
{.}  %
{ }  %
{\thmname{#1}\thmnumber{~#2}}  %

\newtheoremstyle{myexample}  %
{\topsep}{\topsep}  %
{\itshape}  %
{}  %
{\slshape}  %
{.}  %
{ }  %
{\underline{\thmname{#1}\thmnumber{~#2}}}  %

\newtheoremstyle{myclaims}  %
{\topsep}{\topsep}  %
{\slshape}  %
{}  %
{\bfseries\slshape}  %
{.}  %
{ }  %
{\thmname{#1}\thmnumber{~#2}\thmnote{\textnormal{~(#3)}}}  %

{} %
\ifndef{notation}{\theoremstyle{mynotation}}
{}
{\theoremstyle{myremark}

\newtheorem*{myremark*}{\dgCapRemark}
}
{} %
{\theoremstyle{mydefinition}
\ifndef{postulate}{}
\ifndef{postulate*}{\newtheorem*{postulate*}{\dgCapPostulate}}
\ifndef{definition}{}
\ifndef{definition*}{\newtheorem*{definition*}{\dgCapDefinition}}
}
{\theoremstyle{myexample}
\ifndef{example}{}
\ifndef{example*}{\newtheorem*{example*}{\dgCapExample}}
}
{\theoremstyle{myclaims}

\newtheorem*{my_claim*}{\dgCapClaim}
\ifndef{fact}{}
\ifndef{fact*}{\newtheorem*{fact*}{\dgCapFact}}
\ifndef{question}{}
\ifndef{question*}{\newtheorem*{question*}{\dgCapQuestion}}
}
{} %
{} %
{} %

{}
\newtheoremstyle{anystatementst}  %
{\topsep}{\topsep}  %
{\itshape}  %
{}  %
{\bfseries}  %
{.}  %
{ }  %
{#3}  %

{\theoremstyle{anystatementst} }

{} %

\newcommand{\mydef}[2]{\def#1{#2}}

\newcommand{\newident}[3][\MyUniPat]{\ifthenelse{\equal{\MyUniPat}{#1}}
{
\newcommand{#2}[1][]{\ensuremath{\mathit{#3##1}}}
}
{\ifthenelse{\equal{}{#1}}
{
\newcommand{#2}[1][]{\ensuremath{\mathit{#3}}}
}
{
\newcommand{#2}[1][\MyUniPat]{\ifthenelse{\equal{\MyUniPat}{##1}}%
{\ensuremath{\mathit{#1}}}%
{\ensuremath{\mathit{#3}}}}
}
}
}

\newcommand{\newidenT}[3][\MyUniPat]{\ifthenelse{\equal{\MyUniPat}{#1}}
{
\newcommand{#2}[1][\MyUniPat]{\ifthenelse{\equal{\MyUniPat}{##1}}%
{\il{#3}}%
{\ensuremath{\mathit{#3##1}}}}
}
{
\newcommand{#2}[1][\MyUniPat]{\ifthenelse{\equal{\MyUniPat}{##1}}%
{\il{#1}}%
{\ensuremath{\mathit{#3}}}}
}
}

\newcommand{\newmat}[3][\MyUniPat]{\ifthenelse{\equal{\MyUniPat}{#1}}%
{\newcommand{#2}[1][]{#3##1}}%
{\newcommand{#2}[1][]{#3}}%
}

\newcommand{\providemat}[3][\MyUniPat]{\ifthenelse{\equal{\MyUniPat}{#1}}
{\providecommand{#2}[1][]{#3##1}}
{\providecommand{#2}[1][]{#3}}  %
}

\newcommand{\newmatop}[3][\MyUniPat]{\ifthenelse{\equal{\MyUniPat}{#1}}
{
\newcommand{#2}{\operatorname{#3}}
}
{
\newcommand{#2}[1][\MyUniPat]{\ifthenelse{\equal{\MyUniPat}{##1}}%
{\operatorname{#1}}%
{\operatorname{#3}}}
}
}

\makeatletter
\newcommand{\MyMakeTheoMacros}[3]{
\expandafter\newcommand\csname\expandafter\@gobble\string#2NostarNoname@DGaux\endcsname[2][]
{\ifthenelse{\equal{}{##1}}%
{\begin{#1}~##2 \end{#1}}%
{\begin{#1}\label{##1}~##2\end{#1}}%
}
\expandafter\newcommand\csname\expandafter\@gobble\string#2StarNoname@DGaux\endcsname[1]
{\begin{#1*}~##1 \end{#1*}}
\def#2{\expandafter\@ifstar%
\expandafter{\csname\expandafter\@gobble\string#2StarNoname@DGaux\endcsname}%
{\csname\expandafter\@gobble\string#2NostarNoname@DGaux\endcsname}%
}

\expandafter\newcommand\csname\expandafter\@gobble\string#2NostarName@DGaux\endcsname[3][]
{\ifthenelse{\equal{}{##1}}%
{\begin{#1}[\e{##2}]~##3 \end{#1}}%
{\begin{#1}[\e{##2}]\label{##1}~##3\end{#1}}%
}
\expandafter\newcommand\csname\expandafter\@gobble\string#2StarName@DGaux\endcsname[2]
{\begin{#1*}[\e{##1}]~##2 \end{#1*}}
\def#3{\expandafter\@ifstar%
\expandafter{\csname\expandafter\@gobble\string#2StarName@DGaux\endcsname}
{\csname\expandafter\@gobble\string#2NostarName@DGaux\endcsname}%
}
}
\makeatother

\makeatletter
\newtheorem*{rep@theorem}{\rep@title}
\newcommand{\newreptheorem}[2]{%
\newenvironment{rep#1}[1]{%
\def\rep@title{#2 \ref{##1}}%
\begin{rep@theorem}}%
{\end{rep@theorem}}}
\makeatother

\newcommand{\MyMakeDupTheoMacros}[7]{
\MyMakeTheoMacros{#1}{#2}{#3}
\newreptheorem{#1}{#6}
\newcommand{#4}[3]{
\newcommand{##2}{##3}
\begin{#1}\label{##1}~##2\end{#1}}
\newcommand{#5}[4]{
\newcommand{##2}{##4}
\begin{#1}{\e{##3}}\label{##1}~##2\end{#1}}
\newcommand{#7}[2]{\begin{rep#1}{##1}~##2 \end{rep#1}}
}

{}
\newcommand{\MyMakeRefMacros}[3]{\newcommand{#1}[2][]
{\ifthenelse{\equal{}{##1}}{#2~\ref{##2}}{#3~\ref{##1} and~\ref{##2}}}}

\newcommand{\MyMakeEqRefMacros}[3]{\newcommand{#1}[2][]
{\ifthenelse{\equal{}{##1}}{#2~\eqref{##2}}{#3~\eqref{##1} and~\eqref{##2}}}}
{}

{}  %
\newcommand{\bibentry}[8]{
{}\bibitem[\nospell{#8}]{#1} {\textup #3}.{}
\ifthenelse{\equal{}{#6}}
{\newblock \textrm{#4.} \newblock{\em #5}, #7....}
{\newblock \textrm{#4.} \newblock{\em #5, #6}, #7.}
}

{} %

\MyMakeTheoMacros{fact}{\fct}{\nfct}

\MyMakeRefMacros{\fctref}{\dgFact}{\dgFacts}
\MyMakeRefMacros{\Fctref}{\dgCapFact}{\dgCapFacts}

\MyMakeTheoMacros{question}{\quest}{\nquest}

\MyMakeRefMacros{\questref}{\dgQuestion}{\dgQuestions}
\MyMakeRefMacros{\Questref}{\dgCapQuestion}{\dgCapQuestions}

\MyMakeTheoMacros{notation}{\nota}{\nnota}

{}
\MyMakeDupTheoMacros{lemma}
{\lem}{\nlem}{\lemdup}{\nlemdup}{\dgLemma}{\lemrep}
{} %

\MyMakeRefMacros{\lemref}{\dgLemma}{\dgLemmas}
\MyMakeRefMacros{\Lemref}{\dgCapLemma}{\dgCapLemmas}

\MyMakeDupTheoMacros{corollary}
{\crl}{\ncrl}{\crldup}{\ncrldup}{\dgCorollary}{\crlrep}

\MyMakeRefMacros{\crlref}{\dgCorollary}{\dgCorollaries}
\MyMakeRefMacros{\Crlref}{\dgCapCorollary}{\dgCapCorollaries}

\MyMakeTheoMacros{proposition}{\prp}{\nprp}

{}
\newtheorem*{prp*}{\e{\dgCapProposition}}

{} %

\MyMakeRefMacros{\prpref}{\dgProposition}{\dgPropositions}
\MyMakeRefMacros{\Prpref}{\dgCapProposition}{\dgCapPropositions}

\MyMakeDupTheoMacros{my_claim}
{\clm}{\nclm}{\clmdup}{\nclmdup}{\dgClaim}{\clmrep}

\MyMakeRefMacros{\clmref}{\dgClaim}{\dgClaims}
\MyMakeRefMacros{\Clmref}{\dgCapClaim}{\dgCapClaims}

\MyMakeDupTheoMacros{theorem}
{\theo}{\ntheo}{\theodup}{\ntheodup}{\dgTheorem}{\theorep}

\MyMakeRefMacros{\theoref}{\dgTheorem}{\dgTheorems}
\MyMakeRefMacros{\Theoref}{\dgCapTheorem}{\dgCapTheorems}

\MyMakeTheoMacros{postulate}{\postu}{\npostu}

\MyMakeRefMacros{\posturef}{\dgPostulate}{\dgPostulates}
\MyMakeRefMacros{\Posturef}{\dgCapPostulate}{\dgCapPostulates}

\MyMakeTheoMacros{definition}{\defi}{\ndefi}

\MyMakeRefMacros{\defiref}{\dgDefinition}{\dgDefinitions}
\MyMakeRefMacros{\Defiref}{\dgCapDefinition}{\dgCapDefinitions}

\MyMakeTheoMacros{problem}{\prob}{\nprob}

\MyMakeRefMacros{\probref}{\dgProblem}{\dgProblems}
\MyMakeRefMacros{\Probref}{\dgCapProblem}{\dgCapProblems}

\MyMakeTheoMacros{myremark}{\rem}{\nrem}

\MyMakeRefMacros{\remref}{\dgRemark}{\dgRemarks}
\MyMakeRefMacros{\Remref}{\dgCapRemark}{\dgCapRemarks}

\MyMakeTheoMacros{conjecture}{\conj}{\nconj}

\MyMakeRefMacros{\conjref}{\dgConjecture}{\dgConjectures}
\MyMakeRefMacros{\Conjref}{\dgCapConjecture}{\dgCapConjectures}

\renewcommand{\qedsymbol}{$\blacksquare$}

\newcommand{\prfstart}[1][]{\ifthenelse{\equal{}{#1}}%
{\begin{proof}\renewcommand{\qedsymbol}{$\blacksquare$}}%
{\begin{proof}[\dgProofOf\ #1]%
\renewcommand{\qedsymbol}{$\blacksquare_{\mbox{\it{\scriptsize{#1}}}}$}}%
}
\newcommand{\prfend}[1][*]{%
\ifthenelse{\equal{}{#1}}{\renewcommand{\qedsymbol}{$\blacksquare$}}{}%
\ifthenelse{\equal{*}{#1}}{}%
{\renewcommand{\qedsymbol}{$\blacksquare_{\mbox{\it{\scriptsize{#1}}}}$}}%
\end{proof}\renewcommand{\qedsymbol}{$\blacksquare$}%
}

\newcommand{\NewSectLikeDG}[2]{
\NewcommandThreeArgsTwoOpt{#1}
{\ifthenelse{\equal{*}{##2}}{#2*}{#2{##2}}}
{#2{##2}\label{##1}}
{#2{\texorpdfstring{##2}{##3}}}
{#2{\texorpdfstring{##2}{##3}}\label{##1}}
}

\NewSectLikeDG{\sect}{\section}

\NewSectLikeDG{\ssect}{\subsection}

\NewSectLikeDG{\sssect}{\subsubsection}

\newcommand{\para}[2][]{\ifthenelse{\equal{}{#1}}
{\paragraph{#2}}
{\paragraph{#2}\label{#1}}}

\MyMakeRefMacros{\chref}{\dgChapter}{\dgChapters}
\MyMakeRefMacros{\Chref}{\dgCapChapter}{\dgCapChapters}

\MyMakeRefMacros{\sref}{\dgSection}{\dgSections}
\MyMakeRefMacros{\Sref}{\dgCapSection}{\dgCapSections}

\MyMakeRefMacros{\ssref}{\dgSubsection}{\dgSubsections}
\MyMakeRefMacros{\Ssref}{\dgCapSubsection}{\dgCapSubsections}

\MyMakeRefMacros{\sssref}{\dgSubsection}{\dgSubsections}
\MyMakeRefMacros{\Sssref}{\dgCapSubsection}{\dgCapSubsections}

\newcommand{\toct}[1]{\phantomsection\addcontentsline{toc}{section}{#1}}

\MyMakeRefMacros{\figref}{\dgFigure}{\dgFigures}
\MyMakeRefMacros{\Figref}{\dgCapFigure}{\dgCapFigures}

\newcommand{\IfMathMode}[2]{\ifmmode{#1}\else{#2}\fi}

\newcommand{\fbr}[1]{\IfMathMode%
{#1}{$#1$}}                     %

\newcommand{\fnbr}[1]{\mbox{\fbr{#1}}}  %

\newcommand{\fla}[2][*]{\ifthenelse{\equal{}{#1}}{\fbr{#2}}{\fnbr{#2}}}
\newcommand{\f}{\fla}

\newcommand{\bfla}[2][]{\mbox{\fbr{\pmb{#2}}}}
\newcommand{\fb}{\bfla}

\newcommand{\mal}[2][]{\MyChangeMathMargins
{}\delimiterfactor=1001{}%
\ifthenelse{\equal{}{#1}}%
{\begin{align*} #2 \end{align*}}%
{\ifthenelse{\equal{P}{#1}}%
{\allowdisplaybreaks\begin{align*} #2%
\end{align*}\interdisplaylinepenalty=10000}%
{\begin{align}\begin{split}\label{#1} #2 \end{split}\end{align}}%
}\delimiterfactor=901%
}

\newcommand{\m}{\mal}

\newcommand{\mac}{\substack}

\MyMakeEqRefMacros{\equref}{\dgEquation}{\dgEquations}
\MyMakeEqRefMacros{\Equref}{\dgCapEquation}{\dgCapEquations}

\MyMakeEqRefMacros{\expref}{\dgExpression}{\dgExpressions}
\MyMakeEqRefMacros{\Expref}{\dgCapExpression}{\dgCapExpressions}

\MyMakeEqRefMacros{\inequref}{\dgInequality}{\dgInequalities}
\MyMakeEqRefMacros{\Inequref}{\dgCapInequality}{\dgCapInequalities}

\makeatletter

\DeclareRobustCommand\bref{\@ifnextchar[%
{\bref@presq}%
{\bref@nopresq}%
}

\def\bref@presq[#1]{\@ifnextchar[%
{(\ref{#1}), \bref@presq}
{(\ref{#1}) and~\bref@nopresq}
}

\def\bref@nopresq#1{(\ref{#1})}

\makeatother

\makeatletter

\newcommand\Cases{%
\left\{\!\!\!\begin{array}{ll}\Cases@continue%
}

\def\Cases@continue#1#2{\@ifnextchar\bgroup%
{#1 &\txt{#2}\\ \Cases@continue}%
{#1 &\txt{#2}\end{array}\Cases@end}%
}

\def\Cases@end{\@ifnextchar[%
{\Cases@end@postsq}%
{\right.}
}

\def\Cases@end@postsq[#1]{\ifthenelse{\equal{\}}{#1}}
{\!\!\right\}}%
{\right.}
}

\makeatother

\newcommand{\lf}{\mathopen{}\mathclose\bgroup\left}
\newcommand{\rt}{\aftergroup\egroup\right}

\providecommand{\middle}{\big}
\newcommand{\md}{\middle}

\newcommand{\ud}{\vphantom{|_1^1}}

\newcommand{\chs}{\genfrac(){0cm}{}}  %

\newmatop{\plog}{poly-log}
\newmatop{\supp}{supp}   %

\makeatletter
\def\moverlay{\mathpalette\mov@rlay}
\def\mov@rlay#1#2{\leavevmode\vtop{%
\baselineskip\z@skip \lineskiplimit-\maxdimen
\ialign{\hfil$\m@th#1##$\hfil\cr#2\crcr}}}
\makeatother

\newcommand{\NewHLikeDG}[2]{
\NewcommandThreeArgsTwoOpt{#1}
{#2\l(##2\r)}
{#2_{##1}\l(##2\r)}
{#2\l(##2\md|{\ud}##3\r)}
{#2_{##1}\l(##2\md|{\ud}##3\r)}
}

\NewHLikeDG{\h}{\mathop H}

\NewHLikeDG{\hm}{\mathop{H_{\txt{min}}}}

\newcommand{\NewILikeDG}[2]{
\NewcommandFourArgsTwoOpt{#1}
{#2\lf[##2:{\ud}##3\rt]}
{#2_{##1}\lf[##2:{\ud}##3\rt]}
{#2\lf[##2:##3\md|{\ud}##4\rt]}
{#2_{##1}\lf[##2:##3\md|{\ud}##4\rt]}
}

\NewILikeDG{\I}{\mathop{\pmb{I}}}

\newcommand{\NewELikeDG}[2]{
\NewcommandThreeArgsTwoOpt{#1}
{#2\lf[##2\rt]}
{#2_{##1}\lf[##2\rt]}
{#2\lf[##2\md|{\ud}##3\rt]}
{#2_{##1}\lf[##2\md|{\ud}##3\rt]}
}

\NewELikeDG{\PR}{\mathop{\pmb{Pr}}}

\NewELikeDG{\E}{\mathop{\pmb{E}}}

\NewELikeDG{\Var}{\mathop{\pmb{Var}}}

\newcommand{\Cov}[2]{\mathop{\pmb{Cov}}\lf[{#1},\, {#2}\rt]}

\providecommand{\U}{}  %
\renewcommand{\U}[1][]{\ifthenelse{\equal{}{#1}}%
{{\Cl U}}%
{{\Cl U}_{#1}}}

\providemat{\NN}{\mathbb{N}}
\providemat{\RR}{\mathbb{R}}

\newcommand{\wtl}{\widetilde}
\newcommand{\wbr}{\overline}   %

\newcommand{\pss}[1][]{\nospell{\ifthenelse{\equal{}{#1}}%
{\txt{'s}}%
{\fla{#1\txt{'s}}}}}

\newcommand{\pl}[1][]{\nospell{\ifthenelse{\equal{}{#1}}%
{\mskip-6mu\stackrel{\text-}{}\mskip-4mu\txt{s}}%
{\fla{#1\mskip-6mu\stackrel{\text-}{}\mskip-4mu\txt{s}}}}}

\newcommand{\ord}[1][]{\nospell{\ifthenelse{\equal{}{#1}}%
{\txt{'th}}%
{\ifthenelse{\equal{1}{#1}}{$1\txt{'st}$}{\ifthenelse{\equal{2}{#1}}{$2\txt{'nd}$}{\ifthenelse{\equal{3}{#1}}{$3\txt{'rd}$}{\fla{#1\txt{'th}}}}}}}}

\newcommand{\fr}[3][*]{%
\ifthenelse{\equal{*}{#1}}%
{\frac{#2}{#3}}{}%
\ifthenelse{\equal{}{#1}}%
{\dr{#2}{#3}}{}%
\ifthenelse{\equal{/}{#1}}%
{\lf.#2\md/#3\rt.}{}%
\ifthenelse{\equal{p_}{#1}}%
{\lf.\lf(#2\rt)\md/#3\rt.}{}%
\ifthenelse{\equal{_p}{#1}}%
{\lf.#2\md/\lf(#3\rt)\rt.}{}%
\ifthenelse{\equal{pp}{#1}}%
{\lf.\lf(#2\rt)\md/\lf(#3\rt)\rt.}{}%
}

\newcommand{\sq}{\sqrt}

\NewcommandThreeArgsTwoOpt{\set}
{\lf\{#2\rt\}}
{\ClassWarning{My Macros}{Second optional argument missing in "\set"?}}
{\lf\{#2\md|{\ud}#3\rt\}}
{\lf\{#2{\ud}#1{\ud}#3\rt\}}

\newcommand{\s}{\set}

\newcommand{\setc}{\s[:]}

\newcommand{\NewMinLikeDG}[2]{
\NewcommandThreeArgsTwoOpt{#1}
{#2\lf\{##2\rt\}}
{#2_{##1}\lf\{##2\rt\}}
{#2\lf\{##2\md|\ud##3\rt\}}
{#2_{##1}\lf\{##2\md|\ud##3\rt\}}
}

\NewMinLikeDG{\Min}{\min}

\NewMinLikeDG{\Max}{\max}

\newmatop{\argmin}{argmin}
\NewMinLikeDG{\Argmin}{\argmin}

\newmatop{\argmax}{argmax}
\NewMinLikeDG{\Argmax}{\argmax}

\NewMinLikeDG{\Sup}{\sup}

\NewMinLikeDG{\Inf}{\inf}

\newcommand{\NewOLikeDG}[2]{
\newcommand{#1}[2][\MyUniPat]{\ifthenelse{\equal{\MyUniPat}{##1}}%
{\ensuremath{#2\lf(##2\rt)}}%
{#2(##2)}%
}}

\NewOLikeDG{\asO}{O}
\NewOLikeDG{\astO}{\tilde O}
\NewOLikeDG{\aso}{o}
\NewOLikeDG{\asOm}{\Omega}
\NewOLikeDG{\astOm}{\tilde \Omega}
\NewOLikeDG{\asom}{\omega}
\NewOLikeDG{\asT}{\Theta}

\makeatletter

\DeclareRobustCommand\bra{\@ifnextchar[%
{\bra@presq}%
{\bra@nopresq}%
}

\def\bra@presq[#1]{\@ifnextchar[%
{\ensuremath{\lla #1\rt|}\bra@presq}
{\ensuremath{\lla #1\rt|}\bra@nopresq}
}

\def\bra@nopresq#1{\ensuremath{\lla #1\rt|}}

\DeclareRobustCommand\ket{\@ifnextchar[%
{\ket@presq}%
{\ket@nopresq}%
}

\def\ket@presq[#1]{\@ifnextchar[%
{\ensuremath{\lf|#1\rra}\ket@presq}
{\ensuremath{\lf|#1\rra}\ket@nopresq}
}

\def\ket@nopresq#1{\ensuremath{\lf|#1\rra}}

\makeatother

\newcommand{\kbra}[2][]{\ifthenelse{\equal{}{#1}}%
{\lf|#2\md\rangle\hspace{-2.5pt}\md\langle #2\rt|}%
{\lf|#1\md\rangle\hspace{-2.5pt}\md\langle #2\rt|}%
}
\newcommand{\bket}[3][]{\ifthenelse{\equal{}{#1}}%
{\lla #2\md|#3\rra}%
{\lla #2\md|#1\md|#3\rra}%
}

\providecommand{\ip}[2]{\lla #1\,,\,#2\rra}

\newcommand{\sz}[2][]{\ifthenelse{\equal{}{#1}}%
{\lf|#2\rt|}%
{\lf|#2\rt|_{#1}}}

\providecommand{\ceil}[2][*]{\ifthenelse{\equal{}{#1}}%
{\lceil #2 \rceil}%
{\llc #2 \rrc}}
\providecommand{\floor}[2][*]{\ifthenelse{\equal{}{#1}}%
{\lfloor #2 \rfloor}%
{\llf #2 \rrf}}

\newcommand{\bm}{\pmb} %

\newcommand{\txt}[1]{\textrm{#1}}  %

\newcommand{\Cl}{\mathcal}   %

\DeclareMathAlphabet{\mathbfcal}{OMS}{cmsy}{b}{n}

\DeclareMathAlphabet{\mathlowcal}{OT1}{pzc}{m}{it}
\newcommand{\Cll}{\mathlowcal}

\newident[]{\R}{\mathcal R_{#1}}
\newident[]{\QII}{\mathcal Q_{#1}^{\parallel}}
\newidenT{\SMP}{SMP}

\newidenT{\AM}{AM}

\newmat{\mset}{\smin\set}
\newmat{\pset}{\cup\set}

\newcommand{\lla}{\lf\langle}
\newcommand{\rra}{\rt\rangle}
\newcommand{\llc}{\lf\lceil}
\newcommand{\rrc}{\rt\rceil}
\newcommand{\llf}{\lf\lfloor}
\newcommand{\rrf}{\rt\rfloor}

\newcommand{\nin}{\not\in}  %

\newcommand{\Then}{\Longrightarrow}
\newcommand{\IfThen}{\Longleftrightarrow}

\newcommand{\dt}{\cdot}
\newcommand{\tm}{\cdot}
\newcommand{\deq}{\stackrel{\textrm{def}}{=}}
\newcommand{\smin}{\setminus}

\newcommand{\redstar}{\ensuremath{\textcolor{Red}{\divideontimes}}}

\newcommand{\sbseq}{\subseteq}
\newcommand{\sbs}{\subset}

\newcommand{\eps}{\varepsilon}
\newcommand{\es}{\emptyset}

\newcommand{\overlay}[3][0mu]{
\begingroup
\mathchoice{
\bgroup
\ooalign{$\displaystyle#2$\cr
\hidewidth{$\displaystyle\mkern#1#3$}\hidewidth}
\egroup
}{
\bgroup
\ooalign{$\textstyle#2$\cr
\hidewidth{$\textstyle\mkern#1#3$}\hidewidth}
\egroup
}{
\bgroup
\ooalign{$\scriptstyle#2$\cr
\hidewidth{$\scriptstyle\mkern#1#3$}\hidewidth}
\egroup
}{
\bgroup
\ooalign{$\scriptscriptstyle#2$\cr
\hidewidth{$\scriptscriptstyle\mkern#1#3$}\hidewidth}
\egroup
}
\endgroup
}
\newcommand{\unin}{\mathrel{\overlay[1.5mu]{\subset}{\sim}}}

\newcommand{\ds}[1][]
{\ifthenelse{\equal{}{#1}}{\allowbreak\dots}{#1\allowbreak\dots#1}}
\newmat{\dc}{\ds[,]}
\newmat{\dtm}{\cdots}

\mydef{\01}{\set{0,1}}
\mydef{\12}{\set{1,2}}

\mydef{\l(}{\lf(}
\mydef{\r)}{\rt)}

\mydef{\=}{\equiv}
\mydef{\+}{\oplus}

\makeatletter

\let\dgampersand\&

\DeclareRobustCommand\&{%
\new@ifnextchar[%
{\dgsep@reposit}%
{\dgampersand}%
}

\def\dgsep@reposit[#1]{\hspace{#1}&\hspace{-#1}}

\makeatother

\mathchardef\myhyphen="2D
\mydef{\-}{\myhyphen}

\newcommand{\tmin}{{\scalebox{0.5}{\fb-}}}
\newcommand{\tpl}{{\scalebox{0.5}{\fb+}}}

\newcommand{\abstart}{\begin{abstract}}
\newcommand{\abend}{\end{abstract}}

\newenvironment{myepig}
{\par\addtolength{\leftskip}{28mm}\addtolength{\rightskip}{8mm}\noindent\ignorespaces}
{\par}

\newenvironment{myepigsgn}
{\par\addtolength{\leftskip}{84mm}\noindent\ignorespaces}
{\par}

\NewcommandThreeArgsTwoOpt{\epig}
{\begin{myepig} \textit{#2} \end{myepig} \addvspace{\baselineskip}}
{\ClassError{My Macros}{The first optional argument is not expected in "\epig"}{}}
{\begin{myepig} \textit{#2} \end{myepig} \Nopagebreak%
\begin{myepigsgn} -- #3 \end{myepigsgn}\addvspace{\baselineskip}}
{\ClassError{My Macros}{The first optional argument is not expected in "\epig"}{}}

{} %

\newcommand{\itstart}[1][\MyUniPat]{\ifthenelse{\equal{\MyUniPat}{#1}}%
{\begin{itemize}[noitemsep,topsep=3pt]}%
{\begin{itemize}[#1]}%
}
\newcommand{\enstart}[1][\MyUniPat]{\ifthenelse{\equal{\MyUniPat}{#1}}%
{\begin{enumerate}[noitemsep,topsep=3pt]}%
{\begin{enumerate}[#1]}%
}
{}  %

\newcommand{\itend}{\end{itemize}}
\newcommand{\enend}{\end{enumerate}}

\protected \def \dg #1{%
\textcolor{Red}
{
{\normalmarginpar\marginnote{\bl{DG's comment}}}
{\reversemarginpar\marginnote{\bl{DG's comment}}\\}
\IfMathMode{
~~~\txt{#1}~
}{
~\\~~~#1~\\
{\normalmarginpar\marginnote{\bl{\ul{------}}}}
{\reversemarginpar\marginnote{\bl{\ul{------}}}\\}
}
}
\ClassWarning{My Macros}{#1}
}

\newcommand{\fn}[2][]{%
\IfMathMode{}{}%
\ifthenelse{\equal{}{#1}}%
{\footnote{
\ignorespaces #2}}%
{\footnote{\label{#1}
\ignorespaces #2}}%
}

\newcommand{\fnm}{\footnotemark}
\newcommand{\fnt}[2][]{\ifthenelse{\equal{}{#1}}%
{\footnotetext{
\ignorespaces #2}}%
{\footnotetext{\label{#1}
\ignorespaces #2}}%
}

\makeatletter

\newcommand{\Nopagebreak}{\par\nobreak\@afterheading} 

\makeatother

\DeclareTextFontCommand{\bemph}{\bfseries}
\DeclareTextFontCommand{\ibemph}{\bfseries\em}
{} %
\newcommand{\e}{\emph}

\newcommand{\eu}[1]{\ul{\emph{#1}}}

{}  %

\newcommand{\bl}[1]{{\bf #1}} %

\newcommand{\il}[1]{{\it #1}} %

\newcommand{\tb}{\quad}
\newcommand{\tbb}{\qquad}
\newcommand{\tbbb}{\qquad\qquad}
\newcommand{\tbbbb}{\qquad\qquad\qquad\qquad}
\newcommand{\tbbbbb}{\qquad\qquad\qquad\qquad\qquad\qquad}

\makeatother%

{}

\newcommand{\MyChangeMathMargins}{%
\setlength{\abovedisplayskip}{\abovedisplayshortskip + 4pt}%
\setlength{\belowdisplayskip}{\belowdisplayshortskip + 3pt}%
}

{} %

\makeatother%

\newidenT{\GHD}{GHD}
\newident{\GHDd}{GHD_d}
\newidenT{\GHR}{GHR}
\newident{\tGHR}{\wtl{GHR}}

\newmatop{\rw}{rw}

\title{Bare quantum simultaneity versus classical interactivity\\
in communication complexity}

\newcommand{\instDG}{Institute of Mathematics of the Czech Academy of Sciences, \v Zitna 25, Praha 1, Czech Republic.}

\newcommand{\thanksDG}{Partially funded by the grant 19-27871X of GA \v CR.
Part of this work was done while visiting the Centre for Quantum Technologies at the National University of Singapore, and was partially supported by the Singapore National Research Foundation, the Prime Minister's Office and the Ministry of Education under the Research Centres of Excellence programme under grant R 710-000-012-135.}

\author{Dmitry Gavinsky\thanks{\instDG\newline\thanksDG}
}

\begin{document}

\maketitle

\thispagestyle{empty}

\abstart

A relational bipartite communication problem is presented that has an efficient quantum simultaneous-messages protocol, but no efficient classical two-way protocol.
\abend

\sect[s_intro]{Introduction}

The physical universe that we have the pleasure to experiment with may not be classical.
For instance, from the point of view of Popperian methodology~\cite{P34_Log}, the statements \eu{the Lorentz transformation adequately describes the kinematics of the Universe} and \eu{the Bell inequalities can be violated} both are plausible, and together they imply that the Universe is \e{non-local} -- in particular, \e{non-classical}.\fn
{
Despite its apparent formality, the above statement overlooks various \e{hidden assumptions}:\ e.g., that of \e{unpredictable input} in the Bell-violating experiments.
Such assumptions are likely unavoidable in any mathematical treatment of the phenomenal reality.
These problems, albeit of highest interest and importance, are beyond our current scope.
}

Among various non-classical theories, \e{quantum mechanics} is, probably, the most adequate:\ first of all, it is very accurate in predicting the \e{probabilities} of experimental outcomes (and there are reasons to believe that this is the best we can hope for prophesy-wise); on the other hand, quantum mechanics is compatible with our other plausible theories in the \e{regimes that have been tested experimentally} (as well as in those ones that we can expect to be able to test any time soon).
It is interesting to identify and investigate those scenarios where the predictions of quantum mechanics are \e{most non-classical}, as studying physical reality can be entertaining and might be of philosophical significance.\fn
{
The significance of understanding quantum mechanics seems to be at least two-fold.
On the one hand, the theory is among the frontiers of physical research, and analytical investigation of the possibility of a priori physical knowledge has led to some of the deepest ontological and epistemological doctrines so far.
On the other hand, quantum mechanics is intimately related to the problem of causation, which might be not as fundamental as the problem of a priori synthetic knowledge, but is nevertheless very important.
}

This problem makes sense, in particular, in the context of \e{computational complexity theory}:\ e.g., we may ask whether computational devices that have at their disposal the flexibility offered by quantum mechanics are qualitatively stronger than their classical counterparts.
Obviously, the question would sound most natural in the context of \e{Turing machines}, but addressing that seems to be beyond our current understanding.
Viewing this line of research as an endeavour to understand the ``quantum aspect'' of physical reality, we would like to minimise the number of assumptions made in the analysis, in particular, to avoid postulating unproven classical hardness of computational problems.

One of the richest computational models that we already know how to analyse is the setting of \e{communication complexity}, and in this work we will be investigating \e{the superiority of quantum over classical models in the context of two-player communication}.
Here is a brief informal introduction of some concepts of communication complexity.
\itstart
\item
In the model of \e{bipartite} communication there are two \eu{players}, \e{Alice} and \e{Bob}, who receive one portion of input each:\ Alice gets $x$ and Bob gets $y$.
Their goal is to use the allowed type of communication (as described next) in order to compute an answer that would be correct with respect to the received pair $(x,\, y)$.
\item 
The three main bipartite \eu{layouts} are \e{two-way (interactive) communication}, \e{one-way communication} and \e{simultaneous message passing (\SMP)}.
In the first case the players can exchange messages interactively before answering, in the second case only Alice can send a message to Bob (who then answers), in the third case both Alice and Bob send one message each to a third participant -- the \e{referee} (who then answers).
\item 
Communication problems determine which answers are correct.
The three main \eu{types of problems} are \e{total functions}, \e{partial functions} and \e{relations}:\ in the first case there is exactly one correct answer for each possible pair of input values, and the set of those pairs equals the direct product of possible inputs of Alice and possible inputs of Bob; the second case is similar, but the set of possible inputs can be arbitrary; in the third case multiple correct answers for the same pair of input values are allowed.
\item 
An \e{efficient} solution is a communication protocol where the players use at most poly-logarithmic (in the input length) amount of communication and produce a right answer with high confidence.
\item 
Communication models can be strengthened by \e{shared randomness}, which corresponds to allowing the players to use \e{mixed strategies} (this can be helpful only in the weakest among the layouts -- the \SMP), or by \e{shared entanglement}, which allows the players to share any (input-independent) quantum state and use it while running the protocol.\fn
{
Sometimes in this work we call a model \e{bare} to emphasise that it allows no shared resources.
}
\itend

Quantum communication complexity has been an active area of research over the last few decades.
Among numerous results in the field, many of which surprised us a lot when they came out, the most relevant to this work are the following:
\itstart
\item In 1998 a \e{partial function} was demonstrated~\cite{BCW98_Qua} for which in \e{zero-error regime} quantum protocols had exponential advantage over the classical ones (both one-way and interactive).
\item In 1999 a \e{partial function} was demonstrated~\cite{R99_Ex} that had an efficient \e{quantum two-way protocol}, but no efficient \e{classical two-way protocol}.
\item In 2001 a \e{total function} was demonstrated~\cite{BCWW01_Qua} that had an efficient \e{quantum simultaneous-messages protocol without shared randomness}, but no efficient \e{classical simultaneous-messages protocol without shared randomness}.
\item In 2004 a \e{relation} was demonstrated~\cite{BJK04_Exp} that had an efficient \e{quantum simultaneous-messages protocol without shared randomness}, but no efficient \e{classical one-way protocol}.
\item In 2007 a \e{partial function} was demonstrated~\cite{GKKRW08_Ex} that had an efficient \e{quantum one-way protocol}, but no efficient \e{classical one-way protocol}.
\item In 2008 a \e{relation} was demonstrated~\cite{G08_Cla} with an efficient \e{quantum one-way protocol}, but no efficient \e{classical two-way protocol}.
\item In 2010 a \e{partial function} was demonstrated~\cite{KR11_Qua} with an efficient \e{quantum one-way protocol}, but no efficient \e{classical two-way protocol}.
\item In 2016 a \e{partial function} was demonstrated~\cite{G20_En} with an efficient \e{quantum simultaneous-messages protocol with shared entanglement}, but no efficient \e{classical two-way protocol}.
\item In 2017 a \e{partial function} was demonstrated~\cite{G19_Qua} with an efficient \e{quantum simultaneous-messages protocol without shared randomness}, but no efficient \e{classical simultaneous-messages protocol}, even \e{with shared randomness}.
\itend

Looking for the regimes where the predictions of quantum mechanics are as far as possible from those of classical mechanics, we can ask the following questions:
\itstart
\item What is the \e{weakest} quantum communication model that outperforms its classical counterpart?
\item What is the \e{strongest} classical communication model that is outperformed by the quantum counterpart?
\itend
Both of these questions have been answered:\ the weakest quantum model, \SMP, can outperform the classical \SMP~\cite{BCWW01_Qua}, and the strongest classical model, two-way, can be outperformed by the quantum two-way~\cite{R99_Ex}.

Can these two separations be ``united''? 
In other words:
\itstart
\item Can quantum \SMP\ (the quantum weakest) outperform classical two-way communication (the classical strongest)?
\itend
Although some of the separations mentioned earlier come rather close to giving a positive answer, no such example has been known before.
In particular, the strongest classical model known to be outperformed by quantum \SMP\ is classical \SMP\ with shared randomness~\cite{G19_Qua}, and the weakest quantum model known to outperform classical two-way communication is quantum \SMP\ with shared entanglement~\cite{G20_En}.\fn
{
Intuitively, shared entanglement is a significant additional ``quantum resource'':\ the players are supposed to remain \e{entangled} -- that is, to maintain an inherently non-classical condition -- for indefinitely long period of time between preparing the ``set-up'' and actually receiving the input values and executing the protocol.
}

This work demonstrates a relational problem that has a quantum simultaneous-messages protocol (with no shared resources) of cost \asO{\log n}, and whose classical two-way complexity is \asOm{n^{\dr13}}.
In other words, \e{qualitative advantage over the strongest feasible model of classical communication is inherent even in the weakest reasonable quantum model}.

The communication problem will be called \e{Gap Hamming Relation (\GHR)}, and it can be viewed as a relational version of the \e{Gap Hamming Distance (\GHD)} promise function, well known in the context of classical communication complexity.

\sect[s_prelim]{Preliminaries}

We will write $[n]$ to denote the set $\s{1\dc n}\sbs\NN$ and $[n]-1$ to denote $\s{0\dc n-1}$.
Let $(a,\, b)$, $[a,\, b]$, $[a,\, b)$ and $(a,\, b]$ denote the corresponding open, closed and half-open intervals in $\RR$.
We will use ``$|$'' to denote divisibility.

Let $x_i$ address the \ord[i] bit of $x$ for $x\in\01^n$ and $i\in{[n]}$.
Similarly, for $S\sbseq{[n]}$, let $x_S$ denote the $\sz S$-bit string, consisting of (naturally-ordered) bits of $x$, whose indices are in $S$.
To avoid ambiguity, we will sometimes write $x|_S$ ($x|_i$) instead of $x_S$ ($x_i$).
For a set (or a family) $A$, we will write $A|_i$ and $A|_S$ to address, respectively, $\s{x_i}[x\in A]$ and $\s{x_S}[x\in A]$.
For $y\in\01^n$, let $x\+y$ stand for the bit-wise XOR between the two strings.

Let $\sz x$ denote the Hamming weight of $x$.
At times we will assume (the trivial) isomorphisms among the \f n-bit strings, the subsets of $[n]$ and the elements of $[2^n]-1$.
In particular, the notation $\chs{[n]}k$ will stand for $\s{x\in\01^n}[\sz{x}=k]$, and $x\cap y$ will address the set $\s{i\in[n]}[x_i=y_i=1]$.
We will write $\ip xy$ for $\l(\sz{x\+y}\mod2\r)$, letting it extend naturally to the cases of $x$, $y$ or both being elements of $[2^n]-1$.

Let $\sigma_i$ denote the permutation of $[n]$ which is the \ord[i] cyclic shift (i.e., $\sigma_i(j)=i+j$ if $i+j\le n$ and $i+j-n$ otherwise).
For $x\in\01^n$, denote by $\sigma_j(x)$ the \f j-bit cyclic shift of $x$ -- that is, the string of $n$ bits where on \ord[\sigma_j(i)] position is $x_i$ for each $i\in[n]$.

We use square brackets $[\ds]$ to denote events and capital letters from a ``mildly calligraphic'' font to represent random variables.
We will consider probabilities, expectations and so on, conditioned on random variables (sometimes in addition to conditioning on certain events), viewing the corresponding expressions as random variables by themselves.
E.g., if $\Cl X$ and $\Cl Y$ are random variables, then $\PR{\Cl X=1}[\Cl Y=1]\in\RR$, but both $\PR{\Cl X=1}[\Cl Y]$ and $\E{\Cl X=1}[\Cl Y]$ are random variables whose values are determined by that of $\Cl Y$.

Let $\U[A]$ denote the uniform distribution over the elements of $A$.
Sometimes (e.g., in subscripts) we will write ``$\unin A$'' instead of ``$\sim\U[A]$''.
We will write things like $\01^{n+n}$ when the set should be viewed as $\01^n\times\01^n$:\ e.g., $(\Cl X,\Cl Y)\unin\01^{n+n}$ refers to the uniform distribution of two independent \f n-bit strings, and $A\times B\sbseq\01^{n+n}$ will be used when $A\times B$ is a combinatorial rectangle in $\01^n\times\01^n$.

We let $\bot$ and $\top$ denote, respectively, the false and the true values of Boolean predicates.
More generally, we will use the Boolean domain $\s{\bot,\top}$ instead of $\01$ when there is, subjectively, a \e{strong logical flavour} to the values.

We will use Dirac's notation (\e{bra-ket}) for quantum messages and manipulations with them.

\sect[s_CommC]{Communication complexity}

For an excellent survey of classical communication complexity, see~\cite{KN97_Comm}.
Quantum communication models differ from their classical counterparts in two aspects:\ the players are allowed to send \e{quantum messages} (consequently, the complexity is measured in \e{qubits}) and to perform arbitrary \e{quantum operations} locally.\fn
{
We say that a communication model allows \e{shared entanglement} if the players can share any (input-independent) quantum state and use it in the protocol (in the case of simultaneous message passing, entanglement is only allowed between Alice and Bob).
Models with shared entanglement will not be used in this work, but they are mentioned in some discussions. 
}
An excellent survey of quantum communication complexity is~\cite{BCMW10_Non}.

Of central importance to this work is the model of bipartite \e{Simultaneous Message Passing (\SMP)}, where there are three participants:\ \e{players} Alice and Bob, and \e{the referee}.
An \SMP-protocol has the following structure:
Alice receives $x$ and sends her message to the referee; at the same time, Bob receives $y$ and sends his message to the referee; the referee uses the content of the two received messages to compute the answer.

In the \e{quantum version} of \SMP, denoted by \QII, the players can send quantum messages and the referee can apply any quantum measurement to compute the answer.

Another important communication model is that of \e{interactive (two-way) communication}.
We will be primarily interested in the \e{classical randomised version}, denoted by \R:\ here two players, Alice and Bob, receive the two parts of input $x$ and $y$ and interact in order to compute their answer.

A problem is said to have an efficient solution (in a given model) if for any constant $\eps>0$ there exists a protocol that transfers at most \e{poly-logarithmic} (in the input length) number of (qu)bits and produces a correct answer with probability at least $\eps$.
Note that if the considered problem is \e{functional}, then it is sufficient for a protocol to succeed with any probability $\dr12+\asOm1$:\ a constant number of independent repetitions would allow reducing the error below $\eps$.
On the other hand, in the \e{relational} case error reduction by repetition is not necessarily possible, as there can be multiple correct answers; accordingly, in order to claim that a relation has an efficient solution, one has to make sure that there is an efficient protocol for \e{every} constant positive error bound. 

\ssect[ss_GHD]{Gap Hamming Distance problem and its variations}

As usual, we use $n$ to denote the input length of a computational problem.
To avoid unnecessary complication of the notation, in this work we will always assume that $n>1$ and $2|\log n$.

The \e{Gap Hamming Distance (\GHD)} is the following bipartite functional problem:
\m{
&\GHDd:\01^{n+n}\to\01;\\
&\GHDd(x,y)\deq\Cases
{1}{if $|x\+y|\ge\dr n2+d$;}
{0}{if $|x\+y|\le\dr n2-d$;}
{\txt{undefined}}{otherwise.}
}

In the above definition $d\in[\dr n2]$ is a parameter; the most interesting regime (both in general and for us) is that of $d\in\asT{\sq n}$, so we will write \GHD\ to denote $\GHD[_{\sq n}]$.

It has been demonstrated~\cite{CR11_An_Op} that the \R-complexity of \GHD\ is \asOm n; moreover, a nearly-trivial lower bound of \asOm{\sq n} has been known as folklore since the introduction of \GHD\ in~\cite{IW03_Ti}.

In this work we will relax \GHD\ in order to let the model \QII\ solve it efficiently, at the same time preserving -- up to a polynomial -- the \R-hardness of the problem.\fn
{
As we are interested in \e{structural} separation between \QII\ and \R, we only need a lower bound of $n^{\asOm1}$ on the \R-complexity of any problem that admits a feasible solution in \QII\ -- that is, a protocol of complexity $\plog(n)$.
}
Our relaxation will turn the original problem into a relation (the possibility of a functional separation is left as an open problem, cf.~\sref{s_concl}).

Informally, \GHD\ can be viewed as making the players decide whether $x$ and $y$ are slightly closer together or slightly further apart than the expected $\dr n2$; the high-level idea behind turning it into an easier relational problem is to ask them for a \e{transformation} (of some pre-determined form) that maps $x$ to a string, whose distance from $y$ is somehow biased.

Let us have a look at the following candidate problem:
\m{
&\tGHR \sbseq \01^{n+n}\times \01^n;\\
&\tGHR \deq \s{(x,y,\tau)}[|\tau\+x\+y|\le\dr n2-\sq n].
}
That is, the valid answers for the input pair $(x,y)$ are all such $\tau\in\01^n$ that make $\tau\+x$ at most $(\dr n2-\sq n)$-far from $y$.
This relaxation is \e{too radical}, as \tGHR\ is easy even for classical one-way communication with shared randomness (let alone \R, which is a strictly stronger model).
Indeed, Alice and Bob can use shared randomness to generate a sufficiently long sample of uniformly-random \f n-bit strings $\Cl Z_1\dc \Cl Z_t$, then Alice can send to Bob the index of $\Cl Z_{i_0}$ that is closest to $x$ (with high probability the distance will be less than $\dr n2-\sq n$, as $t$ is a large enough constant) and Bob can output $\tau=\Cl Z_{i_0}\+y$.

To make our relation harder than \tGHR\ (in order to preserve the \R-hardness of the original problem), we reduce the number of allowed transformations (that is, of possible answers) to polynomial.
On the other hand, the structure of the allowed transformations makes the new problem more convenient for \QII-protocols.

\nota{
\m[m_nota_aleph]{
&\forall s\in\01^{\log n},\,i\in[n]:\tb
\tau_s|_i \deq \ip si;~\fnm\\
&\forall j\in[n],\,s\in\01^{\log n}:\tb
\Delta_{j,s}(x,y) \deq \sz{\sigma_j\l(\tau_s\+x\r)\+y}.\\
&\aleph(x,y) \deq \Cases
{\top}{
if $\displaystyle\sum_{\mac{j,s:\\\dr n2-\dr{\sq n}2\le\Delta_{j,s}(x,y)\le\dr n2+\dr{\sq n}2}}
\l(\Delta_{j,s}(x,y)-\fr n2\r)^2\le\fr{n^3}{9}$;
}
{\bot}{otherwise.}
}\fnt
{
Recall that $\ip xy$ stands for $\l(\sz{x\+y}\mod2\r)$.
}
}

That is, $\s{\tau_s}$ is the standard Fourier basis for $\01^{\log n}\to\s{\pm1}$ in the ``$\01$-form'' (in particular, $\sz{\tau_{s_1}\+\tau_{s_2}}\=\dr n2$ for all $s_1\ne s_2$), and allowed are the $n^2$ transformations that combine bit-flipping according to some $\tau_s$ with a cyclic rotation $\sigma_j$.
The predicate $\aleph(x,y)$ asserts that both values are, in some sense, \e{typical}.\fn
{
The constants used in the definition of the predicate are such that, on the one hand, $\aleph(\Cl X,\Cl Y)$ is satisfied almost always by \e{uniformly-random $\Cl X$ and $\Cl Y$} (showing this will be important for the analysis), while on the other hand, $\aleph(x,y)$ testifies that our quantum \SMP-protocol exhibits certain ``super-classical'' behaviour when its input is $(x,y)$.
The structure of our communication problem $\GHR(x,y)$ will be such that this behaviour results in a correct answer, while every classical approach to solving \GHR\ is either fruitless or inefficient.
}

\ndefi[d_GHR]{Gap Hamming Relation (\GHR)}{The bipartite relation \GHR\ is a subset of $\01^n\times \01^n\times \l([n]\times\01^{\log n}\r)^{\log n}$ -- that is, the input consists of two \f n-bit strings and the output is a sequence of $\log n$ pairs $(j_i,s_i)$.
The sequence $((j_1,s_1)\dc(j_{\log n},s_{\log n}))$ is a valid answer to $\GHR(x,y)$ if the following holds:
\m{
\Cases
{\sz{\setc i[{\Delta_{j_i,s_i}(x,y)\nin\lf[\dr n2-\dr{\sq n}2,\dr n2+\dr{\sq n}2\rt]}]}\ge\dr{\log n}2}
{if $\aleph(x,y)$;}
{\top}{otherwise.}
}
}

Intuitively, if we ignore conditioning on $\aleph(x,y)$, then a valid answer to $\GHR(x,y)$ is a sequence of $\log n$ transformations, at least half of which map the value of $x$ either \e{somewhat close to} or \e{somewhat far from} the value of $y$. 
We will see that \GHR\ is easy for \QII\ but hard for \R.

\sssect*{Intuition behind the \GHR\ problem}

The players are required to find a short list of transformations of $x$ that are likely to make its distance from $y$ slightly different from the expected $\dr n2$ for $(x,y)\unin\01^{n+n}$.
Intuitively one can view the family of $n^2$ outcomes of the allowed transformations
\m{
\l(\sigma_j\l(\tau_s\+\Cl X\r)\ud\r)_{j\in[n],\,s\in\01^{\log n}}
}
as nearly-independent and uniformly distributed when $\Cl X\unin\01^n$.~\fn
{
Of course, the \f n-bit values themselves are very far from being independent:\ the claimed intuitive independence only makes sense in the context of their distance to a uniformly-random $\Cl Y$.
Formalising and quantifying this intuition will be the technical core of the analysis of \pss[\GHR] complexity.
}
Then the challenge faced by the players is \e{to find among these $n^2$ random strings some that are biased either away from or towards $y$}.
Note that the required bias is $\dr{\sq n}2$, and we expect most of the $n^2$ distances to be around the expected $\dr{n}2$ within the order of magnitude of the standard deviation of the corresponding binomial distribution -- that is, inside $\dr{n}2 \pm \asO{\sq n}$.~\fn
{
And all the $n^2$ distances are, almost certainly, inside $\dr{n}2 \pm \asO{\sq{n\tm \log n}}$.
}

The reason why, intuitively, this problem may have an efficient quantum simultaneous-message protocol is that if we consider the situation when Alice and Bob, respectively, send to the referee $\dr1{\sq n}\tm \sum_{i=1}^n(-1)^{x_i}\tm\ket i$ and $\dr1{\sq n}\tm\sum_{i=1}^n(-1)^{y_i}\tm\ket i$, then a very simple and intuitive multi-outcome quantum measurement exists that combines applying an allowed transformation to the state representing $x$ with estimating the resulting inner product with the state representing $y$:
\m{
\fr{1}n\tm \sum_{i=1}^n(-1)^{\l(\sigma_j\l(\tau_s\+x\r)\ud\r)_i+y_i}\tm\bket ii
.}
As the Hamming distance of $\dr{n}2$ corresponds to the inner product $0$ between the representatives, and, at the same time, the probability of the corresponding measurement outcome is proportional to the square of the inner product, the whole procedure favours those outcomes that correspond to the transformations introducing some bias.
See \sref{s_qua} for details.

The reason why, intuitively, \GHR\ may be hard for classical interactive protocols is that solving the problem requires estimating the Hamming distance between a pair of uniformly-random \f n-bit string, which is known to be hard:\ as mentioned earlier, the \R-complexity of \GHD\ is \asOm n.
We have already seen how this intuition breaks down when the family of allowed transformations is exponentially-large, but allowing only polynomially-many choices indeed preserves the classical hardness of the original \GHD:\ when the number of choices is strictly limited, a protocol \e{must find out more about each possibility}.
See \sref{s_cla} for details.

Hence, \GHR\ is a plausible candidate for revealing qualitative advantage of bare quantum simultaneity over classical interactivity:
On the one hand, an efficient \QII-protocol can sample among the allowed transformations of $x$ those that make its distance from $y$ biased away from the expected $\dr{n}2$; on the other hand, we may expect that the distances from the interval $\dr n2\pm\astO{\sq n}$ between a pair of \f n-bit strings are well obscured from efficient classical protocols.

\sect[s_qua]{Solving \GHR\ with simultaneous quantum messages}[Solving GHR with simultaneous quantum messages]

Let
\m{
\forall j\in[n],\,s\in\01^{\log n}:\:
\ket{u_j^s} \deq \fr1{\sq n}\tm\sum_{i=1}^n(-1)^{\tau_s|_i}\tm\ket i\ket{\sigma_j(i)}
,}
then
\m{
\s{\ket{u_j^s}}[{j\in[n],\,s\in\01^{\log n}}]
}
is an orthonormal basis for $\RR^{n+n}$.~\fn
{
All quantum states considered in this work can be represented as unit vectors in the real space of the corresponding dimension, so we do not lose generality by ignoring the case of complex vectors.
}
Indeed, if $j_1\ne j_2$, then
\m{
\bket{u_{j_1}^{s_1}}{u_{j_2}^{s_2}}
= \fr1n\tm\sum_{i=1}^n(-1)^{\tau_{s_1}(i)+\tau_{s_2}(i)}
\tm\bket ii\tm\bket{\sigma_{j_1}(i)}{\sigma_{j_2}(i)}
=0
;}
otherwise, if $j_1=j_2$ and $s_1\ne s_2$, then
\m{
\bket{u_{j_1}^{s_1}}{u_{j_2}^{s_2}}
= \fr1n\tm\sum_{i=1}^n(-1)^{\tau_{s_1}(i)+\tau_{s_2}(i)}
\tm\bket ii\tm\bket{\sigma_{j_1}(i)}{\sigma_{j_2}(i)}
= \fr1n\tm\sum_{i=1}^n(-1)^{\tau_{s_1}(i)+\tau_{s_2}(i)}
=0
,}
as $\s{\tau_s}$ is the ``$\01$-form'' of a Fourier basis for $\01^{\log n}\to\s{\pm1}$.

\para{A protocol for \GHR.}
Upon receiving the input, Alice and Bob construct, respectively,
\m{
\ket{\phi_{Al}} \deq \fr1{\sq n}\tm\sum_{i=1}^n(-1)^{x_i}\tm\ket i
\txt{~~and~~}
\ket{\phi_{Bo}} \deq \fr1{\sq n}\tm\sum_{i=1}^n(-1)^{y_i}\tm\ket i
,}
and send $\log n$ copies of both the states to the referee.
The referee then independently applies to $\log n$ copies of the state $\ket{\phi_{Al}}\ket{\phi_{Bo}}$ the complete projective measurement 
\m{
\setc{\kbra{u_j^s}}[{j\in[n],\,s\in\01^{\log n}}]
}
and outputs the list of obtained outcomes $\l((j_1,s_1)\dc(j_{\log n},s_{\log n})\r)$ as his answer.

The complexity of this \QII-protocol is \asO{(\log n)^2}.

\para{Correctness.}
Assume that $\aleph(x,y)$ holds with respect to the received input $(x,y)$ -- otherwise any answer would be correct by the definition of \GHR.
For $i\in[\log n]$, let $(\Cl J_i,\Cl S_i)$ be random variables that take values $(j_i,s_i)$ -- that is, the outcome of the \ord[i] referee's measurement.
We want to analyse the probability of the event
\m{
\Cll e_i\deq\lf[\Delta_{\Cl J_i,\Cl S_i}(x,y)\in\lf[\fr n2-\fr{\sq n}2,\fr n2+\fr{\sq n}2\rt]\rt]
,}
computed with respect to the action of the protocol ($x$ and $y$ are fixed, subject to $\aleph(x,y)$).

Fix $j\in[n]$ and $s\in\01^{\log n}$, then for any $i\in[\log n]$:
\m[m_js]{
&\PR{(\Cl J_i,\Cl S_i)=(j,s)}
= \sz{\l( \bra{\phi_{Al}}\bra{\phi_{Bo}} \r) \ket{u_j^s}}^2\\
&\tbb = \fr1{n^3} \tm \sz{ \l( \sum_{k,j=1}^n(-1)^{x_k+y_j}\tm\bra k\bra j \r)
\l( \sum_{k=1}^n(-1)^{\tau_s|_k}\tm\ket k\ket{\sigma_j(k)} \r) }^2\\
&\tbb = \fr1{n^3} \tm \l( \sum_{k=1}^n(-1)^{x_k+\tau_s|_k+y_{\sigma_j(k)}} \r)^2\\
&\tbb = \fr4{n^3} \tm \l( \sz{\sigma_j(x\+\tau_s)\+y} - \fr n2 \r)^2
~=~ \fr4{n^3} \tm \l( \Delta_{j,s}(x,y) - \fr n2 \r)^2
}
and
\m{
\PR{\Cll e_i}
&= \sum_{\mac{j,s:\\\dr n2-\dr{\sq n}2\le\Delta_{j,s}(x,y)\le\dr n2+\dr{\sq n}2}}
\PR{(\Cl J_i,\Cl S_i)=(j,s)}\\
&= \fr4{n^3} \tm
\sum_{\mac{j,s:\\\dr n2-\dr{\sq n}2\le\Delta_{j,s}(x,y)\le\dr n2+\dr{\sq n}2}}
\l( \Delta_{j,s}(x,y) - \fr n2 \r)^2
\le \fr49
,}
where the inequality is the assumption that $\aleph(x,y)$ holds (cf.~\bref{m_nota_aleph}).~\fn
{
Note that the value (of the right-hand side) of~\bref{m_js} is always at most $\dr1n$ -- that is, the measurement outcome is not determined even if the input is \e{perfectly aligned} with one of the allowed transformations.
This is crucial:\ while the states $\ket{u_j^s}$ form an orthonormal basis, the concept of input's perfect alignment is \e{not necessarily} monogamous with respect to the transformations (think about $x=y=\bar0$), and this is reflected by the uncertainty in the outcome.
This is also why we need the power of relations for our separation.
}

As the outcomes of the referee's measurements are independent,
\m{
\&[22pt]\PR{\txt{the protocol errs}}\\
& \le \PR{
\sz{\setc i[{\Delta_{j_i,s_i}(\Cl X,\Cl Y)\in\lf[\fr n2-\fr{\sq n}2,\fr n2+\fr{\sq n}2\rt]}]}
>\fr{\log n}2
}[{\aleph(\Cl X,\Cl Y)}]\\
& = \PR{\txt{more than half of the \pl[\Cll e_i] take place}}[{\aleph(\Cl X,\Cl Y)}]
~\in~ n^{-\asOm1}
,}
as follows from Hoeffding's inequality~(\fctref{f_Hoeff}).

From the above analysis it follows readily that an \QII-protocol of cost \asO{\log n} would suffice for guaranteeing accuracy at least $1-\eps$ for arbitrarily small constant $\eps$:
As the definition of \GHR\ allows repeated elements in the output sequence, the players can send sufficiently large \e{constant} number $t$ of the states $\ket{\phi_{Al}}$ and $\ket{\phi_{Bo}}$ to the referee, who can measure them to generate $\l((j_1,s_1)\dc(j_t,s_t)\r)$, and then output this sequence repeated $\ceil{\dr{\log n}t}$ times, trimming the last copy if necessary.

\crl[crl_GHR_Q_upper]{$\QII(\GHR) \,\in\, \asO{\log n}$.
}

\sect[s_Cher]{Concentration bounds}

Let us start by listing some known bounds that this work uses.

\nfct[f_MaChe]{Markov's and Chebyshev's inequalities}{Let $\Cl X$ be a random variable, then
\m{
\forall t\ge0:\:
\PR{\sz{\Cl X} \ge t} \le \fr{\E{\sz{\Cl X}}}t
~~\txt{and}~~
\PR{\sz{\Cl X - \E {\Cl X}}\ge t} \le \fr{\Var {\Cl X}}{t^2}
.}
}

One of the most convenient formulations of Chernoff-like statements -- that is, a tail bound for the sum of mutually-independent random variables -- is the well-known Hoeffding's inequality.
\nfct[f_Hoeff]{Hoeffding's inequality}{Let $\Cl X_1\dc \Cl X_m$ be mutually-independent random variables, such that $\forall i\in[m]:\: a_i\le \Cl X_i\le b_i$.
Then 
\m{
\forall t\ge0:\:
\PR{\sz{\sum_{i=1}^m\Cl X_i - \E{\sum_{i=1}^m\Cl X_i}}\ge t}
\le 2\tm e^{\fr{-2t^2}{\sum_{j=1}^m(b_j-a_j)^2}}
.}
}

In some situations it will be useful to have a tail-of-sum bound with more relaxed conditions and somewhat tighter guarantees, which we prove next.
Intuitively, the following formulation combines some advantages of the one by Angluin and Valiant~\cite{AV79_Fas} (which beats qualitatively the classical statement in the regime of low expectations) with the universality of Azuma's inequality applied to the Doob martingale (which gets rid of the independence requirement).\fn
{
See~\cite{S18_Ran} for an excellent overview of Chernoff-like concentration bounds.
}

\ntheo[t_Cher]{A relaxed Chernoff bound}{Let $\Cl X_1\dc \Cl X_m$ be random variables with countable supports\fnm, such that $\forall i\in[m]:\: 0\le \Cl X_i\le1$.\fnt
{
The continuous version can be obtained along similar lines under the integrability assumption.
}
Then
\m[m_Cher_1]{
\forall a,\, t\ge0:\:
& \PR{\sum_{i=1}^m\Cl X_i \le a - t}
[{\sum_{i=1}^m\E{\Cl X_i}[\Cl X_1\dc \Cl X_{i-1}]\ge a}]
\le e^{\fr{-t^2}{2a}},\\
& \PR{\sum_{i=1}^m\Cl X_i \ge a + t}
[{\sum_{i=1}^m\E{\Cl X_i}[\Cl X_1\dc \Cl X_{i-1}]\le a}]
\le e^{\fr{-t^2}{2a + t}}
.}
In particular,
\m
{
& \PR{\sum_{i=1}^m\Cl X_i \le a - t} \le e^{\fr{-t^2}{2a}} + \PR{\sum_{i=1}^m\E{\Cl X_i}[\Cl X_1\dc \Cl X_{i-1}]<a},\\
& \PR{\sum_{i=1}^m\Cl X_i \ge a + t} \le e^{\fr{-t^2}{2a + t}} + \PR{\sum_{i=1}^m\E{\Cl X_i}[\Cl X_1\dc \Cl X_{i-1}]>a}
,}
and
\m[m_Cher_2]{
\forall \mu,\, t\ge0:\:
& \sum_{i=1}^m\E{\Cl X_i}[\Cl X_1\dc \Cl X_{i-1}]\ge\mu
~\Then~
\PR{\sum_{i=1}^m\Cl X_i \le \mu - t} \le e^{\fr{-t^2}{2\mu}},\\
& \sum_{i=1}^m\E{\Cl X_i}[\Cl X_1\dc \Cl X_{i-1}]\le\mu
~\Then~
\PR{\sum_{i=1}^m\Cl X_i \ge \mu + t} \le e^{\fr{-t^2}{2\mu + t}}
.}
}

The three formulations are, essentially, equivalent and given only for convenience.
Note that we use a \e{cut-off limit} $a$ ($\mu$ in~\bref{m_Cher_2}), which can lead to loss of tightness in some situations:\fnm\ a more thorough alternative would be to take into account the actual distribution of $\sum\E{\Cl X_i}[\Cl X_1\dc \Cl X_{i-1}]$, which would likely result in a less convenient formulation.\fnt[fn_loss]
{
E.g., it is likely that the bound guaranteed by \lemref{l_ABdecomp} loses a multiplicative log-factor due to this.
}

We prove it via a standard approach to this type of inequalities, due to Bernstein -- namely, by analysing the expectation of the target value's exponent.
The following lemma will be the technical core.

\lem[l_CherE]{Let $\Cl X_1\dc \Cl X_m$ be random variables with countable supports, such that $\forall i\in[m]:\: 0\le \Cl X_i\le1$ and
\m{
\sum_{i=1}^m\E{\Cl X_i}[\Cl X_1\dc \Cl X_{i-1}]\le\mu
.}
Then
\m{
\forall u\ge0:\: 
\E{e^{u\tm\sum_{i=1}^m\Cl X_i}} \le \l((e^u-1)\tm\fr\mu m+1\r)^m
.}
}

\prfstart[\lemref{l_CherE}]
We argue by induction on $m$.

\para{Base case:\: $m=1$.}
\m{
\E{e^{u\tm \Cl X_1}} = \sum_x\PR{\Cl X_1=x}\tm e^{ux}
\le \sum_x\PR{\Cl X_1=x}\tm\l((e^u-1)\tm x+1\r)
\le (e^u-1)\tm\mu + 1
,}
where the summing is over $x\in\supp(\Cl X_1)$ and the first inequality follows from the convexity of $e^{u\tm x}$ on $x\in[0,1]$.

\para{Inductive step:\: $m>1$.}
Let $p_1\deq\E{\Cl X_1}$, then
\m{
\E{e^{u\tm\sum_{i=1}^m\Cl X_i}}
& = \sum_x\PR{\Cl X_1=x} \tm e^{u\tm x}
\tm \E{e^{u\tm\sum_{i=2}^m\Cl X_i}}[{\Cl X_1=x}]\\
& \le \sum_x\PR{\Cl X_1=x} \tm e^{u\tm x}
\tm \l((e^u-1)\tm\fr{\mu-p_1}{m-1}+1\r)^{m-1}\\
& = \E{e^{u\tm \Cl X_1}}
\tm \l((e^u-1)\tm\fr{\mu-p_1}{m-1}+1\r)^{m-1}
,}
as follows from the inductive assumption and the fact that $\sum_{i=2}^m\E{\Cl X_i}[\Cl X_1=x,\, \Cl X_2\dc \Cl X_{i-1}]\le\mu-p_1$ for any $x\in\supp(\Cl X_1)$, and the required
\m{
\ds~
&\le \l((e^u-1)\tm p_1 + 1\r) \tm \l((e^u-1)\tm\fr{\mu-p_1}{m-1}+1\r)^{m-1}\\
&= \l(
\l((e^u-1)\tm p_1 + 1\r)^{\fr1m}
\tm \l((e^u-1)\tm\fr{\mu-p_1}{m-1}+1\r)^{\fr{m-1}m}
\r)^m\\
& \le \l(
\fr{(e^u-1)\tm p_1 + 1}m + \fr{(e^u-1)\tm(\mu-p_1)}m + \fr{m-1}m
\r)^m
= \l((e^u-1)\tm\fr\mu m+1\r)^m
}
follows from the base case and the weighted arithmetic/geometric means inequality.
\prfend

We are ready to prove the relaxed Chernoff bound.

\prfstart[\theoref{t_Cher}]
First suppose
\m{
\sum_{i=1}^m\E{\Cl X_i}[\Cl X_1\dc \Cl X_{i-1}]\le\mu
,}
then
\m{
\forall A\ge\mu,\, u\ge0:\:&
\PR{\sum_{i=1}^m\Cl X_i \ge A}
~\le~ \PR{e^{u\tm\sum_{i=1}^m\Cl X_i} \ge e^{u\tm A}}\\
&\tbb \le~ e^{-u\tm A} \tm \E{e^{u\tm\sum_{i=1}^m\Cl X_i}}
~\le~ e^{-u\tm A} \tm \l((e^u-1)\tm\fr\mu m+1\r)^m
,}
where the last two inequalities are, respectively, Markov's~(\fctref{f_MaChe}) and \lemref{l_CherE}.
Plugging in
\m{
e^u \deq \fr{A\tm (m-\mu)}{\mu\tm (m-A)}
,}
where the right-hand side is at least $1$ (as $A\ge\mu$), we obtain
\m[m_geA]{
\PR{\sum_{i=1}^m\Cl X_i \ge A}
& \le \l(\fr{\mu\tm (m-A)}{A\tm (m-\mu)}\r)^A
\tm \l(\fr{A\tm (m-\mu)-\mu\tm (m-A)}{\mu\tm (m-A)}\tm\fr\mu m+1\r)^m\\
& = \l(\fr{\mu}A\r)^A \tm \l(\fr{m-\mu}{m-A}\r)^{m-A}
,}
which holds whenever $\sum_{i=1}^m\E{\Cl X_i}[\Cl X_1\dc \Cl X_{i-1}]\le\mu$.

Letting $\forall i\in[m]:\: \wbr {\Cl X_i}\=1-\Cl X_i$, we get that if $\sum_{i=1}^m\E{\Cl X_i}[\Cl X_1\dc \Cl X_{i-1}]\ge\mu$, then
\m[m_leA]{
\forall A\le\mu:\:
\PR{\sum_{i=1}^m\Cl X_i \le A}
= \PR{\sum_{i=1}^m\wbr{\Cl X_i} \ge m-A}
\le \l(\fr{m-\mu}{m-A}\r)^{m-A} \tm \l(\fr{\mu}A\r)^A
,}
where the inequality is~\bref{m_geA} (note that $\sum_{i=1}^m\E{\wbr {\Cl X_i}}[\wbr {\Cl X_1}\dc \wbr{\Cl X_{i-1}}]\le m-\mu$).~\fn
{
The identically-looking right-hand sides of~\bref{m_geA} and~\bref{m_leA} are semantically different, as the former corresponds to $[A\ge\mu]$ and the latter to $[A\le\mu]$.
}

Next we are going to use the facts that
\m{
\forall x\ge0:\: \fr{2x}{2+x} \le \ln(1+x) \le x
~~\txt{and}~~
\forall x\in[0,1]:\: \fr{x^2-2x}{2-2x} \le \ln(1-x) \le -x
.}
From~\bref{m_geA} it follows that
\m{
\forall \mu,\, t\ge0:\:&
\sum_{i=1}^m\E{\Cl X_i}[\Cl X_1\dc \Cl X_{i-1}] \le \mu~\Then\\
\ln\l(\PR{\sum_{i=1}^m\Cl X_i \ge \mu + t}\r)
& \le (\mu + t)\tm \ln\l(\fr{\mu}{\mu + t}\r)
+ (m - \mu - t)\tm \ln\l(1+\fr{t}{m - \mu - t}\r)\\
&\hspace{-40pt} \le t - (\mu + t)\tm \ln\l(1+\fr{t}{\mu}\r)
~\le~ t - (\mu + t)\tm\fr{2t}{2\mu + t}
= -\fr{t^2}{2\mu + t}
,}
and from~\bref{m_leA} it follows that
\m{
\forall \mu\ge0,\, t\in[0,\mu]:\:&
\sum_{i=1}^m\E{\Cl X_i}[\Cl X_1\dc \Cl X_{i-1}] \ge \mu~\Then\\
\ln\l(\PR{\sum_{i=1}^m\Cl X_i \le \mu - t}\r)
& \le (\mu - t)\tm \ln\l(\fr{\mu}{\mu - t}\r)
+ (m - \mu + t)\tm \ln\l(1-\fr{t}{m - \mu + t}\r)\\
&\hspace{-40pt} \le -t - (\mu - t)\tm \ln\l(1-\fr{t}{\mu}\r)
~\le~ -t - \l(-t + \fr{t^2}{2\mu}\r)
= -\fr{t^2}{2\mu}
.}
Thus, we have established~\bref{m_Cher_2}.

Let $\eta$ denote the distribution of $\Cl X_1\dc \Cl X_m$, and let $\eta'$ be their distribution, conditioned on $\lf[\sum_{i=1}^m\E[\eta]{\Cl X_i}[\Cl X_1\dc \Cl X_{i-1}]\ge a\rt]$ (assuming that the event's probability is positive).
Then it is easy to see that
\m{
\sum_{i=1}^m\E[\eta']{\Cl X_i}[\Cl X_1\dc \Cl X_{i-1}]\ge a
,}
which, together with~\bref{m_Cher_2}, implies the first inequality in~\bref{m_Cher_1}.
The second one follows by a similar argument.
\prfend

Lastly, there are cases where typical Chernoff-like bounds are insufficient:\ aiming for universality, they usually compromise tightness.
The following ad hoc statement deals with concentration near the expectation under symmetric binomial distributions.

\lem[l_con_bin]{Let $2|m$ and $\Cl X_1\dc \Cl X_m$ be mutually-independent random variables, such that $\forall i\in[m]:\: \PR{\Cl X_i=0} = \PR{\Cl X_i=1} = \dr12$.
Then for every $a < b$ in $[m]\pset0$:
\m{
& \PR{a\le\sum_{i=1}^m\Cl X_i\le b}\\
&\tbbb \ge \sq{\fr2{\pi m}}\tm (b-a)
- \sq{\fr8{9\pi m^3}}\tm \l(\l(b-\fr{m}2\r)^3 - \l(a-\fr{m}2\r)^3\r)
- \asO{\fr1{m}}
.~~\fnm}
}
\fnt
{
Here again, we do not aim for tightness beyond the needs, preferring simplicity and self-containment of the argument:\ in particular, we will use the convenient $[(1-\dr{2d}m)^d \ge 1-\dr{2d^2}m]$ instead of the less wasteful $[(1-\dr{2d}m)^d \ge \exp\l(\dr{-2d^2}{(m-2d)}\r)]$ or directly analysing the left-hand side.
The assumption that $2|m$ is made in order to simplify the notation:\ dropping it would not affect the claimed bound.
}

To prove it, we will use Stirling's approximation in the following form.

\nfct[f_Stir]{Stirling's approximation}{$$n! ~\in~  \sq{2\pi n}\tm \l(\fr{n}e\r)^n\tm \l(1+\asO{\fr1{n}}\r).$$}

\prfstart[\lemref{l_con_bin}]
\m{
\PR{\sum_{i=1}^m\Cl X_i=\fr{m}2}
= \fr[/]{\chs{m}{\dr m2}}{2^m}
= 2^{-m}\tm \fr{m!}{\l((\fr m2)!\r)^2}
\in \sq{\fr2{\pi m}}\tm \l(1\pm\asO{\fr1{m}}\r)
,}
where the containment is \fctref{f_Stir}.
For any $d\in[\dr{m}2]\pset0$,
\m{
\fr{\PR{\sum \Cl X_i=\fr{m}2+d}}{\PR{\sum \Cl X_i=\fr{m}2}}
= \fr{\PR{\sum \Cl X_i=\fr{m}2-d}}{\PR{\sum \Cl X_i=\fr{m}2}}
& = \fr{\l((\fr m2)!\r)^2}{(\fr m2-d)!\tm (\fr m2+d)!}\\
&\hspace{-128pt} = \fr{(\fr m2-d+1)\dtm(\fr{m}2)}{(\fr m2+1)\dtm(\fr m2+d)}
> \l(\fr{\fr m2-d}{\fr m2}\r)^d 
= \l(1-\fr{2d}m\r)^d
\ge 1-\fr{2d^2}m
.}

Then
\m{
\PR{a\le\sum_{i=1}^m\Cl X_i\le b}
\&[-40pt] \ge \PR{\sum \Cl X_i=\fr{m}2}
\tm \int_{a-\fr{m}2}^{b-\fr{m}2} \l(1-\fr{2x^2}m\r) dx\\
& \ge \l(\sq{\fr2{\pi m}}\tm \l(1-\asO{\fr1{m}}\r)\r)
\tm \l(b-a - \fr2{3m}\tm \lf.x^3\rt|_{a-\fr{m}2}^{b-\fr{m}2}\r)\\
& \ge \sq{\fr2{\pi m}}\tm (b-a)
- \sq{\fr8{9\pi m^3}}\tm \l(\l(b-\fr{m}2\r)^3 - \l(a-\fr{m}2\r)^3\r)
- \asO{\fr1{m}}
,}
as desired.
\prfend

\sect[s_cla]{The \R-complexity of \GHR:\ a lower bound}[The R-complexity of GHR: a lower bound]

The definition of \GHR\ relies on the predicate $\aleph(\dt,\dt)$, and it will be useful to know how likely $\aleph(\Cl X,\Cl Y)$ is to hold with respect to our main input distribution for \GHR\ -- the uniform.

\theo[t_aleph]{$$\PR[(\Cl X,\Cl Y)\unin\01^{n+n}]{\aleph(\Cl X,\Cl Y)} \in 1-\asO{\fr{(\ln n)^2}{\sq n}}.$$
}

To prove it, we will use several auxiliary lemmas.
The first one is a concentration statement for the bit-wise XOR of a uniformly-random \f n-bit string with its own cyclic shift.

\lem[l_sXsiX]{Let $i\in[n-1]$, $s\in\01^n$ and ${\Cl A}\sim\U[\01^n]$.
Then
\m{
\forall t\ge0:\:
\PR{\sz{\sz{{\Cl A}\+\sigma_i({\Cl A})\+s} - \fr n2} \ge t}
\le 4\tm e^{\fr{-t^2}{2n}}
.}
}

\prfstart[\lemref{l_sXsiX}]
Let $k|n$ be the orbit length of $\sigma_i$ in $[n]$ and $\s{o_i}[{i\in[\dr{n}k]}]$ be the corresponding partition of $[n]$ into orbits.
For each $i\in[\dr{n}k]$, let $c_i$ be the lexicographically-first element of $o_i$ and $o_i'\deq o_i\smin\s{c_i}$ (note that the sets $\cup_{i} o_i'$ and $\s{c_1\dc c_{\dr{n}k}}$ partition $[n]$).

Then
\m{
\l({\Cl A}\+\sigma_i({\Cl A})\r)|_{\cup o_i'}
\sim \U[\01^{\fr{n\tm (k-1)}k}]
,}
and therefore,
\m{
\PR{\sz{\sz{\l({\Cl A}\+\sigma_i({\Cl A})\+s\r)|_{\cup o_i'}} - \fr{n\tm (k-1)}{2k}} \ge \fr t2}
\le 2\tm e^{\fr{-t^2}{2n}}
}
by Hoeffding's inequality~(\fctref{f_Hoeff}).
Similarly, $\l({\Cl A}\+\sigma_i({\Cl A})\r)|_{\s{c_i}} \unin \01^{\dr{n}k}$ and
\m{
\PR{\sz{\sz{\l({\Cl A}\+\sigma_i({\Cl A})\+s\r)|_{\s{c_i}}} - \fr{n}{2k}} \ge \fr t2}
\le 2\tm e^{\fr{-t^2}{2n}}
.}
By the union bound,
\m{
\PR{\sz{\sz{{\Cl A}\+\sigma_i({\Cl A})\+s} - \fr n2} \ge t}
& \le \PR{\sz{\sz{\l({\Cl A}\+\sigma_i({\Cl A})\+s\r)|_{\cup o_i'}} - \fr{n\tm (k-1)}{2k}} \ge \fr t2}\\
&\tbb + \PR{\sz{\sz{\l({\Cl A}\+\sigma_i({\Cl A})\+s\r)|_{\s{c_i}}} - \fr{n}{2k}} \ge \fr t2}\\
& \le 4\tm e^{\fr{-t^2}{2n}}
,}
as desired.
\prfend

Second, we need a simple bound on the expectation with respect to ``anti-correlating'' probability distribution.

\lem[l_E_mono]{Let $\mu$ be a probability distribution with finite support\fnm~$A$ and $f:A\to\RR$, such that $\forall a_1,\, a_2\in A:\: f(a_1) \le f(a_2) \IfThen \mu(a_1) \ge \mu(a_2)$.\fnt
{
The continuous version can be obtained along similar lines under the integrability assumption.
}
Then
\m{
\E[\Cl X\sim\mu]{f(\Cl X)} \le \E[\Cl X\unin A]{f(\Cl X)}
.}
}

\prfstart[\lemref{l_E_mono}]
\m{
\E[\Cl X\sim\mu]{f(\Cl X)}
& = \sum_{x\in A} f(x)\tm\mu(x)\\
& = \fr{1}{\sz{A}}\tm
\l(
\l(\sum_{x\in A} f(x)\r) \tm \l(\sum_{x\in A}\mu(x)\r)
+\fr12\tm \sum_{x_1\ne x_2} \l(f(x_1)-f(x_2)\r)\tm \l(\mu(x_1)-\mu(x_2)\r)
\r)\\
& = \E[\Cl X\unin A]{f(\Cl X)}
+ \fr{1}{2\tm\sz{A}}\tm\sum_{x_1\ne x_2} \l(f(x_1)-f(x_2)\r)\tm \l(\mu(x_1)-\mu(x_2)\r)\\
& \le \E[\Cl X\unin A]{f(\Cl X)}
,}
as
\m{
\forall x_1,\, x_2\in A:\: 
\l(f(x_1)-f(x_2)\r)\tm \l(\mu(x_1)-\mu(x_2)\r) \le 0
}
by assumption.
\prfend

The third auxiliary lemma bounds the dependence between the Hamming weight of a uniformly-random \f n-bit string $\Cl A$ and that of $\Cl A\+s$ for some fixed $s\in\01^n$.

\lem[l_ABdecomp]{Let $s\in\01^n$, $\sz{\sz{s}-\ud\dr n2}\le d$ and $\Cl A\sim\U[\01^n]$.
Then there exists a random variable $\wtl{\Cl A}$, such that
\m{
\I{|{\Cl A}\+\wtl {\Cl A}\+s|}{\sz{{\Cl A}}} = 0
}
and
\m{
\forall t\ge0:\:
\PR{\sz{\wtl {\Cl A}} \ge \fr{16d^2}n + t}
\le 4\tm e^{\ln n - \fr{t}{4\ln n}}
+ 2\tm e^{\fr{-t^2n}{224\tm t\tm n\tm \ln n + 2048\tm d^2\tm \ln n}}
.}
More specifically,
\m{
&\forall t,\, l\ge0:\\
&\tb \PR{\sz{\wtl {\Cl A}}\ge \fr{(2d+2l)^2}n + t}[{\sz{\sz {\Cl A}-\ud\dr n2}\le l}]
\le 2\tm \l(e^{\fr{-t^2n}{32\tm (2d+2l)^2\tm \ln n+80tn}}
+ e^{\ln n - \fr{t}{2\ln n}}\r)
.}
}

As we will be dealing with the uniform distribution over \f n-bit strings, typical for us will be the regime of $l,\, d\in\asO{\sq n}$, in which case the above statement reads as follows: \e{the value of $\sz{{\Cl A}\+s}$ is poly-logarithmically-close\fnm\ in absolute distance to being independent from $\sz{{\Cl A}}$} -- clearly, this is a very strong bound on the dependence between a pair of random values whose standard deviations are in $\asT{\sq n}$.\fnt
{
The above lemma gives meaningful upper bound for $t\in\asOm{(\ln n)^2}$, which is sufficient for our needs.
We would expect the actual $\sz{\wtl {\Cl A}}$ -- as constructed in the following proof -- to be almost always in $\asO{\ln n}$, or even in $\aso{\ln n}$:\ cf.~Footnote~\ref{fn_loss}.
}

\prfstart[\lemref{l_ABdecomp}]
Assume $\sz{s}\ge\dr n2$ and let $d \deq \sz{s}-\ud\dr n2$ (the case of $\sz{s}<\dr n2$ can be treated similarly).
Let $a_1\dc a_{\dr n2+d}$ be the coordinates where $s(a_i)=1$ and $b_1\dc b_{\fr n2-d}$ be the coordinates where $s(b_i)=0$.

We will construct the distribution of $\wtl{\Cl A}$ bit by bit, then argue that $\sz{{\Cl A}\+\wtl {\Cl A}\+s}$ doesn't depend on $\sz{\Cl A}$.
The high-level intuition behind the construction will be to handle one by one the pairs of coordinates $(a_i,b_j)$ of $\wtl{\Cl A}$:\ in this case the fact that $s(a_i)=1$ and $s(b_j)=0$ allows for a very efficient symmetric argument, which will be described in Stage $2$ below.
But due to the fact that the numbers of \f0- and \f1-coordinates in the string $s$ can be different, we have to treat the ``excess'' separately, and that will be the purpose of Stage $1$ (if we were not assuming for simplicity that $\sz{s}\ge\dr n2$, we would end up with two very symmetric versions of Stage $1$, while Stage $2$ would remain unchanged).

\para{Stage 1.}
Here we describe the distribution of $\wtl{\Cl A}_{a_1\dc a_{2d}}$ (skipping to Stage $2$ if $d=0$).

Let $k_0$ be the value of $\sz{\Cl A}$.
By the bit-symmetry of both the distribution of ${\Cl A}$ and the assumption $[\sz{\Cl A}=k_0]$, it holds that
\m{
\forall i\in[n]:\:
\PR{{\Cl A}_i=1}[{\sz{\Cl A}=k_0}] = \fr{k_0}n
;}
in particular, $\PR{{\Cl A}_{a_1}=1}[{\sz{\Cl A}=k_0}] = \dr{k_0}n$.
Let:
\m{
\PR{\wtl {\Cl A}_{a_1} = 1}
 \deq \Cases
{1-\fr{n}{2k_0}}
{if $2k_0>n$ and ${\Cl A}_{a_1}=1$;}
{1-\fr{n}{2n-2k_0}}
{if $2k_0<n$ and ${\Cl A}_{a_1}=0$;}
{0}{otherwise.}
}

Note that \e{conditioned on $[\sz {\Cl A}=k_0]$}, it is the case that
\m[m_1_1/2]{
\&[16pt]\PR{{\Cl A}_{a_1}\+\wtl{\Cl A}_{a_1}=1}\\
&= \PR{{\Cl A}_{a_1} = 1} \tm \PR{\wtl {\Cl A}_{a_1} = 0}[{{\Cl A}_{a_1} = 1}]
+ \PR{{\Cl A}_{a_1} = 0} \tm \PR{\wtl {\Cl A}_{a_1} = 1}[{{\Cl A}_{a_1} = 0}]\\
&= \Cases
{\fr{k_0}n \tm \fr{n}{2k_0}}
{if $2k_0>n$,}
{\fr{k_0}n + \l(1-\fr{k_0}n\r) \tm \l(1-\fr{n}{2n-2k_0}\r)}
{if $2k_0<n$,}
{\fr{k_0}n}
{if $2k_0=n$}
[\}]
= \fr12
.}

The distribution of $\wtl{\Cl A}_{a_i}$ for each $i\in[2d]$ will depend on $\sz{\Cl A}=k_0$ and on $\sz{{\Cl A}_{a_1\dc a_{i-1}}}$:\ namely, on $k_{i-1}\deq k_0-\sum_{j=1}^{i-1}{\Cl A}_{a_j}~\l(=\sz{{\Cl A}_{[n]\mset{a_1\dc a_{i-1}}}}\r)$.
Additionally, it will depend on the value of ${\Cl A}_{a_i}$.

Let $n_{i-1}\deq n-i+1$; note that for every $t\in[n]\mset{a_1\dc a_{i-1}}$ the probability of $[{\Cl A}_t=1]$, conditioned on the content of ${\Cl A}_{a_1\dc a_{i-1}}$ and on $[\sz{\Cl A}=k_0]$, equals $\fr{k_{i-1}}{n_{i-1}}$.
Generalising the case of $\wtl{\Cl A}_{a_1}$, define:
\m[m_wAa_1_n]{
\forall i\in[2d]:\:
\PR{\wtl {\Cl A}_{a_i} = 1}
 \deq \Cases
{1-\fr{n_{i-1}}{2k_{i-1}}}
{if $2k_{i-1}>n_{i-1}$ and ${\Cl A}_{a_i}=1$;}
{1-\fr{n_{i-1}}{2n_{i-1}-2k_{i-1}}}
{if $2k_{i-1}<n_{i-1}$ and ${\Cl A}_{a_i}=0$;}
{0}{otherwise.}
}

Let $\wbr {\Cl A}\deq {\Cl A}\+\wtl{\Cl A}$.
Then, similarly to \bref{m_1_1/2}, it hods that
\m[m_1_1-2d]{
\PR{\wbr {\Cl A}_{a_i}=1}[{{\Cl A}_{a_1\dc a_{i-1}},\, \sz{\Cl A}}] \= \fr12
.}

Let $i\in[2d]$.
As $\wbr{\Cl A}_{a_1\dc a_i}={\Cl A}_{a_1\dc a_i}\+\wtl{\Cl A}_{a_1\dc a_i}$ and the distribution of $\wtl{\Cl A}_{a_1\dc a_i}$ only depends on $\sz{\Cl A}$ and ${\Cl A}_{a_1\dc a_i}$,
\m{
\I{\wbr {\Cl A}_{a_1\dc a_i}}{{\Cl A}}[{\Cl A}_{a_1\dc a_i},\, \sz{\Cl A}] = 0
,}
and therefore,
\m[m_hA]{
\h{\wbr {\Cl A}_{a_1\dc a_i}}[{{\Cl A}_{a_1\dc a_i},\, \sz{\Cl A}}]
= \h{\wbr {\Cl A}_{a_1\dc a_i}}[{{\Cl A}_{a_1\dc a_i},\, \sz{\Cl A},\, {\Cl A}}]
= \h{\wbr {\Cl A}_{a_1\dc a_i}}[{\Cl A}]
.}

As the distribution of $\wtl{\Cl A}$ only depends on ${\Cl A}$ and there are no other ``inter-bit'' dependencies,
\m{
\I{\wtl {\Cl A}_{a_i}}{\wtl{\Cl A}_{a_1\dc a_{i-1}}}[{\Cl A}] = \I{\wbr {\Cl A}_{a_i}}{\wbr{\Cl A}_{a_1\dc a_{i-1}}}[{\Cl A}] = 0
,}
and therefore,
\m{
\h{\wbr {\Cl A}_{a_1\dc a_{i-1}}}[\wbr{\Cl A}_{a_i},\, {\Cl A}]
= \h{\wbr {\Cl A}_{a_1\dc a_{i-1}}}[{\Cl A}]
= \h{\wbr {\Cl A}_{a_1\dc a_{i-1}}}[{\Cl A}_{a_1\dc a_{i-1}},\, \sz{\Cl A}]
,}
where the latter equality is \bref{m_hA}.
So,
\m[m_1_Ia]{
0
& = \h{\wbr {\Cl A}_{a_1\dc a_{i-1}}}[{{\Cl A}_{a_1\dc a_{i-1}},\, \sz{\Cl A}}] - \h{\wbr {\Cl A}_{a_1\dc a_{i-1}}}
[{\wbr{\Cl A}_{a_i},\, {\Cl A},\, {\Cl A}_{a_1\dc a_{i-1}},\, \sz{\Cl A}}]\\
& = \I{\wbr {\Cl A}_{a_1\dc a_{i-1}}}{\wbr{\Cl A}_{a_i},\, {\Cl A}}[{\Cl A}_{a_1\dc a_{i-1}},\, \sz{\Cl A}]
\ge \I{\wbr {\Cl A}_{a_1\dc a_{i-1}}}{\wbr{\Cl A}_{a_i}}[{\Cl A}_{a_1\dc a_{i-1}},\, \sz{\Cl A}]
.}
Therefore,
\m{
\h{\wbr {\Cl A}_{a_i}}[{\Cl A}_{a_1\dc a_{i-1}},\, \sz{\Cl A}]
& = \h{\wbr {\Cl A}_{a_i}}[{{\Cl A}_{a_1\dc a_{i-1}},\, \sz{\Cl A},\, \wbr{\Cl A}_{a_1\dc a_{i-1}}}]\\
& \le \h{\wbr {\Cl A}_{a_i}}[{\wbr{\Cl A}_{a_1\dc a_{i-1}},\, \sz{\Cl A}}]
}
and
\m[m_hwA_1_2d]{
\h{\wbr {\Cl A}_{a_1\dc a_{2d}}}[\sz{\Cl A}]
& = \sum_{i=1}^{2d} \h{\wbr {\Cl A}_{a_i}}[{\wbr{\Cl A}_{a_1\dc a_{i-1}},\, \sz{\Cl A}}]\\
& \ge \sum_{i=1}^{2d} \h{\wbr {\Cl A}_{a_i}}[{{\Cl A}_{a_1\dc a_{i-1}},\, \sz{\Cl A}}]
= 2d
,}
where the concluding equality is \bref{m_1_1-2d}.

Let us estimate $\sz{\wtl {\Cl A}_{a_1\dc a_{2d}}}$.
If we denote $l_i\deq\sz{k_{i}-\dr{n_{i}}2}$, then from~\bref{m_wAa_1_n} it follows that
\m{
\forall i\in[2d]:\:
\PR{\wtl {\Cl A}_i = 1}[{\wtl{\Cl A}_{[i-1]}}] \in \s{\fr{2l_{i-1}}{n_{i-1}+2l_{i-1}},0}
}
always.
As $0\le l_i\le \sz{\sz {\Cl A}-\ud\dr n2}+\dr i2$ always,
\m{
\PR{\wtl {\Cl A}_i = 1}[{\wtl{\Cl A}_{[i-1]}}] \le \fr{2l_0+2d}{n}
,}
where $l_0=\sz{\sz {\Cl A}-\ud\dr n2}$, and therefore,
\m{
\sum_{i=1}^{2d}\E{\wtl {\Cl A}_i}[\wtl{\Cl A}_{[i-1]}]\le\fr{4dl_0+4d^2}n
.}
It follows from the relaxed Chernoff bound (\theoref{t_Cher}) that
\m[m_tail_wA_1_2d]{
\forall t,\, l\ge0:\:
\PR{\sz{\wtl {\Cl A}_{a_1\dc a_{2d}}}\ge \fr{4dl+4d^2}n + t}[{\sz{\sz {\Cl A}-\ud\dr n2}=l}]
\le e^{\fr{-t^2\tm n}{8dl+8d^2+tn}}
.}
Note that in the typical regime of $t,\, l\in\asO{\sq n}$ the right-hand side of \bref{m_tail_wA_1_2d} is $e^{-\asOm t}$.

\para{Stage 2.}
Here we describe the distribution of $\wtl{\Cl A}_{a_{2d+1}\dc a_{\fr n2+d},b_1\dc b_{\fr n2-d}}$.

Let $k_{2d}$ denote the Hamming weight of the remaining bits of ${\Cl A}$ after the completion of Stage $1$, namely
\m{
k_{2d}\deq\sz{{\Cl A}_{a_{2d+1}\dc a_{\fr n2+d},b_1\dc b_{\fr n2-d}}}
}
(note that if $d=0$ and Stage $1$ has been skipped, then $k_{2d}=k_0=\sz{\Cl A}$).
As before, $\forall i\in[n]-1:\: n_i=n-i$.

Our tactics will be to handle one by one the $\fr n2-d$ pairs $(a_{2d+i},b_i)$ and toss $\wtl{\Cl A}_{a_{2d+i},b_i}$ in a way that will guarantee – conditional on all the earlier choices – that the distribution of every $\sz{{\Cl A}_{a_{2d+i},b_i}\+\wtl{\Cl A}_{a_{2d+i},b_i}\+s_{a_{2d+i},b_i}}$ is the same as the distribution of $\Cl B_1+\Cl B_2$ for $(\Cl B_1,\Cl B_2)\unin\01^2$.

Conditional on $\lf[\sz{{\Cl A}_{a_{2d+1}\dc a_{\fr n2+d},b_1\dc b_{\fr n2-d}}} = k_{2d}\rt]$, the distribution of $({\Cl A}_{a_{2d+1}},{\Cl A}_{b_1})$ (as well as of any other pair of coordinates) is perfectly symmetric -- although most likely not independent.
Accordingly, conditional on $\lf[{\Cl A}_{a_{2d+1}}\ne{\Cl A}_{b_1}\rt]$, the events $\lf[\sz{{\Cl A}_{a_{2d+1},b_1}\+s_{a_{2d+1},b_1}}=0\rt]$ and $\lf[\sz{{\Cl A}_{a_{2d+1},b_1}\+s_{a_{2d+1},b_1}}=2\rt]$ are equiprobable -- just like the corresponding events with respect to $\sz{{\Cl B}_1+{\Cl B}_2}$ -- so, we only need to use $\wtl{\Cl A}_{a_{2d+1},b_1}$ in order to balance the probabilities of $\lf[{\Cl A}_{a_{2d+1}}\+\wtl{\Cl A}_{a_{2d+1}}= {\Cl A}_{b_1}\+\wtl{\Cl A}_{b_1}\rt]$ and $\lf[{\Cl A}_{a_{2d+1}}\+\wtl{\Cl A}_{a_{2d+1}}\ne{\Cl A}_{b_1}\+\wtl{\Cl A}_{b_1}\rt]$ (trying to keep the Hamming weight of $\wtl{\Cl A}$ as small as possible).

It holds that
\m{
\PR{{\Cl A}_{a_{2d+1}} \ne{\Cl A}_{b_1}}
= \fr{2k_{2d}\tm(n_{2d}-k_{2d})}{n_{2d}\tm(n_{2d}-1)}
.}
First let us define a random variable ${\Cl Z}_1$, such that $\PR{{\Cl A}_{a_{2d+1}} \+ {\Cl A}_{b_1} \+ {\Cl Z}_1=1}=\dr12$, afterwards we will use it to construct the distribution of $\wtl{\Cl A}_{a_{2d+1},b_1}$.
Let:
\m[m_def_Z1]{
&\PR{{\Cl Z}_1 = 1}\\
&\tbb \deq \Cases
{1 - \fr{n_{2d}\tm(n_{2d}-1)}{4k_{2d}\tm(n_{2d}-k_{2d})} }
{if $\sz{\fr{n_{2d}}2-k_{2d}} < \fr{\sq{n_{2d}}}{2}$ and ${\Cl A}_{a_{2d+1}} \ne{\Cl A}_{b_1}$;}
{1 - \fr{n_{2d}\tm(n_{2d}-1)}{2n_{2d}\tm(n_{2d}-1) - 4k_{2d}\tm(n_{2d}-k_{2d})} }
{if $\sz{\fr{n_{2d}}2-k_{2d}} > \fr{\sq{n_{2d}}}{2}$ and ${\Cl A}_{a_{2d+1}} = {\Cl A}_{b_1}$;}
{0}{otherwise.}
}
Note that
\m{
n_{2d}\tm(n_{2d}-1) ~=~ 4k_{2d}\tm(n_{2d}-k_{2d})
~~\IfThen~~
\sz{\fr{n_{2d}}2-k_{2d}} ~=~ \fr{\sq{n_{2d}}}{2}
,}
and therefore the probability of $[{\Cl Z}_1 = 1]$ is well-defined.

Then
\m[m_2_1/2]{
\PR{{\Cl A}_{a_{2d+1}} \+ {\Cl A}_{b_1} \+ {\Cl Z}_1=1}
&= \PR{{\Cl A}_{a_{2d+1}} \ne{\Cl A}_{b_1}} \tm \PR{{\Cl Z}_1 = 0}[{{\Cl A}_{a_{2d+1}} \ne{\Cl A}_{b_1}}]\\
&\tbbb + \PR{{\Cl A}_{a_{2d+1}} = {\Cl A}_{b_1}} \tm \PR{{\Cl Z}_1 = 1}[{{\Cl A}_{a_{2d+1}} = {\Cl A}_{b_1}}]\\
&= \Cases
{\fr{2k_{2d}\tm(n_{2d}-k_{2d})}{n_{2d}\tm(n_{2d}-1)}
\tm \fr{n_{2d}\tm(n_{2d}-1)}{4k_{2d}\tm(n_{2d}-k_{2d})}}
{if $\sz{\fr{n_{2d}}2-k_{2d}} < \fr{\sq{n_{2d}}}{2}$,}
{\redstar}{if $\sz{\fr{n_{2d}}2-k_{2d}} > \fr{\sq{n_{2d}}}{2}$,}
{\fr{2(\fr{n_{2d}}2-\fr{\sq{n_{2d}}}{2})(\fr{n_{2d}}2+\fr{\sq{n_{2d}}}{2})}{n_{2d}\tm(n_{2d}-1)}}
{if $\sz{\fr{n_{2d}}2-k_{2d}} = \fr{\sq{n_{2d}}}{2}$}\\
&= \fr12
,}
where $\redstar$ stands for
\m{
&\fr{2k_{2d}\tm(n_{2d}-k_{2d})}{n_{2d}\tm(n_{2d}-1)}\\
&\tbbb + \l(1-\fr{2k_{2d}\tm(n_{2d}-k_{2d})}{n_{2d}\tm(n_{2d}-1)}\r)
\tm \l(1 - \fr{n_{2d}\tm(n_{2d}-1)}{2n_{2d}\tm(n_{2d}-1) - 4k_{2d}\tm(n_{2d}-k_{2d})}\r)\\
& = 1
- \fr{n_{2d}\tm(n_{2d}-1)} {2n_{2d}\tm(n_{2d}-1) - 4k_{2d}\tm(n_{2d}-k_{2d})}
+ \fr{2k_{2d}\tm(n_{2d}-k_{2d})} {2n_{2d}\tm(n_{2d}-1) - 4k_{2d}\tm(n_{2d}-k_{2d})}
= \fr12
.}

Let us convert the content of ${\Cl Z}_1$ into a distribution of $\wtl{\Cl A}_{a_{2d+1}}$ and $\wtl{\Cl A}_{b_1}$:
\m{
& [{\Cl Z}_1 = 0] ~\Then~ [\wtl{\Cl A}_{a_{2d+1},b_1} \= (0,0)];\\
& \PR{\wtl {\Cl A}_{a_{2d+1},b_1} = (0,1)}[{{\Cl Z}_1 = 1}] = \PR{\wtl {\Cl A}_{a_{2d+1},b_1} = (1,0)}[{{\Cl Z}_1 = 1}] = \fr12.
}
Now it is easy to see that the distribution of $\sz{{\Cl A}_{a_{2d+1},b_1}\+\wtl{\Cl A}_{a_{2d+1},b_1}\+s_{a_{2d+1},b_1}}$, conditioned on $\lf[\sz{{\Cl A}_{a_{2d+1}\dc a_{\fr n2+d},b_1\dc b_{\fr n2-d}}}=k_{2d}\rt]$, is the same as the distribution of ${\Cl B}_1+{\Cl B}_2$ for $({\Cl B}_1,{\Cl B}_2)\unin\01^2$.~\fn
{
Note that the distribution of $\l({\Cl A}_{a_{2d+1},b_1}\+\wtl{\Cl A}_{a_{2d+1},b_1}\r)$ can be non-uniform, and this can be intuitively viewed as the \e{resource} that allows savings in terms of $\sz{\wtl {\Cl A}}$ in Stage $2$ in comparison to Stage $1$.
Now we do not have to worry if the probabilities of $\lf[{\Cl A}_{a_{2d+1},b_1}\+\wtl{\Cl A}_{a_{2d+1},b_1}=(0,0)\rt]$ and $\lf[{\Cl A}_{a_{2d+1},b_1}\+\wtl{\Cl A}_{a_{2d+1},b_1}=(1,1)\rt]$ are different from each other, which indeed happens in the above construction if $\sz{\fr{n_{2d}}2-k_{2d}}$ is larger than $\fr{\sq{n_{2d}}}{2}$. 
}
By the symmetry of all the involved distributions with respect to the bit-positions $a_{2d+1}$ and $b_1$, we only need to check that
\m{
\PR{{\Cl A}_{a_{2d+1}}\+\wtl{\Cl A}_{a_{2d+1}}= {\Cl A}_{b_1}\+\wtl{\Cl A}_{b_1}}
~=~
\PR{{\Cl A}_{a_{2d+1}}\+\wtl{\Cl A}_{a_{2d+1}}\ne{\Cl A}_{b_1}\+\wtl{\Cl A}_{b_1}}
,}
and this follows trivially from~\bref{m_2_1/2} and the symmetry of \ord[\wtl{\Cl A}] distribution with respect to those two bit-positions.
In other words,
\m[m_2_1/4]{
\sz{{\Cl A}_{a_{2d+1},b_1}\+\wtl{\Cl A}_{a_{2d+1},b_1}\+s_{a_{2d+1},b_1}} =
\Cases
{0}{with probability $\dr14$;}
{1}{with probability $\dr12$;}
{2}{with probability $\dr14$.}
}

The distribution of ${\Cl Z}_i$ for each $i\in[\dr n2-d]$ will depend, naturally, on $\sz{{\Cl A}_{a_{2d+1}\dc a_{\fr n2+d},b_1\dc b_{\fr n2-d}}}=k_{2d}$, as well as on $\sz{{\Cl A}_{a_{2d+1}\dc a_{2d+i-1},b_1\dc b_{i-1}}}$:\ namely, on
\m{
k_{2d+2i-2}\deq k_{2d}
- \sum_{j=1}^{i-1}\l({\Cl A}_{a_{2d+j}} + {\Cl A}_{b_j}\r)
\tb\l(=~\sz{{\Cl A}_{a_{2d+i}\dc a_{\fr n2+d},b_i\dc b_{\fr n2-d}}}\r)
.}
Additionally, it will depend on ${\Cl A}_{a_{2d+i}} \+ {\Cl A}_{b_i}$.

Note that 
\m{
&\forall i\in[\dr n2-d],\, j\in\s{a_{2d+i}\dc a_{\fr n2+d},b_i\dc b_{\fr n2-d}}:\\
&\tbb \PR{{\Cl A}_j=1}[{\sz{{\Cl A}_{a_{2d+i}\dc a_{\fr n2+d},b_i\dc b_{\fr n2-d}}}=k_{2d+2i-2}}]
= \fr{k_{2d+2i-2}}{n_{2d+2i-2}}
}
and -- as ${\Cl A}_{a_{2d+i}}$ and ${\Cl A}_{b_i}$ are not, in general, independent --
\m{
\PR{{\Cl A}_{a_{2d+i}} \ne{\Cl A}_{b_i}}[{\sz{{\Cl A}_{a_{2d+i}\dc a_{\fr n2+d},b_i\dc b_{\fr n2-d}}}=k_{2d+2i-2}}]
= \fr{2k_{2d+2i-2}\tm(n_{2d+2i-2}-k_{2d+2i-2})}{n_{2d+2i-2}\tm(n_{2d+2i-2}-1)}
.}
Generalising the case of ${\Cl Z}_1$~\bref{m_def_Z1}, define for $\redstar\deq 2d+2i-2$:
\m[m_def_Zi]{
&\forall i\in[\dr n2-d]:\\
&\tb \PR{{\Cl Z}_i = 1}
\deq \Cases
{1 - \fr{n_{\redstar}\tm(n_{\redstar}-1)}{4k_{\redstar}\tm(n_{\redstar}-k_{\redstar})} }
{if $\sz{\fr{n_{\redstar}}2-k_{\redstar}} < \fr{\sq{n_{\redstar}}}{2}$ and ${\Cl A}_{a_{2d+i}} \ne{\Cl A}_{b_i}$;}
{1 - \fr{n_{\redstar}\tm(n_{\redstar}-1)}{2n_{\redstar}\tm(n_{\redstar}-1) - 4k_{\redstar}\tm(n_{\redstar}-k_{\redstar})} }
{if $\sz{\fr{n_{\redstar}}2-k_{\redstar}} > \fr{\sq{n_{\redstar}}}{2}$ and ${\Cl A}_{a_{2d+i}} = {\Cl A}_{b_i}$;}
{0}{otherwise.}
}
Then, similarly to \bref{m_2_1/2}, it holds that
\m[m_2_Z_all]{
\PR{{\Cl A}_{a_{2d+i}} \+ {\Cl A}_{b_i} \+ {\Cl Z}_i=1}[{{\Cl A}_{a_1\dc a_{2d+i-1},b_1\dc b_{i-1}},\, \sz{\Cl A}}] \= \fr12
.}

Define the distribution of $\wtl{\Cl A}_{a_{2d+i}}$ and $\wtl{\Cl A}_{b_i}$:
\m{
& [{\Cl Z}_i = 0] ~\Then~ [\wtl{\Cl A}_{a_{2d+i},b_i} \= (0,0)];\\
& \PR{\wtl {\Cl A}_{a_{2d+i},b_i} = (0,1)}[{{\Cl Z}_i = 1}] = \PR{\wtl {\Cl A}_{a_{2d+i},b_i} = (1,0)}[{{\Cl Z}_i = 1}] = \fr12.
}
Similarly to \bref{m_2_1/4}, conditional on ${\Cl A}_{a_1\dc a_{2d+i-1},b_1\dc b_{i-1}}$ and $\sz{\Cl A}$ it holds that
\m[m_2_1/4_all]{
\sz{\l({\Cl A}\+\wtl {\Cl A}\+s\r)_{a_{2d+i},b_i}} = \Cases
{0}{with probability $\dr14$;}
{1}{with probability $\dr12$;}
{2}{with probability $\dr14$.}
}

From now and until the end of Stage $2$, all the distributions are assumed to be conditioned on
\m[m_2_cond]{
\lf[\sz{{\Cl A}_{a_{2d+1}\dc a_{\fr n2+d},b_1\dc b_{\fr n2-d}}}=k_{2d}\rt]
}
for some $k_{2d}\in[n-2d]\pset0$, unless stated otherwise.

Let us argue that the distribution of
\m{
\sz{\l({\Cl A}\+\wtl {\Cl A}\+s\r)_{a_{2d+1}\dc a_{\fr n2+d},b_1\dc b_{\fr n2-d}}}
}
equals the distribution of $\sz{{\Cl B}}$ for ${\Cl B}\sim\U[\01^{n-2d}]$.
We prove it by induction.

Let $i\in[\dr n2-d]$, we claim that the distribution of $\sz{\l({\Cl A}\+\wtl {\Cl A}\+s\r)_{a_{2d+1}\dc a_{2d+i},b_1\dc b_{i}}}$ equals that of $\sz{{\Cl B}_{[2i]}}$.
For $i=1$ the statement follows readily from~\bref{m_2_1/4_all}, so assume that $i>1$ and the distribution of $\sz{\l({\Cl A}\+\wtl {\Cl A}\+s\r)_{a_{2d+1}\dc a_{2d+i-1},b_1\dc b_{i-1}}}$ equals that of $\sz{{\Cl B}_{[2i-2]}}$.

By the argument, fully analogous to that behind~\bref{m_1_Ia}, it holds that
\m{
\I{\l({\Cl A}\+\wtl {\Cl A}\+s\r)_{a_{2d+1}\dc a_{2d+i-1},b_1\dc b_{i-1}}}{\l({\Cl A}\+\wtl {\Cl A}\+s\r)_{a_{2d+i},b_i}}[{\Cl A}_{a_{2d+1}\dc a_{2d+i-1},b_1\dc b_{i-1}}] = 0
.}
Then~\bref{m_2_1/4_all} implies that for any $w\in\01^n$, the distribution of $\l({\Cl A}\+\wtl {\Cl A}\+s\r)_{a_{2d+i},b_i}$ conditioned on $\sz{\l({\Cl A}\+\wtl {\Cl A}\+s\r)_{a_{2d+1}\dc a_{2d+i-1},b_1\dc b_{i-1}}}$ equals that of $\sz{{\Cl B}_{[2]}}$, and therefore the distribution of
\m{
&\sz{\l({\Cl A}\+\wtl {\Cl A}\+s\r)_{a_{2d+1}\dc a_{2d+i},b_1\dc b_{i}}}\\
&\tbbb = \sz{\l({\Cl A}\+\wtl {\Cl A}\+s\r)_{a_{2d+1}\dc a_{2d+i-1},b_1\dc b_{i-1}}} + \sz{\l({\Cl A}\+\wtl {\Cl A}\+s\r)_{a_{2d+i},b_i}}
}
equals that of $\sz{{\Cl B}_{[2i]}}$, as required for the induction step.

\para{Completion.}
Let us no longer implicitly assume~\bref{m_2_cond}.
We have just seen that the distribution of
\m{
\sz{\l({\Cl A}\+\wtl {\Cl A}\+s\r)_{a_{2d+1}\dc a_{\fr n2+d},b_1\dc b_{\fr n2-d}}}
,}
conditioned on
\m{
\sz{{\Cl A}_{a_{2d+1}\dc a_{\fr n2+d},b_1\dc b_{\fr n2-d}}}
= \sz{{\Cl A}} - \sz{{\Cl A}_{a_1\dc a_{2d}}}
,}
equals that of $\sz{{\Cl B}}$ for ${\Cl B}\sim\U[\01^{n-2d}]$.
It is clear from the construction and the symmetry of the considered distributions of ${\Cl A}$, that additional conditioning on the content of ${\Cl A}_{a_1\dc a_{2d}}$ doesn't affect the above statement.

From~\bref{m_hwA_1_2d}:
\m{
\h{\l({\Cl A}\+\wtl {\Cl A}\r)_{a_1\dc a_{2d}}}[\sz{\Cl A}] = 2d
~\Then~
\h{\l({\Cl A}\+\wtl {\Cl A}\+s\r)_{a_1\dc a_{2d}}}[\sz{\Cl A}] = 2d 	
,}
and therefore the distribution of
\m{
\sz{\l({\Cl A}\+\wtl {\Cl A}\+s\r)_{a_1\dc a_{2d}}}
,}
conditioned on $\sz{\Cl A}$, equals that of $\sz{{\Cl B}}$ for ${\Cl B}\sim\U[\01^{2d}]$.

By the argument, fully analogous to that behind~\bref{m_1_Ia}, it holds that
\m{
\I{\l({\Cl A}\+\wtl {\Cl A}\+s\r)_{a_1\dc a_{2d}}}{\l({\Cl A}\+\wtl {\Cl A}\+s\r)_{a_{2d+1}\dc a_{\fr n2+d},b_1\dc b_{\fr n2-d}}}[{\Cl A}_{a_1\dc a_{2d}},\, \sz{{\Cl A}}] = 0
.}
Therefore, conditioned on $[\sz{\Cl A}=k_0]$,
\m{
\sz{{\Cl A}\+\wtl {\Cl A}\+s}
= \sz{{\Cl A}_{a_{2d+1}\dc a_{\fr n2+d},b_1\dc b_{\fr n2-d}}}
+ \sz{\l({\Cl A}\+\wtl {\Cl A}\+s\r)_{a_1\dc a_{2d}}}
}
is distributed like $\sz{{\Cl B}}$ for ${\Cl B}\sim\U[\01^n]$ -- in particular, the fact that the above distribution doesn't depend on $k_0$ implies that
\m{
\I{\sz{{\Cl A}\+\wtl {\Cl A}\+s}}{\sz{{\Cl A}}} = 0
,}
as required.

It remains to argue that $\sz{\wtl {\Cl A}}$ is likely to be small.
Towards estimating $\sz{\wtl {\Cl A}_{a_1\dc a_{2d}}}$ we will use~\bref{m_tail_wA_1_2d}, now let us analyse $\sz{\wtl {\Cl A}_{a_{2d+1}\dc a_{\fr n2+d},b_1\dc b_{\fr n2-d}}}$, as constructed during Stage $2$.

Let $\forall j\in[n]-1:\: l_j\deq\sz{k_{j}-\dr{n_{j}}2}$ and $\forall i\in[\dr n2-d]:\: \redstar\deq 2d+2i-2$, then it follows from~\bref{m_def_Zi} that
\m[P]{
\txt{if $l_{\redstar} = \fr{\sq{n_{\redstar}}}{2}$, then }
\PR{{\Cl Z}_i = 1} &= 0;\\
\txt{if $l_{\redstar} < \fr{\sq{n_{\redstar}}}{2}$, then }
\PR{{\Cl Z}_i = 1}
& \le 1 - \fr{n_{\redstar}\tm(n_{\redstar}-1)}{4k_{\redstar}\tm(n_{\redstar}-k_{\redstar})}
= 1 - \fr{n_{\redstar}^2-n_{\redstar}}{4\l(\fr{n_{\redstar}}2-l_{\redstar}\r)\l(\fr{n_{\redstar}}2+l_{\redstar}\r)}\\
& = \fr{n_{\redstar}-4l_{\redstar}^2}{n_{\redstar}^2-4l_{\redstar}^2}
\le \fr{1}{n_{\redstar}};\\
\txt{if $l_{\redstar} > \fr{\sq{n_{\redstar}}}{2}$, then }
\PR{{\Cl Z}_i = 1}
& \le 1 - \fr{n_{\redstar}^2-n_{\redstar}}{n_{\redstar}^2-2n_{\redstar}+4l_{\redstar}^2}\\
& = \fr{4l_{\redstar}^2-n_{\redstar}}{n_{\redstar}^2-2n_{\redstar}+4l_{\redstar}^2}
\le \fr{4l_{\redstar}^2}{n_{\redstar}^2-n_{\redstar}+4l_{\redstar}^2}
< \fr{4l_{\redstar}^2}{n_{\redstar}^2}
.}
Therefore, $\forall i\in\lf[\dr n2-d\rt]:$
\m{
\PR{{\Cl Z}_i = 1}
\le \fr{1}{n_{2d+2i-2}} + \fr{4l_{2d+2i-2}^2}{n_{2d+2i-2}^2}
= \fr{1}{n_{2d+2i-2}} + \l(\fr{2k_{2d+2i-2}}{n_{2d+2i-2}}-1\r)^2
.}
By construction, subject to the assignment in~\bref{m_def_Zi}, every ${\Cl Z}_i$ is independent from all other variables -- in particular,
\m[m_bou_Zi]{
\forall i\in\lf[\fr n2-d\rt]:\: 
\PR{{\Cl Z}_i = 1}[{{\Cl Z}_{[i-1]}}]
\le \fr{1}{n_{2d+2i-2}} + \l(\fr{2k_{2d+2i-2}}{n_{2d+2i-2}}-1\r)^2
.}

For all $i\in[n]-1$, let $K_i$ be a random variable that takes the value of $k_{i}$ (in particular, $K_0=\sz{{\Cl A}}$).
For $u\ge0$, denote by $\bm e_u$ the event
\m{
\lf[
\forall i\in[n]-1:\:
\sz{K_i - \sz{{\Cl A}}\tm \fr{n_i}n} \le u\tm \sq{n_i}
\rt]
.}
Then
\m[m_not_eu]{
\PR{\neg \bm e_u}
\le\sum_{i=0}^{n-1} \PR{\sz{K_i - \sz{{\Cl A}}\tm \fr{n_i}n} > u\tm \sq{n_i}}
\le\sum_{i=0}^{n-1} 2\tm e^{\fr{-2u^2n_i}{n_i}}
= 2n\tm e^{-2u^2}
,}
where the second inequality is Hoeffding's (\fctref{f_Hoeff}).

Assuming $\bm e_u$ and letting $l\deq\sz{\sz {\Cl A}-\ud\dr n2}$,
\m{
\forall i\in[n]-1:\:
\sz{\fr{2k_i}{n_i}-1}
\le \sz{\fr{2\tm \sz{{\Cl A}}}n-1} + \fr{u}{\sq{n_i}}
= \fr{2l}n + \fr{u}{\sq{n_i}}
,}
and therefore~\bref{m_bou_Zi} gives
\m{
\forall i\in\lf[\fr n2-d\rt]:\: 
\PR{{\Cl Z}_i = 1}[{{\Cl Z}_{[i-1]},\, \bm e_u}]
\le \fr{1+2u^2}{n_{2d+2i-2}} + \fr{8l^2}{n^2}
}
and
\m{
\sum_{i\in\lf[\fr n2-d\rt]} \PR{{\Cl Z}_i = 1}[{{\Cl Z}_{[i-1]},\, \bm e_u}]
& \le \fr{4l^2}n + \l(1+2u^2\r)\tm \sum_{i\in\lf[\fr n2-d\rt]} \fr1{n_{2d+2i-2}}\\
& \le \fr{4l^2}n + \l(1+2u^2\r)\tm\ln n
.}
Combined with~\bref{m_not_eu}, this gives
\m{
\PR{
\sum_{i\in\lf[\fr n2-d\rt]} \PR{{\Cl Z}_i = 1}[{{\Cl Z}_{[i-1]}}] > \fr{4l^2}n + \l(1+2u^2\r)\tm\ln n
}
\le \PR{\neg \bm e_u}
\le 2n\tm e^{-2u^2}
.}
Applying the relaxed Chernoff bound (\theoref{t_Cher}) with $a\deq\fr{4l^2}n + \l(1+2u^2\r)\tm\ln n$ gives
\m{
\forall u,\, t,\, l\ge0:\hspace{-40pt}&\\
& \PR{\sz{{\Cl Z}}\ge \fr{4l^2}n + \l(1+2u^2\r)\tm\ln n + t}[{\sz{\sz {\Cl A}-\ud\dr n2}=l}]\\
&\tbbb \le e^{\fr{-t^2}{\dr{8l^2}n+(2+4u^2)\tm \ln n+t}} + 2\tm e^{\ln n - 2u^2}
.}
As $\sz{\wtl {\Cl A}_{a_{2d+1}\dc a_{\fr n2+d},b_1\dc b_{\fr n2-d}}}\=\sz{{\Cl Z}}$ by construction, combined with~\bref{m_tail_wA_1_2d}, this gives
\m{
\forall u,\, t,\, l\ge0:\hspace{-40pt}&\\
& \PR{\sz{\wtl {\Cl A}}\ge \fr{4dl+4d^2+4l^2}n + \l(1+2u^2\r)\tm\ln n + 2t}[{\sz{\sz {\Cl A}-\ud\dr n2}=l}]\\
&\tbbb \le e^{\fr{-t^2\tm n}{8dl+8d^2+tn}}
+ e^{\fr{-t^2\tm n}{8l^2+(2+4u^2)\tm n\tm \ln n+tn}} + 2\tm e^{\ln n - 2u^2}
.}

Let us denote $T\deq 4t\sq{\ln n}$ and choose $u^2\deq\fr{t}{\sq{\ln n}}=\fr{T}{4\ln n}$.
Assume for now that $T\ge4\ln n$, then
\m{
\fr{4dl+4d^2+4l^2}n + \l(1+2u^2\r)\tm\ln n + 2t
\le \fr{(2d+2l)^2}n + T
,}
and so,
\m{
\forall l\ge0,\, T\ge4\ln n:\: 
\PR{\sz{\wtl {\Cl A}}\ge \fr{(2d+2l)^2}n + T}[{\sz{\sz {\Cl A}-\ud\dr n2}=l}]
\le{\redstar}
,}
where
\m{
\redstar
& = e^{\fr{-t^2\tm n}{8dl+8d^2+tn}}
+ e^{\fr{-t^2\tm n}{8l^2+(2+4u^2)\tm n\tm \ln n+tn}} + 2\tm e^{\ln n - 2u^2}\\
& \le 2\tm e^{\fr{-T^2n}{32\tm (2d+2l)^2\tm \ln n+80Tn}}
+ 2\tm e^{\ln n - \fr{T}{2\ln n}}
.}
As we always assume that $n\ge4$, the condition $[T\ge4\ln n]$ can be removed, as otherwise $2\tm e^{\ln n - \fr{T}{2\ln n}}>1$ and the inequality holds trivially.
So,
\m{
&\forall l,\, T\ge0:\\
&\tb \PR{\sz{\wtl {\Cl A}}\ge \fr{(2d+2l)^2}n + T}[{\sz{\sz {\Cl A}-\ud\dr n2}=l}]
\le 2\tm \l(e^{\fr{-T^2n}{32\tm (2d+2l)^2\tm \ln n+80Tn}}
+ e^{\ln n - \fr{T}{2\ln n}}\r)
.}
As a function of $s$ and $l$, the above bound on $\PR{\sz{\wtl {\Cl A}}>s}[{\sz{\sz {\Cl A} - \dr n2\ud}=l}]$ is monotonically increasing in $l$, and therefore,
\m{
\PR{\sz{\wtl {\Cl A}}\ge \fr{(2d+2l)^2}n + T}[{\sz{\sz {\Cl A}-\ud\dr n2}\le l}]
\le 2\tm \l(e^{\fr{-T^2n}{32\tm (2d+2l)^2\tm \ln n+80Tn}}
+ e^{\ln n - \fr{T}{2\ln n}}\r)
,}
as required.

To complete the proof, observe that Hoeffding's inequality~(\fctref{f_Hoeff}) implies that
\m{
\forall l\ge0:\:
\PR{\sz{\sz {\Cl A}-\fr n2}\ge l} \le 2\tm e^{\fr{-2l^2}n}
.}
Letting $l\deq\Max{d,\fr{\sq{Tn}}4}~\Then~(2d+2l)^2\le16d^2+Tn$, we obtain
\m[P]{
\forall T\ge0:\&[16pt]
\PR{\sz{\wtl {\Cl A}} \ge \fr{16d^2}n + 2T}&\\
&\le \PR{\sz{\wtl {\Cl A}}\ge \fr{(2d+2l)^2}n + T}[{\sz{\sz {\Cl A}-\fr n2}\le l}] + \PR{\sz{\sz {\Cl A}-\fr n2}\ge l}\\
&\le 2\tm e^{\fr{-T}2} + 2\tm e^{\ln n - \fr{T}{2\ln n}}
+ 2\tm e^{\fr{-T^2n}{32\tm (16d^2+Tn)\tm \ln n+80Tn}}\\
&\le 4\tm e^{\ln n - \fr{T}{2\ln n}}
+ 2\tm e^{\fr{-T^2n}{112\tm \ln n\tm T\tm n + 512\tm d^2\tm \ln n}}
,}
and the result follows by substituting $t\deq2T$ and noticing that increasing $d$ leads to shrinkage of the event whose probability is bounded and, at the same time, growing of the claimed upper bound (therefore the assumption $\lf[\sz{\sz{s}-\ud\dr n2}= d\rt]$ may be replaced by $\lf[\sz{\sz{s}-\ud\dr n2}\le d\rt]$).
\prfend

We are ready to investigate the likelihood of $\aleph(\Cl X,\Cl Y)$.

\prfstart[\theoref{t_aleph}]
Recall the definition of $\Delta_{j,s}$, as given in~\bref{m_nota_aleph}, and let us define ``weighted indicators'':
\m{
&w:[n]\pset0\to\lf[0,\dr14\rt];\;
w(x)\deq \Cases
{\fr1n\tm\l(x-\fr n2\r)^2}
{if $\fr n2-\fr{\sq n}2\le x\le\fr n2+\fr{\sq n}2$;}
{0}{otherwise.}\\
&\forall j\in[n],\,s\in\01^{\log n}:\:  {\Cl W}_{j,s} \deq w\l(\Delta_{j,s}(\Cl X,\Cl Y)\ud\r).
}
Then we want to show that
\m[m_W]{
\PR{\sum_{j\in[n],\,s\in\01^{\log n}} {\Cl W}_{j,s} > \fr{n^2}{9}}
\in \asO{\fr{(\ln n)^2}{\sq n}}
.}

Towards understanding the behaviour of the sum of \pl[{\Cl W}_{j,s}] we analyse the dependencies among them.
Let $(j_1,s_1)\ne(j_2,s_2)$ and define random variables:
\m{
& \txt{for } i\in\12:\: {\Cl D}_i \deq \Delta_{j_i,s_i}(\Cl X,\Cl Y);\\
& {\Cl Z} \deq \sigma_{j_1}\l(\tau_{s_1}\+\Cl X\r) \+ \sigma_{j_2}\l(\tau_{s_2}\+\Cl X\r);\\
& {\Cl V} \deq \sigma_{j_1}\l(\tau_{s_1}\+\Cl X\r) \+ \Cl Y
.}
Note that
\m{
{\Cl D}_1 \= \sz{{\Cl V}},\, {\Cl D}_2 \= \sz{{\Cl Z}\+{\Cl V}},\, 
{\Cl W}_{j_1,s_1} \= w({\Cl D}_1)
\txt{ and }
{\Cl W}_{j_2,s_2} \= w({\Cl D}_2)
.}
As $(\Cl X,\Cl Y)\sim\U[\01^{n+n}]$,
\m{
{\Cl V}\sim\U[\01^n]
\txt{ and }
\I{\Cl V}{\Cl Z}=0
.}

Apply \lemref{l_ABdecomp} to define a random variable $\wtl{\Cl V}$, such that
\m{
\I{\sz{{\Cl V}}}{\sz{{\Cl Z}\+{\Cl V}\+\wtl{\Cl V}}}[{\Cl Z}] = 0
}
and
\m[m_tlV_bou]{
\forall c,\, d\ge0:\: 
& \PR{\sz{\wtl {\Cl V}} \ge \fr{16d^2}n + c\tm (\ln n)^2}[{\sz{\sz{{\Cl Z}}-\fr n2}\le d}]\\
&\tbbbbb \le 4\tm e^{(1 - \dr{c}{4})\tm \ln n}
+ 2\tm e^{\fr{-c^2\tm n\tm (\ln n)^3}{224\tm c\tm n\tm (\ln n)^2 + 2048\tm d^2}}
.}
Then
\m{
\I{\sz{{\Cl V}}}{\sz{{\Cl Z}\+{\Cl V}\+\wtl{\Cl V}}}
& \le \I{\sz{{\Cl V}}}{{\Cl Z},\, \sz{{\Cl Z}\+{\Cl V}\+\wtl{\Cl V}}}\\
& = \I{\sz{{\Cl V}}}{{\Cl Z}} + \I{\sz{{\Cl V}}}{\sz{{\Cl Z}\+{\Cl V}\+\wtl{\Cl V}}}[{\Cl Z}]
=0
.}

As ${\Cl D}_2 = \sz{{\Cl Z}\+{\Cl V}}$,
\m{
{\Cl D}_2 \in \lf[\sz{{\Cl Z}\+{\Cl V}\+\wtl{\Cl V}}-\sz{\wtl {\Cl V}},\, \sz{{\Cl Z}\+{\Cl V}\+\wtl{\Cl V}}+\sz{\wtl {\Cl V}}\rt]
.}
Define
\m{
\wtl{\Cl D}_2 \deq \sz{{\Cl Z}\+{\Cl V}\+\wtl{\Cl V}} - {\Cl D}_2
,}
then
\m{
\I{{\Cl D}_1}{{\Cl D}_2 + \wtl{\Cl D}_2} = \I{\sz{{\Cl V}}}{\sz{{\Cl Z}\+{\Cl V}\+\wtl{\Cl V}}} = 0
}
and
\m[m_CoW12_bou]{
\Cov{{\Cl W}_{j_1,s_1}}{{\Cl W}_{j_2,s_2}}
& = \Cov{w({\Cl D}_1)}{w({\Cl D}_2 + \wtl{\Cl D}_2) + \l(w({\Cl D}_2) - w({\Cl D}_2 + \wtl{\Cl D}_2)\r)}\\
&\hspace{-40pt} = \Cov{w({\Cl D}_1)}{w({\Cl D}_2 + \wtl{\Cl D}_2)}
+ \Cov{w({\Cl D}_1)}{w({\Cl D}_2) - w({\Cl D}_2 + \wtl{\Cl D}_2)}\\
& = \Cov{w({\Cl D}_1)}{w({\Cl D}_2) - w({\Cl D}_2 + \wtl{\Cl D}_2)}\\
& \le \fr14\tm \E{\sz{w({\Cl D}_2) - w({\Cl D}_2 + \wtl{\Cl D}_2) - \E{w({\Cl D}_2) - w({\Cl D}_2 + \wtl{\Cl D}_2)}}}\\
& \le \fr12\tm \E{\sz{w({\Cl D}_2) - w({\Cl D}_2 + \wtl{\Cl D}_2)}}
.}

Note that $\sz{\wtl {\Cl D}_2} \le \sz{\wtl {\Cl V}}$ by construction (notice the difference in the meanings of $\sz{\ds}$), therefore $\sz{\wtl {\Cl D}_2}$ is also subject to~\bref{m_tlV_bou}.
Let us make the bound more comfortable by analysing the behaviour of $\sz{\sz{{\Cl Z}}-\fr n2}$.
If $j_1=j_2$, then $s_1\ne s_2$ and we have
\m{
{\Cl Z} = \sigma_{j_1}\l(\tau_{s_1}\+\Cl X\r) \+ \sigma_{j_1}\l(\tau_{s_2}\+\Cl X\r)
= \sigma_{j_1}\l(\tau_{s_1}\+\tau_{s_2}\r)
,}
which implies $\sz{{\Cl Z}}\=\fr n2$ by the construction of \pl[\tau_s].
If, on the other hand, $j_1\ne j_2$, then
\m{
{\Cl Z} = \sigma_{j_1}\l(\tau_{s_1}\r) \+ \sigma_{j_2}\l(\tau_{s_2}\r)
\+ \sigma_{j_1}(\Cl X) \+ \sigma_{j_2}(\Cl X)
= \sigma_{j_1}\l(\tau_{s_1}\r) \+ \sigma_{j_2}\l(\tau_{s_2}\r)
\+ \Cl X' \+ \sigma_{j_2-j_1}(\Cl X')
}
for $\Cl X'\deq \sigma_{j_1}(\Cl X)\sim\U[\01^n]$, and therefore,
\m{
\PR{\sz{{\sz{{\Cl Z}} - \fr n2}} \ge \sq n\tm\ln n} \le 4\tm e^{\fr{-(\ln n)^2}2}
}
by \lemref{l_sXsiX}.
Combining this with~\bref{m_tlV_bou}, we obtain that $\forall c\ge0$:
\m{
\&[22pt] \PR{\sz{\wtl {\Cl D}_2} \ge (16 + c)\tm (\ln n)^2}\\
& \le \PR{\sz{\wtl {\Cl V}} \ge (16 + c)\tm (\ln n)^2}[{\sz{\sz{{\Cl Z}}-\fr n2}\le \sq n\tm\ln n}]
+ \PR{\sz{{\sz{{\Cl Z}} - \fr n2}} > \sq n\tm\ln n}\\
& \le 4\tm e^{(1 - \dr{c}{4})\tm \ln n}
+ 2\tm e^{\fr{-c^2\tm \ln n}{224\tm c + 2048}}
+ 4\tm e^{\fr{-(\ln n)^2}2}
,}
and choosing $c\deq121$ gives
\m[m_tlD2_bou]{
\PR{\sz{\wtl {\Cl D}_2} \ge 137\tm (\ln n)^2} < \fr{10}{\sq n}
.}

We are ready to analyse $\E{\sz{w({\Cl D}_2) - w({\Cl D}_2 + \wtl{\Cl D}_2)}}$.
Let $\xi \deq 137\tm (\ln n)^2$ and
\m{
& I_{\xi}
\deq \lf[\fr{n-\sq n}2-\xi,\, \fr{n-\sq n}2+\xi\rt]
\cup \lf[\fr{n+\sq n}2-\xi,\, \fr{n+\sq n}2+\xi\rt],\\
& J_{\xi}
\deq \lf[\fr{n-\sq n}2+\xi,\, \fr{n+\sq n}2-\xi\rt]
.}
Then by the construction of $w(\ds)$:
\m{
\E{\sz{w({\Cl D}_2) - w({\Cl D}_2 + \wtl{\Cl D}_2)}}\hspace{-80pt}&\\
& \le \l(\PR{\sz{\wtl {\Cl D}_2} > \xi} + \PR{{\Cl D}_2 \in I_{\xi}}\r) \tm \fr14\\
&\tbbbb + \E{w({\Cl D}_2) - w({\Cl D}_2 + \wtl{\Cl D}_2)}[\sz{\wtl {\Cl D}_2} \le \xi,\, {\Cl D}_2 \in J_{\xi}]\\
& < \fr{3}{\sq n} + \fr{\PR{{\Cl D}_2 \in I_{\xi}}}4
+ \E{w({\Cl D}_2) - w({\Cl D}_2 + \wtl{\Cl D}_2)}[\sz{\wtl {\Cl D}_2} \le \xi,\, {\Cl D}_2 \in J_{\xi}]
,}
where the latter inequality is~\bref{m_tlD2_bou}.
\m{
\PR{{\Cl D}_2 \in I_{\xi}}
& \le \PR{{\Cl D}_2 \in I_{\xi}}
[{{\Cl D}_2 \in \lf[\fr{n-\sq n}2-\xi,\, \fr{n+\sq n}2+\xi\rt]}]\\
& \le \fr{\sz{I_{\xi}}}{\sz{\lf[\fr{n-\sq n}2-\xi,\, \fr{n+\sq n}2+\xi\rt]}}
~~<~~ \fr{4\xi}{\sq n}
,}
where the second inequality is due to the fact that ${\Cl D}_2$ comes from the symmetric binomial distribution centred (and therefore maximised) at $\dr{n}2$ and $I_{\xi}$ is a union of two symmetric edges of the considered interval $\lf[\fr{n-\sq n}2-\xi,\, \fr{n+\sq n}2+\xi\rt]$.
\m{
\E{w({\Cl D}_2) - w({\Cl D}_2 + \wtl{\Cl D}_2)}
[\sz{\wtl {\Cl D}_2} \le \xi,\, {\Cl D}_2 \in J_{\xi}]\hspace{-80pt}&\\
& \le \Sup[{x_0\in\lf[\fr{n-\sq n}2,\, \fr{n+\sq n}2\rt]}] {\sz{\fr{dw(x)}{dx}[x_0]}}
\tm \xi
~~=~~ \fr{\xi}{\sq n}
.}
Putting this together gives
\m{
\E{\sz{w({\Cl D}_2) - w({\Cl D}_2 + \wtl{\Cl D}_2)}}
< \fr{3\xi}{\sq n}
= \fr{411\tm (\ln n)^2}{\sq n}
}
and then from~\bref{m_CoW12_bou}:
\m[m_CoW12_bou_fin]{
\forall (j_1,s_1)\ne(j_2,s_2):\: 
\Cov{{\Cl W}_{j_1,s_1}}{{\Cl W}_{j_2,s_2}}
\le \fr{206\tm (\ln n)^2}{\sq n}
.}

Now -- still towards understanding the behaviour of the sum of \pl[{\Cl W}_{j,s}] -- let us estimate their expectations.
\m{
\forall (j,s):\: 
\E{{\Cl W}_{j,s}}
& = \PR{\Delta_{j,s}(\Cl X,\Cl Y) \in \lf[\fr n2-\fr{\sq n}2,\, \fr n2+\fr{\sq n}2\rt]}\\
&\tbb \tm \E{\fr1n\tm \l(\Delta_{j,s}(\Cl X,\Cl Y) - \fr n2\r)^2}
[{\Delta_{j,s}(\Cl X,\Cl Y) \in \lf[\fr n2-\fr{\sq n}2,\, \fr n2+\fr{\sq n}2\rt]}]\\
& \le \fr1n\tm \E{\l(\Delta_{j,s}(\Cl X,\Cl Y) - \fr n2\r)^2}
[{\Delta_{j,s}(\Cl X,\Cl Y) \in \lf[\fr n2-\fr{\sq n}2,\, \fr n2+\fr{\sq n}2\rt]}]
,}
where $\Delta_{j,s}(\Cl X,\Cl Y) = \sz{\sigma_j\l(\tau_s\+\Cl X\r)\+\Cl Y}$ comes from the symmetric binomial distribution $B(n,\dr12)$.
Let $S\=\Delta_{j,s}(\Cl X,\Cl Y)$, then
\m{
n\tm\E{{\Cl W}_{j,s}}
\le \E[S\sim B(n,\dr12)]{\l(S - \fr n2\r)^2}
[{S \in \lf[\fr n2-\fr{\sq n}2,\, \fr n2+\fr{\sq n}2\rt]}]
.}
As for any $a_1,\, a_2\in \s{\dr n2-\dr{\sq n}2\dc \dr n2+\dr{\sq n}2}$ it, obviously, holds that
\m{
\PR{S=a_1} \le \PR{S=a_2}
~~\IfThen~~
(a_1 - \dr n2)^2 \ge (a_2 - \dr n2)^2
,}
\lemref{l_E_mono} implies that
\m{
\E[S\sim B(n,\dr12)]{\l(S - \fr n2\r)^2}
[{S \in \lf[\fr n2-\fr{\sq n}2,\, \fr n2+\fr{\sq n}2\rt]}]
\le \E[{S\unin\s{\fr n2-\fr{\sq n}2\dc \fr n2+\fr{\sq n}2}}]{\l(S - \fr n2\r)^2}
.}

So,
\m{
\E{{\Cl W}_{j,s}}
& \le \fr1n\tm \E[{S\unin\s{\fr n2-\fr{\sq n}2\dc \fr n2+\fr{\sq n}2}}]{\l(S - \fr n2\r)^2}\\
& = \fr1n\tm \E[{S\unin\s{-\fr{\sq n}2\dc \fr{\sq n}2}}]{S^2}
~~<~~ \fr{n+3\sq n+2}{12n} 
~~\le~~ \fr1{11} 
,}
as $n$ is large enough, and
\m{
\E{\sum_{j\in[n],\,s\in\01^{\log n}} {\Cl W}_{j,s}}
< \fr{n^2}{11} 
.}
Combining this with~\bref{m_CoW12_bou_fin} and applying Chebyshev's inequality~(\fctref{f_MaChe}), we get
\m[P]{
&\PR{\sum_{j\in[n],\,s\in\01^{\log n}} {\Cl W}_{j,s} > \fr{n^2}{9}}
\le \PR{ \sz{\sum_{j,s} {\Cl W}_{j,s} - \E{\sum_{j,s} {\Cl W}_{j,s}}} > \fr{n^2}{50}}\\
&\tbbb \le 2500\tm \fr{\Var{\sum_{j,s} {\Cl W}_{j,s}}}{n^4}\\
&\tbbb = 2500\tm
\fr{ \sum_{j,s}\Var{{\Cl W}_{j,s}} + \sum_{(j_1,s_1)\ne(j_2,s_2)}\Cov{{\Cl W}_{j_1,s_1}}{{\Cl W}_{j_2,s_2}} }{n^4}\\
&\tbbb \le 2500\tm 
\fr{ n^2\tm \fr1{16} + n^4\tm \fr{206\tm (\ln n)^2}{\sq n} }{n^4}
~\le~ \fr{517500 \tm (\ln n)^2}{\sq n}
,}
which implies the required~\bref{m_W}.
\prfend

Now that we have satisfied our curiosity concerning the likelihood of $\aleph(\Cl X,\Cl Y)$ with respect to the uniform distribution, let us analyse the \R-complexity of \GHR.

\ssect[s_sp_rect]{Spectral stability of large rectangles}

Lower-bounding the \R-complexity of a communication problem is often based on showing that it admits no large nearly-monochromatic rectangles with respect to certain input distribution.\fn
{
A set of inputs to a relational problem is called (nearly-)monochromatic if some answer is correct for (almost) all of the set elements.
}
We will use this approach in our analysis of \GHR, and the input distribution that we consider will be the uniform.

A core structural property of large rectangles that will be used in the analysis is their stability with respect to the spectrum of Hamming distances between the two input values.
Namely, we will claim that if $(\Cl X,\Cl Y)$ is sampled at uniform from a large rectangle, then $\sz{\Cl X\+\Cl Y}$ is distributed, in certain sense, similarly to how it is distributed when $(\Cl X,\Cl Y)\unin \01^{n+n}$.

Note that large rectangles can fully control the \e{parity} of $\sz{\Cl X\+\Cl Y}$:\ say, if both $A$ and $B$ are the set of strings of even Hemming weight, then $2\big|\sz{x\+y}$ for every $(x,y)\in A\times B$.
Also it is relatively easy to control the \e{marginal} values of $\sz{\Cl X\+\Cl Y}$:\ say, an efficient protocol can use binary search to find a poly-logarithmic number of coordinates where the pair of inputs disagree and then test the equality over the rest of coordinates, thus checking whether the distance between the two input values is at most poly-logarithmic.
On the other hand, it is intuitively obvious that, other than knowing the parity of $\sz{\Cl X\+\Cl Y}$, a large rectangle cannot have a good command of this value's distribution near $\dr{n}2$.
The following statement formalises this intuition.~\fn
{
We've tried to give \theoref{t_spec} a modular formulation, but did not struggle to optimise it beyond our requirements, so some of the parameters can possibly be improved.
}

\theo[t_spec]{Let $A\times B\sbseq\01^{n+n}$ be a rectangle, $k\in[n]-1$ and $\mu_{A\times B}\deq\log \l( \fr{n}{\U[\01^{n+n}](A\times B)} \r)$.
If $\sq{n\tm \mu_{A\times B}}\le \dr{n}{14}$ and $\sz{\dr n2-k}\le \dr{n}{14}$, then
\m{
&1 - \asO{ \fr{\l(\mu_{A\times B}\r)^{\fr32}}{\sq n}
+ \sz{\fr 12-\fr{k}n} \tm \mu_{A\times B} }\\
&\tbbb \le~ \fr{\PR[(\Cl X,\Cl Y)\unin A\times B]{\sz{\Cl X\+ \Cl Y}\in\s{k,\, k+1}}}
{\PR[(\Cl X,\Cl Y)\unin \01^{n+n}]{\sz{\Cl X\+ \Cl Y}\in\s{k,\, k+1}}}\\
&\tbbb\tbbb \le~ 1 + \asO{ \l( \sq{\fr{\log n}n} + \sz{\fr 12-\fr{k}n} \r)
\tm \mu_{A\times B} }
,}
where the right-hand side holds only as long as it is $1+\aso1$.~\fn
{
More precisely, the upper bound holds if the expression inside \asO{\dt} is less than some universal constant, otherwise not necessarily.
}
}

In other words, large rectangles are not too biased with respect to the likely values of $\sz{\Cl X\+ \Cl Y}$, as long as we consider at least pairs of consecutive points in the spectrum (this is necessary since singletons are affected by the parity control, as discussed above).

We will use the following result due to Razborov~\cite{R92_On_th}.

\fct[f_Razb]{Let $\mu_j$ be the uniform distribution over $\s{(x,y)\in\chs{[4l-1]}l\times\chs{[4l-1]}l}[\sz{x\cap y}=j]$ for $j\in\01$.
Then for every $A,\, B\sbseq\01^{4l-1}$:
\m{
\mu_1(A\times B) \ge \fr1{45}\tm \mu_0(A\times B) - 2^{-\asOm l}
.}
}

Intuitively, it says that only very small rectangles -- those of size $2^{-\asOm l}$ -- can contain much more disjoint pairs than pairs intersecting over exactly one element (this was used in~\cite{R92_On_th} to show that the \R-complexity of the bipartite disjointness function was \asOm n).

\prfstart[\theoref{t_spec}]
For any $s\sbseq[n]\pset0$, define the \e{relative weight} of $s$ (with respect to $A\times B$) as
\m{
\rw(s) =
\fr{\PR[(\Cl X,\Cl Y)\unin A\times B]{\sz{\Cl X\+ \Cl Y}\in s}}{\PR[(\Cl X,\Cl Y)\unin \01^{n+n}]{\sz{\Cl X\+ \Cl Y}\in s}}
,}
and for every $i\in[n]\pset0$ let $\rw(i)\deq\rw(\s i)$.

Assuming that $A\times B$ exhibits spectral irregularity -- that is, a deviation of $\rw(\dt)$ from $1$ -- we will find a \e{representative pair of coordinates} that, on the one hand, are both not too far from $\dr n2$, and on the other hand, whose relative weights are somewhat apart from each other:\ in combination with \fctref{f_Razb} this will lead to an upper bound on the size of $A\times B$.

Our way of finding a representative pair will depend on $\rw(\s{k,\, k+1})$.

\para{The case of $\rw(\s{k,\, k+1})>1$.}
For
\m{
\lambda\tpl \deq
\Min{\rw(\s{k,\, k+1})-1,\, 1}
,}
let $t_\tpl$ be the minimal distance from $\dr n2$ of a pair of coordinates with relative weight at most $1+\dr{\lambda_\tpl}3$ and same parity pattern as in $\s{k,\, k+1}$:
\m{
t_\tpl \deq
\Min[i]{\sz{\fr{n}2-i}}
[\rw(\s{i,\, i+1})\le1+\fr{\lambda_\tpl}3,\, 2|(i+k)]
.}

The integer interval
\m{
T_\tpl \deq \s{\fr{n}2-t_\tpl+2\dc\fr{n}2+t_\tpl-1}
}
partitions into pairs $(\dr{n}2-t_\tpl+2,\, \dr{n}2-t_\tpl+3)\dc(\dr{n}2+t_\tpl-2,\, \dr{n}2+t_\tpl-1)$, each satisfying $\lf[\rw(\s{j,\, j+1})>1+\dr{\lambda_\tpl}3\rt]$.
Accordingly,
\m{
\rw\l(T_\tpl\r) > 1 + \fr{\lambda_\tpl}3
}
and
\m{
1
& \ge \PR[(\Cl X,\Cl Y)\unin A\times B]{(\Cl X,\Cl Y)\in T_\tpl}\\
& \ge \l(1+\fr{\lambda_\tpl}3\r)\tm \PR[(\Cl X,\Cl Y)\unin \01^{n+n}]{(\Cl X,\Cl Y)\in T_\tpl}\\
& \ge \l(1+\fr{\lambda_\tpl}3\r)\tm \PR[(\Cl X,\Cl Y)\unin \01^{n+n}]{\sz{\sz{\Cl X\+ \Cl Y}-\fr{n}2}<t_\tpl-1}
,}
as $\lf[\sz{\sz{\Cl X\+ \Cl Y}-\dr{n}2}<t_\tpl-1\rt]$ implies $\lf[(\Cl X,\Cl Y)\in T_\tpl\rt]$.
By Hoeffding's inequality~(\fctref{f_Hoeff}),
\m{
\PR[(\Cl X,\Cl Y)\unin \01^{n+n}]{\sz{\sz{\Cl X\+ \Cl Y}-\fr{n}2}\ge t_\tpl-1}
< 2\tm e^{\fr{4t_\tpl-2t_\tpl^2}n}
,}
and therefore
\m{
\l(1+\fr{\lambda_\tpl}3\r)\tm \l(1-2\tm e^{\fr{4t_\tpl-2t_\tpl^2}n}\r) < 1
,}
which, due to ${\lambda_\tpl}\in(0,1]$, implies that for large enough $n$,
\m[m_tpl_boun]{
1-2\tm e^{\fr{4t_\tpl-2t_\tpl^2}n}
< \fr1{1+\fr{\lambda_\tpl}3}
\le 1-\fr{\lambda_\tpl}4
~~\Then~~
e^{\fr{4t_\tpl-2t_\tpl^2}n} > \fr{\lambda_\tpl}8
~~\Then~~
t_\tpl <
\sq{n\tm \ln\l(\fr1{\lambda_\tpl}\r)}
.}

Let $a\in\s{\dr{n}2-t_\tpl,\, \dr{n}2+t_\tpl}$ be such that $\rw(\s{a,\, a+1})\le1+\dr{\lambda_\tpl}3$.
As in the case that we've just considered $\rw(\s{k,\, k+1})\ge1+\lambda_\tpl$ and $0<\lambda_\tpl\le 1$, it holds that $(1+\Lambda)\tm \rw(\s{a,\, a+1}) \le \rw(\s{k,\, k+1})$ for $\Lambda\deq\dr{\lambda_\tpl}3>0$.
Note that
\m[m_on_a_L_pl]{
\sz{\dr{n}2-a} = t_\tpl,\,
\Lambda\in\asT{\lambda_\tpl}
}
and $2|(a+k)$ hold by construction.

\para{The case of $\rw(\s{k,\, k+1})<1$.}
For
\m{
\lambda_\tmin \deq 1-\rw(\s{k,\, k+1})
,}
let $t_\tmin$ be the minimal distance from $\dr n2$ of a pair of coordinates with relative weight at least $1-\dr{\lambda_\tmin}2$ and same parity pattern as in $\s{k,\, k+1}$:
\m{
t_\tmin \deq
\Min[i]{\sz{\fr{n}2-i}}
[\rw(\s{i,\, i+1})\ge1-\fr{\lambda_\tmin}2,\, 2|(i+k)]
.}
The integer interval
\m{
T_\tmin \deq \s{\fr{n}2-t_\tmin+2\dc\fr{n}2+t_\tmin-1}
}
partitions into pairs $(\dr{n}2-t_\tmin+2,\, \dr{n}2-t_\tmin+3)\dc(\dr{n}2+t_\tmin-2,\, \dr{n}2+t_\tmin-1)$, each satisfying $\lf[\rw(\s{j,\, j+1}) < 1 - \dr{\lambda_\tmin}2\rt]$.
Accordingly,
\m{
\rw\l(T_\tmin\r) < 1 - \fr{\lambda_\tmin}2
.}

To bound $t_\tmin$ from above, let us use the fact that the subset of $A\times B$ where $\rw(\sz{x+y})\ge 1-\dr{\lambda_\tmin}2$ must \e{compensate} the missing $\dr{\lambda_\tmin}2$-portion in the rest of the rectangle:
\m{
\PR[(\Cl X,\Cl Y)\unin A\times B]{\sz{\sz{\Cl X\+ \Cl Y}-\fr{n}2} \nin T_\tmin}
& = \sum_{i \nin T_\tmin} \PR[(\Cl X,\Cl Y)\unin A\times B]{\sz{\sz{\Cl X\+ \Cl Y}-\fr{n}2}=i}\\
&\hspace{-128pt} \ge \sum_{i \nin T_\tmin}
\l(
\PR[(\Cl X,\Cl Y)\unin A\times B]{\sz{\sz{\Cl X\+ \Cl Y}-\fr{n}2}=i}
- \PR[(\Cl X,\Cl Y)\unin \01^{n+n}]{\sz{\sz{\Cl X\+ \Cl Y}-\fr{n}2}=i}
\r)\\
&\hspace{-128pt} = \sum_{i \in T_\tmin}
\l(
\PR[(\Cl X,\Cl Y)\unin \01^{n+n}]{\sz{\sz{\Cl X\+ \Cl Y}-\fr{n}2}=i}
- \PR[(\Cl X,\Cl Y)\unin A\times B]{\sz{\sz{\Cl X\+ \Cl Y}-\fr{n}2}=i}
\r)\\
& \ge \fr{\lambda_\tmin}2\tm 
\PR[(\Cl X,\Cl Y)\unin \01^{n+n}]{\sz{\sz{\Cl X\+ \Cl Y}-\fr{n}2} \in T_\tmin}\\
& \ge \fr{\lambda_\tmin}2\tm 
\PR[(\Cl X,\Cl Y)\unin A\times B]{\sz{\sz{\Cl X\+ \Cl Y}-\fr{n}2} \in T_\tmin}
,}
from where by adding $\fr{\lambda_\tmin}2\tm \PR[A\times B]{\sz{\sz{\Cl X\+ \Cl Y}-\fr{n}2} \nin T_\tmin}$ to both the sides we get
\m{
\fr{\lambda_\tmin}2\tm
& \le \fr32\tm \PR[(\Cl X,\Cl Y)\unin A\times B]{\sz{\sz{\Cl X\+ \Cl Y}-\fr{n}2} \nin T_\tmin}\\
& = \fr32\tm \PR[(\Cl X,\Cl Y)\unin \01^{n+n}]{\sz{\sz{\Cl X\+ \Cl Y}-\fr{n}2} \nin T_\tmin}
[(\Cl X,\Cl Y)\in A\times B]\\
& \le \fr32\tm \fr{\PR[(\Cl X,\Cl Y)\unin \01^{n+n}]{\sz{\sz{\Cl X\+ \Cl Y}-\fr{n}2} \nin T_\tmin}}
{\U[\01^{n+n}](A\times B)}\\
& \le \fr32\tm \fr{\PR[(\Cl X,\Cl Y)\unin \01^{n+n}]{\sz{\sz{\Cl X\+ \Cl Y}-\fr{n}2} \ge t_\tmin-1}}
{\U[\01^{n+n}](A\times B)}
,}
as $\lf[\sz{\sz{\Cl X\+ \Cl Y}-\dr{n}2\ud} \nin T_\tmin\rt]$ implies $\lf[\sz{\sz{\Cl X\+ \Cl Y}-\dr{n}2\ud} \ge t_\tmin-1\rt]$.
By Hoeffding's inequality~(\fctref{f_Hoeff}),
\m{
\PR[(\Cl X,\Cl Y)\unin \01^{n+n}]{\sz{\sz{\Cl X\+ \Cl Y}-\fr{n}2}\ge t_\tmin-1}
< 2\tm e^{\fr{4t_\tmin-2t_\tmin^2}n}
,}
and so, for large enough $n$,
\m[m_tmin_boun]{
e^{\fr{4t_\tmin-2t_\tmin^2}n}
> \fr{\lambda_\tmin}6\tm \U[\01^{n+n}](A\times B)
~~\Then~~
t_\tmin <
\sq{n\tm \ln\l(\fr1{\lambda_\tmin\tm \U[\01^{n+n}](A\times B)}\r) }
.}

Let $a\in\s{\dr{n}2-t\tmin,\, \dr{n}2+t\tmin}$ be such that $\rw(\s{a,\, a+1})\ge1-\dr{\lambda_\tmin}2$.
As in the case that we've just considered $\rw(\s{k,\, k+1})=1-\lambda_\tmin$ for $\lambda_\tmin>0$, it holds that $(1+\Lambda)\tm \rw(\s{k,\, k+1}) \le \rw(\s{a,\, a+1})$ for $\Lambda\deq\dr{\lambda_\tmin}2>0$.
Note that
\m[m_on_a_L_min]{
\sz{\dr{n}2-a} = t_\tmin,\,
\Lambda\in\asT{\lambda_\tmin}
}
and $2|(a+k)$ hold by construction.

\para{Completion.}
We need a pair of coordinates $b_1,\, b_2\in[n]-1$ with the following properties:
\m[m_cases_b]{
\begin{cases}
b_1 < b_2 ~\txt{and}~ 2|(b_2-b_1);\\
(1+\Lambda)\tm \rw(\s{b_1,\, b_1+1}) \le \rw(\s{b_2,\, b_2+1}) ~\txt{for some}~ \Lambda\in(0,1];\\
\rw(\s{b_2,\, b_2+1}) \ge \dr12.
\end{cases}
}

The pair of coordinates that we have identified in the previous stage ($k$ and $a$) almost satisfy these requirements:\ it may happen that \e{the ordering is wrong} (say, $k>a$ but $\rw(\s{k,\, k+1})<\rw(\s{a,\, a+1})$).
However, this situation can be addressed by a simple transformation of the rectangle under consideration:\ if we let
\m{
B' \deq \s{\neg b}[b\in B]
,}
where ``$\neg$'' stands for the bit-wise negation of a binary string, then, obviously,
\m{
\forall i\in[n]\pset0:\: 
\PR[(\Cl X,\Cl Y)\unin A\times B]{\sz{\Cl X\+ \Cl Y}=i}
= \PR[(\Cl X,\Cl Y)\unin A\times B']{\sz{\Cl X\+ \Cl Y}=n-i}
.}
That is, the distributions of $\sz{\Cl X\+ \Cl Y}$ corresponding to the rectangles $A\times B$ and $A\times B'$ are symmetrically opposite, and we will get all the three conditions of~\bref{m_cases_b} satisfied either for the rectangle $A\times B$ with the coordinates $k$ and $a$, or for the rectangle $A\times B'$ with the coordinates $n-k$ and $n-a$.

For notational convenience (and without loss of generality), let us assume the former case:
That is, we are assuming that a pair of points $b_1,\, b_2\in[n]-1$ satisfy the conditions listed at~\bref{m_cases_b} with parameters provided by the appropriate version of our pair-picking algorithm (note that the conditions are implicitly referring to the rectangle $A\times B$, as the relative weight $\rw(\tm)$ is defined with respect to it). 

For the following argument we will use a pair of coordinates:\ either $(b_1,\, b_2)$ or $(b_1+1,\, b_2+1)$, let us make the choice.
By the nature of the binomial distribution,
\m{
\forall i\in[n]-1:\: 
\PR[(\Cl X,\Cl Y)\unin \01^{n+n}]{\sz{\Cl X\+ \Cl Y}=i+1}
= \fr{i+1}{n-i}\tm \PR[(\Cl X,\Cl Y)\unin \01^{n+n}]{\sz{\Cl X\+ \Cl Y}=i}
,}
and therefore,
\m[P]{
\rw(\s{i,\, i+1})
& = \fr{\PR[A\times B]{\sz{\Cl X\+ \Cl Y}=i} + \PR[A\times B]{\sz{\Cl X\+ \Cl Y}=i+1}}
{\PR[\01^{n+n}]{\sz{\Cl X\+ \Cl Y}=i} + \PR[\01^{n+n}]{\sz{\Cl X\+ \Cl Y}=i+1}}\\
& = \fr{\PR[A\times B]{\sz{\Cl X\+ \Cl Y}=i}}
{\PR[\01^{n+n}]{\sz{\Cl X\+ \Cl Y}=i}\tm \l(1+\fr{i+1}{n-i}\r) }\\
&\tbbbb + \fr{\PR[A\times B]{\sz{\Cl X\+ \Cl Y}=i+1}}
{\PR[\01^{n+n}]{\sz{\Cl X\+ \Cl Y}=i+1}\tm \l(1+\fr{n-i}{i+1}\r)}\\
& = \rw(i) \tm \fr{n-i}{n+1} + \rw(i+1) \tm \fr{i+1}{n+1}
.}
Accordingly, our assumptions regarding $b_1$ and $b_2$ imply that $\dr 12<\rw(b_2) + \rw(b_2+1)$ and
\m{
& (1+\Lambda)\tm \rw(b_1)\tm (n-b_1)
+ (1+\Lambda)\tm \rw(b_1+1)\tm (b_1+1)\\
&\tbbbbb \le \rw(b_2)\tm (n-b_2) + \rw(b_2+1)\tm (b_2+1)
.}
If we let
\m{
\delta_{max} \deq \fr{\Max{\sz{n-2\tm b_1},\, \sz{n-2\tm b_2}\ud} + 2}n
\tb\l(=\fr{\Max{\sz{n-2\tm a},\, \sz{n-2\tm k}\ud} + 2}n\r)
,}
then
\m{
& (1+\Lambda)\tm \rw(b_1)\tm (1-\delta_{max})
+ (1+\Lambda)\tm \rw(b_1+1)\tm (1-\delta_{max})\\
&\tbb \le \rw(b_2)\tm (1+\delta_{max}) + \rw(b_2+1)\tm (1+\delta_{max})
}
and
\m[m_b_sym]{
(1+\Lambda)\tm (1-2\delta_{max})\tm \l(\rw(b_1) + \rw(b_1+1)\ud\r)
\le \rw(b_2) + \rw(b_2+1)
.}

Recall that $\rw(b_2) + \rw(b_2+1)>\dr 12$ and $\Lambda\in(0,1]$, consider the following cases.
If $\rw(b_2+1)<\dr{\Lambda}7$, then $\rw(b_2) > \dr5{14}$ and from~\bref{m_b_sym}:
\m{
(1+\Lambda)\tm (1-2\delta_{max})\tm \l(\rw(b_1) + \rw(b_1+1)\ud\r)
< \rw(b_2) + \fr{\Lambda}7
< \rw(b_2)\tm \l(1 + \fr{2\Lambda}5\r) 
,}
implying
\m{
\rw(b_2) > \fr{1+\Lambda}{1+\fr{2\Lambda}5} \tm (1-2\delta_{max}) \tm \rw(b_1)
> \l(1+\fr\Lambda5\r) \tm (1-2\delta_{max}) \tm \rw(b_1)
.}
Similarly, if $\rw(b_2)<\dr{\Lambda}7$, then $\rw(b_2+1) > \dr5{14}$ and
\m{
\rw(b_2+1) > \l(1+\fr\Lambda5\r) \tm (1-2\delta_{max}) \tm \rw(b_1+1)
.}
Lastly, if $\rw(b_2),\, \rw(b_2+1)\ge \dr{\Lambda}7$, then it follows from~\bref{m_b_sym} that at least one of
\m{
\rw(b_2) \ge (1+\Lambda)\tm (1-2\delta_{max})\tm \rw(b_1)
,~
\rw(b_2+1) \ge (1+\Lambda)\tm (1-2\delta_{max})\tm \rw(b_1+1)
}
must hold.

In every case, we have identified a pair of coordinates $c_1,\, c_2\in[n]\pset0$, such that the following holds:
\m[m_cases_c]{
\begin{cases}
c_2 > c_1
~\txt{and}~ 2|(c_2-c_1);\\
\sz{\dr n2-c_1},\, \sz{\dr n2-c_2}\le \Max{\sz{\dr n2-a},\, \sz{\dr n2-k}\ud} + 1;\\
\rw(c_2) \ge (1+\dr\Lambda5-4\delta_{max})\tm \rw(c_1)
~\txt{for}~ \delta_{max} = \dr{\l(\Max{\sz{n-2a},\, \sz{n-2k}\ud} + 2\r)}n;\\
\rw(c_2) \ge \dr{\Lambda}7.
\end{cases}
}

Now let
\m{
l \deq \floor{\fr{c_2}{4(c_2-c_1)}}
}
and assume that
\m[m_cc_ass]{
3\tm c_2 \le 4\tm c_1
}
-- that is, $l\ge 1$ (in the end we will make sure that this holds).
Suppose that we are given some $(x,y)\in\chs{[4l-1]}l\times\chs{[4l-1]}l$ and the goal is to distinguish the case of $[x\cap y=\es]$ from that of $[\sz{x\cap y}=1]$, using the assumed irregularity of $A\times B$ (as witnessed by the coordinates $c_1$ and $c_2$) -- that will let us derive the desired conclusion from the hardness of distinguishing the two cases (as expressed by \fctref{f_Razb}).

Consider the following randomised algorithm that we will denote by $\Xi$.
\enstart
\item Map $x$ to $\Cl X_1\in\01^{12l-3}$ by converting every $x_i$, $i\in[4l-1]$ into $\Cl X_1|_{3i-2\dc3i}\in\01^3$ as follows:
\m{
\Cl X_1|_{3i-2\dc3i} = \Cases
{100}{if $x_i=0$;}
{010}{if $x_i=1$.}
}
\item Map $y$ to $\Cl Y_1\in\01^{12l-3}$ by mapping every $y_i$, $i\in[4l-1]$ to $\Cl Y_1|_{3i-2\dc3i}\in\01^3$ as follows:
\m{
\Cl Y_1|_{3i-2\dc3i} = \Cases
{001}{if $x_i=0$;}
{010}{if $x_i=1$.}
}
\item Let $\Cl X_2,\, \Cl Y_2\in\01^{\fr{(12l-3)\tm (c_2-c_1)}2}$ be the concatenations of $\fr{c_2-c_1}2$ copies of, respectively, $\Cl X_1$ and $\Cl Y_1$.
\item Let $\Cl X_3\in\01^n$ be the concatenations of $\Cl X_2$ with $0^{n-\fr{(12l-3)\tm (c_2-c_1)}2}$.
\item Let $\Cl Y_3\in\01^n$ be the concatenations of $\Cl Y_2$ with $1^{c_2-(4l-1)(c_2-c_1)}$ and with $0^{n-c_2-\fr{(4l-1)(c_2-c_1)}2}$.
\item\label{step_S} Let $\Cl S$ be a uniformly-random permutation of $[n]$ and $\Cl X_4,\, \Cl Y_4\in\01^{n}$ contain the outcomes of, respectively, $S(\Cl X_3)$ and $S(\Cl Y_3)$.
\item\label{step_T} Let $\Cl T\sim\U[\01^{n}]$ and $\Cl X_5,\, \Cl Y_5\in\01^{n}$ contain the outcomes of, respectively, $\Cl T\+\Cl X_4$ and $\Cl T\+\Cl Y_4$.
\item\label{step_acc_rej} \e{Accept} if $(\Cl X_5,\, \Cl Y_5)\in A\times B$ and \e{reject} otherwise.
\enend

In the above description of the procedure $\Xi$, let us view $\Cl S$, $\Cl T$, \pl[\Cl X_i] and \pl[\Cl Y_i] as random variables:\ some of them hold values that are fixed by those of $x$ and $y$, others possess entropy of their own.
Note however that once the values of $\Cl S$ and $\Cl T$ are fixed, $\Xi$ becomes a deterministic algorithm with input $(x,y)$.

We claim that if $x\cap y=\es$, then $(\Cl X_5,\, \Cl Y_5)\sim\U[\01^{n+n}]$ subject to $\sz{\Cl X_5\+ \Cl Y_5}=c_2$, and if $\sz{x\cap y}=1$, then $(\Cl X_5,\, \Cl Y_5)\sim\U[\01^{n+n}]$ subject to $\sz{\Cl X_5\+ \Cl Y_5}=c_1$.
Indeed, suppose that $x\cap y=\es$, then the distance between every pair of corresponding bit-triples of $\Cl X_1$ and $\Cl Y_1$ is exactly $2$ and therefore $\sz{\Cl X_1\+ \Cl Y_1}=8l-2$.
Consequently, $\sz{\Cl X_2\+ \Cl Y_2}=(c_2-c_1)(4l-1)$ and
\m{
\sz{\Cl X_3\+ \Cl Y_3} = (c_2-c_1)(4l-1) + c_2 - (4l-1)(c_2-c_1) = c_2
.}
If, on the other hand, $\sz{x\cap y}=1$, then the distance between all but one pair of corresponding bit-triples of $\Cl X_1$ and $\Cl Y_1$ is $2$ and the remaining pair of triples both equal $010$, and therefore $\sz{\Cl X_1\+ \Cl Y_1}=8l-4$, $\sz{\Cl X_2\+ \Cl Y_2}=(c_2-c_1)(4l-2)$ and
\m{
\sz{\Cl X_3\+ \Cl Y_3} = (c_2-c_1)(4l-2) + c_2 - (4l-1)(c_2-c_1) = c_2 - (c_2-c_1) = c_1
.}
The above statement about the conditional distribution of $(\Cl X_5,\, \Cl Y_5)$ follows by noticing that steps \bref[step_S]{step_T} map every $(\Cl X_3,\, \Cl Y_3)$ to uniformly-random $(\Cl X_5,\, \Cl Y_5)$, subject to $[\sz{\Cl X_5\+ \Cl Y_5}=\sz{\Cl X_3\+ \Cl Y_3}]$.

As
\m{
\forall i\in[n]\pset0:\: 
& \PR[{\U[\01^{n+n}]}]{(\Cl X,\, \Cl Y)\in A\times B}[\sz{\Cl X\+ \Cl Y}=i]\\
&\hspace{-40pt} = \PR[{\U[\01^{n+n}]}]{\sz{\Cl X\+ \Cl Y}=i}[(\Cl X,\, \Cl Y)\in A\times B]
\tm \fr{\PR[{\U[\01^{n+n}]}]{(\Cl X,\, \Cl Y)\in A\times B}}
{\PR[{\U[\01^{n+n}]}]{\sz{\Cl X\+ \Cl Y}=i}}\\
&\tb = \rw(i)\tm \U[\01^{n+n}](A\times B)
,}
it follows by~\bref{m_cases_c} that
\m{
\fr{\PR{\txt{$\Xi$ accepts}}[x\cap y=\es]}
{\PR{\txt{$\Xi$ accepts}}[\sz{x\cap y}=1]}
&= \fr{\PR[{\U[\01^{n+n}]}]{(\Cl X,\, \Cl Y)\in A\times B}[\sz{\Cl X\+ \Cl Y}=c_2]}
{\PR[{\U[\01^{n+n}]}]{(\Cl X,\, \Cl Y)\in A\times B}[\sz{\Cl X\+ \Cl Y}=c_1]}\\
&\ge 1+\dr\Lambda5-4\delta_{max}
}
and
\m{
\PR{\txt{$\Xi$ accepts}}[x\cap y=\es]
~\ge~ \fr{\Lambda\tm \U[\01^{n+n}](A\times B)}7
.}

Note that conditioned on the values of $\Cl S$ and $\Cl T$, the set of pairs $(x,y)$ accepted by $\Xi$ forms a combinatorial rectangle in $\chs{[4l-1]}l\times\chs{[4l-1]}l$.
Moreover, if we run $k$ independent instances of $\Xi$ for the same input $(x,y)$ and \e{accept if and only if all $k$ executions have accepted}, even then the set of accepted pairs still forms a combinatorial rectangle (conditions on the values of both $\Cl S$ and $\Cl T$ in each instance), as the intersection of many rectangles is a rectangle.

For the version with $k$ repetitions the following holds, due to the mutual independence of the instances:
\m{
&\forall (x,\, y)\in\chs{[4l-1]}l\times\chs{[4l-1]}l:\\
&\tbbb \fr{\PR{\txt{\e{Accept}}}[x\cap y=\es]}
{\PR{\txt{\e{Accept}}}[\sz{x\cap y}=1]}
~\ge~ \l(1+\fr\Lambda5-4\delta_{max}\r)^k
}
and
\m{
\PR{\txt{\e{Accept}}}[x\cap y=\es]
~\ge~ \l(\fr{\Lambda\tm \U[\01^{n+n}](A\times B)}7\r)^k
.}

Let $\Cl X$ and $\Cl Y$ be random variables that respectively take the values $x$ and $y$, and let $\U$ denote their uniform distribution (i.e., $\U[\chs{[4l-1]}l\times\chs{[4l-1]}l]$).
Let $\Xi_{s_1\dc s_k,\, t_1\dc t_k}^k$ denote the \f k-repetition version of $\Xi$, where the \ord[i] instance runs with $\Cl S=s_i,\, \Cl T=t_i$.
Then the standard counting argument implies that there exist such values of $s_1\dc s_k,\, t_1\dc t_k$ that the (deterministic) procedure $\Xi_{s_1\dc s_k,\, t_1\dc t_k}^k$ satisfies both
\m{
\fr{\PR[(\Cl X,\Cl Y)\sim\U]{\txt{$\Xi_{s_1\dc s_k,\, t_1\dc t_k}^k(\Cl X,\Cl Y)$ accepts}}[\Cl X\cap \Cl Y=\es]}
{\PR[(\Cl X,\Cl Y)\sim\U]{\txt{$\Xi_{s_1\dc s_k,\, t_1\dc t_k}^k(\Cl X,\Cl Y)$ accepts}}[\sz{\Cl X\cap \Cl Y}=1]}
~\ge~ \fr12\tm \l(1+\fr\Lambda5-4\delta_{max}\r)^k
}
and
\m{
\PR[(\Cl X,\Cl Y)\sim\U]{\txt{$\Xi_{s_1\dc s_k,\, t_1\dc t_k}^k(\Cl X,\Cl Y)$ accepts}}[\Cl X\cap \Cl Y=\es]
~\ge~ \fr12\tm \l(\fr{\Lambda\tm \U[\01^{n+n}](A\times B)}7\r)^k
.}
As the acceptance region of $\Xi_{s_1\dc s_k,\, t_1\dc t_k}^k$ forms a rectangle, \fctref{f_Razb} implies that
\m{
&\PR[(\Cl X,\Cl Y)\sim\U]{\txt{$\Xi_{s_1\dc s_k,\, t_1\dc t_k}^k(\Cl X,\Cl Y)$ accepts}}[\sz{\Cl X\cap \Cl Y}=1]\\
&\tbbb \ge \fr1{45}\tm
\PR[(\Cl X,\Cl Y)\sim\U]{\txt{$\Xi_{s_1\dc s_k,\, t_1\dc t_k}^k(\Cl X,\Cl Y)$ accepts}}[\Cl X\cap \Cl Y=\es]
-2^{-\asOm l}
,}
and so,
\m{
&\fr{90\tm \PR[(\Cl X,\Cl Y)\sim\U]{\txt{$\Xi_{s_1\dc s_k,\, t_1\dc t_k}^k(\Cl X,\Cl Y)$ accepts}}[\Cl X\cap \Cl Y=\es]}
{\l(1+\dr\Lambda5-4\delta_{max}\r)^k}\\
&\tbbb \ge
\PR[(\Cl X,\Cl Y)\sim\U]{\txt{$\Xi_{s_1\dc s_k,\, t_1\dc t_k}^k(\Cl X,\Cl Y)$ accepts}}[\Cl X\cap \Cl Y=\es]
-2^{-\asOm l}
.}

In other words,
\m{
\PR[(\Cl X,\Cl Y)\sim\U]{\txt{$\Xi_{s_1\dc s_k,\, t_1\dc t_k}^k(\Cl X,\Cl Y)$ accepts}}[\Cl X\cap \Cl Y=\es]
\tm \l(1 - \fr{90}{\l(1+\dr\Lambda5-4\delta_{max}\r)^k}\r)
\le 2^{-\asOm l}
,}
and so,
\m{
\fr12\tm \l(\fr{\Lambda\tm \U[\01^{n+n}](A\times B)}7\r)^k
\tm \l(1 - \fr{90}{\l(1+\dr\Lambda5-4\delta_{max}\r)^k}\r)
\le 2^{-\asOm l}
.}
Assuming $\Lambda > 20\tm \delta_{max}$, we can pick sufficiently large
\m{
k \in \asO{\fr1{\log (1+\dr\Lambda5-4\delta_{max})}}
= \asO{\fr1{\Lambda - 20\tm \delta_{max}}}
,}
such that
\m{
\log \l(\fr1\Lambda\r) + \log \l(\fr1{\U[\01^{n+n}](A\times B)}\r)
\in \asOm{\fr{l}k}
= \asOm{l\tm (\Lambda - 20\tm \delta_{max})\ud}
.}
Substituting the definition of $l$, we get
\m{
\log \l(\fr1\Lambda\r) + \log \l(\fr1{\U[\01^{n+n}](A\times B)}\r)
\in \asOm{ \fr{c_2}{c_2-c_1} \tm (\Lambda - 20\tm \delta_{max}) }
.}

Let
\m[m_d_max]{
d_{max}\deq\Max{\sz{\fr n2-a},\, \sz{\fr n2-k}\ud}
,}
then
\m{
\Lambda - 20\tm \delta_{max}
= \Lambda - 20\tm \fr{\Max{\sz{n-2\tm a},\, \sz{n-2\tm k}\ud} + 2}n
= \Lambda - 40\tm \fr{d_{max}+1}n
}
and the assumption
\m[m_dmax_cond]{
d_{max} \le \fr{\Lambda\tm n}{41}
}
guarantees that $\Lambda - 20\tm \delta_{max} \in \asOm\Lambda$ and $\dr1{\Lambda}<n$ (note that $d_{max} \ge 1$ is implied by its definition, as $a\ne k$), and therefore,
\m[m_AB_up]{
\log \l(\fr{n}{\U[\01^{n+n}](A\times B)}\r)
~\in~ \asOm{ \fr{c_2\tm \Lambda}{c_2-c_1} }
~\sbseq~ \asOm{ \fr {n\tm \Lambda}{d_{max}}}
.}
As $\Lambda$ must either violate~\bref{m_dmax_cond} or satisfy~\bref{m_AB_up} (or both), we may conclude
\m[m_Lambda_up]{
\Lambda \in
\asO{\fr{d_{max}}n\tm \log \l(\fr{n}{\U[\01^{n+n}](A\times B)}\r)}
.}

If $\rw(\s{k,\, k+1})>1$, then \bref[m_tpl_boun][m_on_a_L_pl][m_d_max]{m_Lambda_up} imply
\m{
\lambda_\tpl
& \in \asO{\fr{t_\tpl + \sz{\fr n2-k}}n
\tm \log \l(\fr{n}{\U[\01^{n+n}](A\times B)}\r)}\\
& \sbseq \asO{ \l( \sq{\fr{\log n}n} + \sz{\fr 12-\fr{k}n} \r)
\tm \log \l( \fr{n}{\U[\01^{n+n}](A\times B)} \r) }
,}
and if $\rw(\s{k,\, k+1})<1$, then \bref[m_tmin_boun][m_on_a_L_min][m_d_max]{m_Lambda_up} imply
\m[P]{
\lambda_\tmin
& \in \asO{\fr{t_\tmin + \sz{\fr n2-k}}n
\tm \log \l(\fr{n}{\U[\01^{n+n}](A\times B)}\r)}\\
& \sbseq \asO{
\fr{\sq{n\tm \ln\l(\fr1{\lambda_\tmin\tm \U[\01^{n+n}](A\times B)}\r) } + \sz{\fr n2-k}}n
\tm \log \l(\fr{n}{\U[\01^{n+n}](A\times B)}\r)}\\
& \sbseq \asO{
\fr{\l(\log\fr{n}{\U[\01^{n+n}](A\times B)}\r)^{\fr32}}{\sq n}
+ \sz{\fr 12-\fr{k}n} \tm \log \l( \fr{n}{\U[\01^{n+n}](A\times B)} \r)}
,}
as required.
Lastly, for~\bref{m_cc_ass} to hold it is enough for $d_{max} \le \dr{n}{14}$ to hold, which is guaranteed by the assumed $\lf[\sq{n\tm \mu_{A\times B}},\, \sz{\dr n2-k}\le \dr{n}{14}\rt]$.
\prfend

It remains to see how spectral stability of large rectangles implies hardness of \GHR\ for classical interactive protocols.

\theo[t_GHR_R_lower]{$\R(\GHR) \,\in\, \asOm{n^{\dr13}}$.}

\prfstart[\theoref{t_GHR_R_lower}]
If there is a randomised protocol of cost $c$ that solves \GHR\ with error at most $\dr18$, then for any fixed distribution some choice of the random seed guarantees at most the same error.
So, a deterministic protocol $\Upsilon$ of cost $c$ solves \GHR\ with error at most $\dr18$ when $(\Cl X,\Cl Y)\sim\U[\01^{n+n}]$.
As there are at most $n^{2\log n}$ different possible answers, $\Upsilon$ partitions $\01^{n+n}$ into at most $2^{c+2(\log n)^2}$ rectangles, each marked with some answer value.\fn
{
As the answer length is poly-logarithmic in the case of \GHR, we could assume without loss of generality that the protocol's last message always contains the answer (then a deterministic protocol of cost $c$ would partition the input space into at most $2^c$ rectangles).
We avoid that assumption in order to admit efficient protocols for communication problems with long answers.
}

Let $V_1\sbseq\01^{n+n}$ be the union of the rectangles of $\Upsilon$ whose size is at least $2^{-c-2(\log n)^2-4}$ (with respect to $\U[\01^{n+n}]$), then $\U[\01^{n+n}](V_1)\ge\dr{15}{16}$.
Let $V_2\sbseq\01^{n+n}$ be the union of such rectangles $C\times D$ of $\Upsilon$ that satisfy
\m{
\PR[(\Cl X,\Cl Y)\unin C\times D]{\neg\, \aleph(\Cl X,\Cl Y)}
~\le~ 16\tm \PR[(\Cl X,\Cl Y)\unin\01^{n+n}]{\neg\, \aleph(\Cl X,\Cl Y)}
,}
then $\U[\01^{n+n}](V_2)\ge\dr{15}{16}$ and $\U[\01^{n+n}](V_1\cap V_2)\ge\dr{7}{8}$.
Accordingly,
\m{
\PR[{(\Cl X,\Cl Y)\sim\U[\01^{n+n}]}]
{\txt{$\Upsilon(\Cl X,\Cl Y)$ is right}}
[(\Cl X,\Cl Y)\in V_1\cap V_2]
\ge 1 - \fr[/]{\fr18}{\fr78} = \fr67
.}

This implies existence of a rectangle $A\times B\sbseq V_1\cap V_2$, where, due to~\theoref{t_aleph},
\m{
\PR[(\Cl X,\Cl Y)\unin A\times B]{\neg\, \aleph(\Cl X,\Cl Y)}
\in \asO{\fr{(\ln n)^2}{\sq n}}
.}
Let us denote by ``$((j_1,s_1)\dc(j_{\log n},s_{\log n}))$'' the answer of $\Upsilon$ on $A\times B$:\ we know that it is correct with probability at least $\dr67$; by the definition of \GHR, this means
\m{
\fr67
& \le \PR[{(\Cl X,\Cl Y)\unin A\times B}]
{\sz{\s[:]{i\in[\log n]}
[{\Delta_{j_i,s_i}(\Cl X,\Cl Y)\nin\lf[\fr n2-\fr{\sq n}2,\, \fr n2+\fr{\sq n}2\rt]}]}
\ge\fr{\log n}2}\\
&\tbbb + \PR[(\Cl X,\Cl Y)\unin A\times B]{\neg\, \aleph(\Cl X,\Cl Y)}
.}
For large enough $n$,
\m[P]{
\fr34
& \le \PR[{(\Cl X,\Cl Y)\unin A\times B}]
{\sz{\s[:]{i\in[\log n]}
[{\Delta_{j_i,s_i}(\Cl X,\Cl Y)\nin\lf[\fr n2-\fr{\sq n}2,\, \fr n2+\fr{\sq n}2\rt]}]}
\ge\fr{\log n}2}\\
& \le \fr2{\log n} \tm \E[{(\Cl X,\Cl Y)\unin A\times B}]
{\sz{\s[:]{i\in[\log n]}
[{\Delta_{j_i,s_i}(\Cl X,\Cl Y)\nin\lf[\fr n2-\fr{\sq n}2,\, \fr n2+\fr{\sq n}2\rt]}]}}\\
& = \fr2{\log n} \tm \sum_{i=1}^{\log n}
\PR[{(\Cl X,\Cl Y)\unin A\times B}]
{\Delta_{j_i,s_i}(\Cl X,\Cl Y)\nin\lf[\fr n2-\fr{\sq n}2,\, \fr n2+\fr{\sq n}2\rt]}
,}
and therefore for some $(j_0,s_0)\in[n]\times\01^{\log n}$:
\m{
\PR[{(\Cl X,\Cl Y)\unin A\times B}]
{\sz{\sigma_{j_0}\l(\tau_{s_0}\+\Cl X\r)\+\Cl Y}\nin\lf[\fr n2-\fr{\sq n}2,\, \fr n2+\fr{\sq n}2\rt]}
~\ge~ \fr38
.}

Let $A'\deq\s{\sigma_{j_0}\l(\tau_{s_0}\+a\r)}[a\in A]$, then
\m[m_ApB_cond]{
\PR[{(\Cl X,\Cl Y)\unin A'\times B}]
{\sz{\Cl X\+\Cl Y}\in\lf[\fr n2-\fr{\sq n}2,\, \fr n2+\fr{\sq n}2\rt]}
~\le~ \fr58
}
and, since $\sz{A'}=\sz{A}$ and $A\times B\sbseq V_1$,
\m[m_Pr_ApB]{
\U[\01^{n+n}](A'\times B) \ge 2^{-c-2(\log n)^2-4}
.}

Aiming to benefit from \theoref{t_spec}, now we need a lower bound on
\m{
& \PR[(\Cl X,\Cl Y)\unin\01^{n+n}]
{\sz{\Cl X\+\Cl Y}\in\lf[\fr n2-\fr{\sq n}2,\, \fr n2+\fr{\sq n}2-1\rt]}\\
&\tbbbbb = \PR[\Cl X\unin\01^n]
{\sz{\Cl X}\in\lf[\fr n2-\fr{\sq n}2,\, \fr n2+\fr{\sq n}2-1\rt]}
.}
The bound has to be above $\dr58$, and a typical Chernoff-like estimation would lack the required tightness, so we use \lemref{l_con_bin}:
\m{
\PR[(\Cl X,\Cl Y)\unin\01^{n+n}]
{\sz{\Cl X\+\Cl Y}\in\lf[\fr n2-\fr{\sq n}2,\, \fr n2+\fr{\sq n}2-1\rt]}
\ge \sq{\fr2{\pi}} - \sq{\fr1{18\pi}} - \aso{1}
> \fr{21}{32}
}
for large enough $n$.
Combining this with~\bref{m_ApB_cond} and noting that $\lf[\dr n2-\dr{\sq n}2,\, \dr n2+\dr{\sq n}2-1\rt]$ can be partitioned into pairs $\s{\dr n2-\dr{\sq n}2,\, \dr n2-\dr{\sq n}2+1}\dc \s{\dr n2+\dr{\sq n}2-2,\, \dr n2+\dr{\sq n}2-1}$, we conclude that for some $k\in\s{\dr n2-\dr{\sq n}2\dc \dr n2+\dr{\sq n}2-2}$:
\m{
\fr{\PR[{(\Cl X,\Cl Y)\unin A'\times B}]{\sz{\Cl X\+\Cl Y}\in\s{k,\, k+1}}}
{\PR[(\Cl X,\Cl Y)\unin\01^{n+n}]{\sz{\Cl X\+\Cl Y}\in\s{k,\, k+1}}}
< \fr[/]{\fr58}{\fr{21}{32}}
=\fr{20}{21}
.}

In the terminology of the spectral stability statement (\theoref{t_spec}) this implies
\m{
\fr{\l(\mu_{A'\times B}\r)^{\dr32}}{\sq n}
+ \sz{\fr 12-\fr{k}n} \tm \mu_{A'\times B}
\in \asOm1
,}
that is,
\m{
\log \l( \fr{n}{\U[\01^{n+n}](A'\times B)} \r)
~=~ \mu_{A'\times B}
~\in~ \asOm{n^{\dr13}}
,}
and the required $\lf[c \in \asOm{n^{\dr13}}\rt]$ follows from~\bref{m_Pr_ApB}.
\prfend

\sect[s_concl]{Conclusions}

From \crlref{crl_GHR_Q_upper} and \theoref{t_GHR_R_lower}:
\crl[crl_conc]{
\m{
\QII(\GHR) \,\in\, \asO{\log n}
\tb\txt{and}\tb
\R(\GHR) \,\in\, \asOm{n^{\dr13}}
.}
}

This demonstrates that quantum \SMP, which is the weakest reasonable quantum model, can outperform classical two-way communication, which is the strongest model of feasible classical communication.

What interesting problems are still open?
One possible way to complicate the challenge would be to ask for examples where a quantum model outperforms a substantially reinforced classical one (say, \AM, where protocols can use both randomness and \e{non-determinism}).
Two possible objections come to mind:
First, it seems that most of the known ``supermodels'' either can be handled by, essentially, the same techniques as \R\ (and therefore are of limited theoretical interest), or are still beyond the analytical reach (like \AM) -- in the latter case making an excellent subject of investigation besides the context of quantum mechanics.
Second, comparing a model that cannot be viewed as classically-feasible to a quantum model would lack the philosophical significance and the appeal of reflecting the \e{super-classical} nature of quantum mechanics.\fn
{
These arguments are rather subjective, and intentionally so.
For instance, one could view possible advantage of quantum two-way communication over \AM\ as a very natural indirect comparison between the quantum and the classical worlds (as \AM\ is known to be stronger than \R); on the other hand, comparing \QII\ to \R\ can be viewed as too indirect and too unnatural, as it mixes very different communication layouts.
}

We think that the most interesting remaining problem in this area is to analyse the possibility of outperforming classical communication by quantum \SMP\ on \e{more elementary levels} than relational problems.
Recall the hierarchy of \e{problem types} that we mentioned earlier:\ total functions, partial functions and relations.
Obviously, in this context total functions are more elementary than partial functions, while relations are the most sophisticated.
Viewed as \e{tools for demonstrating the advantage of one communication model over another}, total functions are the weakest and relations are the strongest -- accordingly, the arguments based on total functions are the hardest to obtain (maybe even impossible in some cases), while the arguments based on relations (like the one presented in this work) are the easiest.

In terms of the \e{known} super-classical merits of \QII, the three types indeed form a hierarchy:
\itstart
\item A \e{total function} is known that has an efficient \QII-protocol, but no efficient classical simultaneous-messages protocol without shared randomness~\cite{BCWW01_Qua}.\fn
{
This is the only known example of qualitative advantage of quantum communication for a total function.
}
\item A \e{partial function} is known that has an efficient \QII-protocol, but no efficient classical simultaneous-messages protocol, even with shared randomness~\cite{G19_Qua}.
\item Our \e{relation} \GHR\ has an efficient \QII-protocol, but no efficient classical two-way protocol.
\itend
Can the first two positions be improved?
Namely:
\itstart
\item Can a quantum simultaneous-messages protocol outperform classical simultaneous-messages protocols with shared randomness over a total function?
\item Can a quantum simultaneous-messages protocol outperform classical one-way protocols over a partial function?
\itend

\toct{Acknowledgements}

\sect*{Acknowledgements}

I am very grateful to Ronald de Wolf for his helpful comments.
A number of very useful suggestions have been received from anonymous reviewers.

\toct{References}

\newcommand{\etalchar}[1]{$^{#1}$}

\end{document}